\documentclass[9pt,twocolumn,twoside]{IEEEtran}
\pdfoutput=1
 \newcommand{\DRAFT}[1]{}

 \usepackage{amsmath,amssymb}
\usepackage{cite}
 \usepackage{graphicx}
 \usepackage{pstricks}
 \usepackage{psfrag}
 \usepackage{footmisc}
\usepackage[linesnumbered]{algorithm2e} 
\usepackage{xspace,bbm}
\input alphabet
\def\norm#1{\left\|#1\right\|}

\def\AND{\text{and\:}}           
\def\OR{\text{or\:}}

\newsavebox{\fminibox}
\newlength{\fminilength}

  \def\+{^\dagger}

\def\nequiv{\not\kern-.05em\equiv}
\def\egal{\kern-.5em=\kern-.5em}        
\def\propt{\kern-.2em\propto\kern-.2em} 
\def\wh#1{\widehat{#1}}                 

\def\argmin{\mathop{\mathrm{arg\,min}}} 

\def\intdouble{\int\kern-0.3em\int}
\def\inttriple{\int\kern-0.3em\int\kern-0.3em\int}

\def\rond#1{\overset{\kern-0.33em~_\circ}{#1}}
\def\rondit[#1]#2{\overset{\kern#1~_\circ}{#2}}

 \def\pardef{:=}
 \def\ie{\textit{i.e.}\XS}

\def\FFT{FFT\XS}

\def\prox#1{\Pc_{#1}}

\usepackage{rotating}
\def\cc#1{\setlength{\tabcolsep}{0pt}\begin{tabular}{c}#1\end{tabular}}
\newcommand{\figc}[2][]
   {\setlength{\tabcolsep}{0pt}\begin{tabular}{c}\includegraphics[#1]{#2}\end{tabular}}
\def\ylabel#1{\cc{\begin{sideways}#1\end{sideways}}}

\usepackage{color}

\begin{document}

\title{Joint reconstruction strategy for structured illumination microscopy with unknown illuminations}

\author{\textbf{\today} -
  Simon~Labouesse, \IEEEmembership{}
  Awoke~Negash, \IEEEmembership{}
  J\'er\^ome~Idier, \IEEEmembership{Member,~IEEE, }
  S\'ebastien~Bourguignon, \IEEEmembership{}
  Thomas Mangeat, \IEEEmembership{}
  Penghuan~Liu, \IEEEmembership{}
  Anne~Sentenac, \IEEEmembership{}
  and~Marc~Allain\IEEEmembership{}
  \thanks{The authors acknowledge partial financial support for this
    paper from the GdR 720 ISIS and the Agence Nationale de la Recherche 
    (ANR-12-BS03-0006).}
  \thanks{S. Labouesse, M. Allain, A. Negash  and A. Sentenac are with 
    Universit\'e Aix-Marseille, Centrale Marseille and the CNRS at Institut Fresnel (CNRS UMR 7249) 
    Campus de St J\'er\^ome, F-13013 Marseille, France. E-mail:
    firstname.name@fresnel.fr.}
  \thanks{J.~Idier, S.~Bourguignon and P.~Liu are with \'Ecole Centrale de Nantes 
    and CNRS at the Laboratoire des Sciences du Numérique de Nantes (LS2N, CNRS UMR 6004), 
    \DRAFT{1 rue de la No\"e, BP 92101,}%
    F-44321 Nantes, France.
    \DRAFT{Tel: (+33)-2 40 37 69 09, Fax: (+33)-2 40 37 69 30.}%
    E-mail: firstname.name@irccyn.ec-nantes.fr.}
  \thanks{{%
      T. Mangeat is with LBCMCP, Centre de Biologie Int\'egrative 
      (CBI), Universit\'e de Toulouse, CNRS, UPS, France. 
      E-mail: thomas.mangeat@univ-tlse3.fr.}}
}

\maketitle

\begin{abstract} 
The \textit{blind} structured illumination microscopy (SIM) 
strategy proposed in \cite{Mudry12} is fully re-founded in this paper,  
unveiling the central role of the sparsity of the illumination patterns
in the mechanism that drives super-resolution in the
method. A numerical analysis shows that the resolving power 
of the method can be further enhanced with optimized one-photon 
or two-photon speckle illuminations.   
A much improved numerical implementation is provided for 
the reconstruction problem under the image positivity constraint. This 
algorithm rests on a new preconditioned proximal iteration faster than 
existing solutions, paving the way to 3D and real-time 2D reconstruction. 
\end{abstract}


\section{Introduction}
In Structured Illumination Microscopy (SIM),  the sample, characterized by
its fluorescence density $\rho$, is illuminated successively by $M$ 
distinct inhomogeneous illuminations $I_m$. Fluorescence
light emitted by the sample is collected by a microscope objective 
and recorded on a camera to form an image $y_m$. In the linear regime, 
and with a high photon counting rate\footnote{{In practice, the photon counting 
rate is expected to be high enough so that this fluctuation source
behaves as an additive  Gaussian process. Measurements plagued 
by low photon counting  rates can nevertheless be addressed within a
statistical framework, by replacing the usual least-squares fitting 
function \eqref{critere2a} by the Poissonian neg-log likelihood 
function instead, see for instance \cite[Chap. 6]{Bertero98}.   
}}, the dataset $\{y_m\}_{m=1}^M$
is related to the sample $\rho$ \textit{via} 
\cite{Goodman05}
\begin{equation}
  y_m  =  \Hc  \otimes (\rho \times I_m) + \varepsilon_m, \qquad m = 1\cdots M,
  \label{eq:observation}
\end{equation}
where $\otimes$ is the convolution operator, $\Hc$ is the microscope
point spread function (PSF) and $\varepsilon_m$ is a perturbation term accounting for (electronic) noise
in the detection and modeling errors. 
Since the spatial spectrum of the PSF [\textit{i.e.,} the optical transfer function (OTF)] is strictly bounded 
by its cut-off frequency, say, $\nu_{\text{psf}}$, if the illumination
pattern $I_m$ is homogeneous, then the spatial spectrum of $\rho$ that
can be retrieved from the image $y_m$ is restricted to frequencies
below  $\nu_{\text{psf}}$.
When the illuminations are inhomogeneous, frequencies beyond
$\nu_{\text{psf}}$ can be recovered from the low resolution images because
the illuminations, acting as carrier waves, downshift part of the
spectrum inside the OTF support \cite{Heintzmann99,Gustafsson00}.
Standard SIM  resorts to harmonic illumination patterns for which the 
reconstruction of the super-resolved image can be easily done by
solving a linear system in the Fourier domain. In this case, the gain
in resolution depends on the OTF support, the illumination 
cut-off frequency and the available signal-to-noise ratio (SNR).
The main drawback of SIM is that it requires the knowledge of the 
illumination patterns and thus a stringent control of the experimental
setup. If these patterns are not known with sufficient 
accuracy~\cite{Mudry12,Ayuk13}, severe artifacts appear in 
the reconstruction. Specific estimation techniques have been developed for 
retrieving the parameters of the periodic patterns from the images \cite{Orieux12,Wicker13a,Wicker13b}, 
but they can fail if the SNR is too low or if the excitation patterns are 
distorted, \textit{e.g.,} by inhomogeneities in the sample refraction index. 
The Blind-SIM strategy~\cite{Mudry12,Ayuk13,Negash16} has been proposed 
to tackle this key issue, the principle being to retrieve the sample
fluorescence density without the knowledge of the illumination patterns. 
In addition, {\it speckle} illumination patterns are promoted instead of 
harmonic ones, the latter being much more difficult 
to generate and control.   
From the methodological viewpoint, this strategy relies on the \textit{simultaneous} 
(joint) reconstruction  of the fluorescence density and of the illumination patterns. 
More precisely, joint reconstruction is achieved through the iterative resolution of
a constrained least-squares problem. However, the computational time of such a scheme clearly restricts 
the applicability of the method. 

{\indent
This paper  provides a global re-foundation of the joint Blind-SIM
strategy. More specifically, our work develops two specific, 
yet complementary, contributions:}


\begin{itemize}
\item {%
The joint Blind-SIM reconstruction problem is first revisited, 
resulting in an improved numerical implementation with  execution  
times decreased by several orders of magnitude. Such an 
acceleration relies on two technical contributions. Firstly, 
we show that the problem proposed in \cite{Mudry12} is equivalent 
to a \textit{fully  separable} constrained minimization problem,
hence bringing the original (large-scale) problem to $M$ 
sub-problems with smaller scales. Then, we introduce a new 
\textit{preconditioned proximal iteration} (denoted PPDS) to 
efficiently solve each sub-problem. The PPDS strategy is an important 
contribution of this article: it is provably convergent~\cite{Condat13}, easy to implement and, for our specific problem, 
we empirically observe a super-linear asymptotic convergence rate. With these elements, the joint Blind-SIM 
reconstruction proposed in this paper is fast and can be highly parallelized, 
opening the way to real-time reconstructions. 
}
\\

\item 
{%
Beside these algorithmic issues, the mechanism driving 
super-resolution (SR) in this blind context is investigated, 
and a connection is established with the well-known 
``Near-black object'' effect introduced in Donoho's seminal  
contribution \cite{Donoho92a}.  We show that the SR relies on  sparsity and  positivity constraints 
enforced by the unknown illumination patterns. 
This finding helps to understand in which situation 
super-resolved reconstructions may be provided or not. A 
significant part of this work is 
then dedicated to numerical simulations aiming at 
illustrating how the SR effect can be enhanced. In this perspective,
our simulations show that two-photon speckle illuminations 
potentially increase the SR power of the proposed method.} 
%
%
\end{itemize}

{The pivotal role played by  sparse 
illuminations in this SR mechanism also draws a connexion between joint 
Blind-SIM and other \textit{random activation} strategies like 
PALM \cite{betzig06} or STORM \cite{rust06}; see also 
\cite{Mukamel12,Min14} for explicit sparsity methods applied to
STORM. With PALM/STORM, unparalleled 
resolutions result from an activation
process that is massively sparse and mostly localized on the marked 
structures. With the joint Blind-SIM strategy, the illumination
pattern playing the role of the activation process is not that 
``efficient'' and lower  resolutions are obviously expected. 
} 
{Joint Blind-SIM however provides SR as long as
  the illumination patterns enforce many zero (or almost zero)  
  values in the product $\rho \times I_m$: the sparser the 
  illuminations, the higher the expected resolution gain with joint
  Blind-SIM. Such super resolution can be induced by either deterministic 
  or random patterns. Let us mention that random illuminations are easy
  and cheap to generate, and that a 
  few recent contributions advocate the use of speckle illuminations 
  for super-resolved imaging, either in fluorescence \cite{Min13,Oh13} or 
  in photo-acoustic \cite{Chaigne16} microscopy.
}  
%
In these contributions, however, the reconstruction strategies 
are derived from the statistical modeling of the speckle, hence, 
relying on the random character of the illumination patterns. In
comparison, our approach only requires  that the illuminations 
cancel-out the fluorescent object and that their sum is known with sufficient accuracy. 
Finally, we also note that \cite{labouesse:hal-01264977} corresponds to an early version of this work.
Compared to \cite{labouesse:hal-01264977}, several important contributions are presented here, 
mainly: the super-resolving power of Blind-SIM is now studied 
in details, and a comprehensive presentation of the proposed 
PPDS algorithm includes a tuning strategy for the algorithm parameter that 
allows a substantial reduction of the computation time.

The remainder of the paper is organized as follows. In Section~\ref{Reformulation}, the original Blind-SIM 
formulation is introduced and further simplified; this reformulation is then used 
to get some insight on the mechanism that drives the SR
in the method. Taking advantage of this analysis, a penalized Blind-SIM
strategy is proposed and characterized with synthetic data in
Section~\ref{penalize}. Finally, the PPDS algorithm developed to cope 
with the minimization problem is presented and tested in 
Section~\ref{algo}, and conclusions are drawn in Section~\ref{conclusion}.

\section{Super-resolution with joint Blind-SIM estimation}
\label{Reformulation}
In the sequel, we focus on a discretized formulation of the observation model 
\eqref{eq:observation}.
Solving the two-dimensional (2D) Blind-SIM reconstruction 
problem is equivalent to finding a \textit{joint} solution  $(\widehat\rhob,\{\widehat{\Ib}_m\}_{m=1}^M)$ to the
following constrained minimization problem~\cite{Mudry12}:
\begin{subequations}
  \label{critere1}
  \begin{align}
  \label{critere1a}
   \min_{\rhob ,\{\Ib_m\}} \quad  { \textstyle \sum_{m=1}^{M}}\, \left\| \yb_m -  \Hb\textbf{diag}(\rhob)\,\Ib_m\right\|^2&&\\[.3em]
  \label{critere1b}
   \text{subject to}   \qquad {\textstyle \sum_m} \,\Ib_{m} \, =  \, M
   \times \Ib_{0} \qquad \qquad\\[.3em]
   \label{critere1c}
    \text{and}\quad\, \rho_n \geq 0, \quad I_{m;n} \geq 0, \qquad  \forall m,n 
  \end{align}
\end{subequations}
with $\Hb\in\eR^{P\times N}$ the 2D convolution matrix built from the
discretized PSF. We also denote $\rhob=\textbf{vect}(\rho_n)\in \eR^N$
the discretized fluorescence density,  $\yb_m=\textbf{vect} (y_{m;n})\in \eR^P$ 
the $m$-th recorded image, and  $\Ib_m=\textbf{vect} (I_{m;n})\in \eR^N$ the $m$-th
illumination with expected spatial intensity  $\Ib_0=\textbf{vect} (I_{0;n})\in
\eR_+^{N}$ (this latter quantity may be spatially inhomogeneous but it 
is supposed  to be known). 
Let us remark that \eqref{critere1} is a \textit{biquadratic}
problem. Block coordinate descent alternating between the object 
and the  illuminations could be a possible minimization strategy, 
relying on cyclically solving $M+1$ quadratic programming 
problems \cite{Jost15}. In~\cite{Mudry12}, a more efficient but 
more complex scheme is proposed. However, the minimization 
problem \eqref{critere1} has a very specific structure, yielding 
a fast and simple strategy, as shown below.

\subsection{Reformulation of the optimization problem}
\label{Reformulationbis}
According to \cite{labouesse:hal-01264977}, let us first consider problem \eqref{critere1} without the equality
constraint \eqref{critere1b}. It is equivalent to $M$ independent quadratic minimization problems:
\begin{subequations}
  \label{critere2}
  \begin{align}
    \label{critere2a}
    {\textstyle\min_{\qb_m}} 		\left\| \yb_m -  \Hb\qb_m\right\|^2 &&\\[0em]
    \label{critere2b}
    \qquad\qquad \text{subject to} \quad \qb_m\geq 0, \quad
  \end{align}
\end{subequations}
where we set $\qb_m \pardef \textbf{vect}(\rho_n \times I_{m;n})$.
Each minimization problem \eqref{critere2} can 
be solved in a simple and efficient way 
(see Sec.~\ref{algo}), hence providing a set of global minimizers 
$\{\widehat{\qb}_m\}_{m=1}^M$. Although the latter set corresponds 
to an infinite number of solutions
$(\widehat\rhob,\{\widehat\Ib_m\}_{m=1}^M)$, 
the equality constraint  \eqref{critere1b} defines a unique solution such that 
$\widehat{\qb}_{m}= \textbf{vect}(\widehat{\rho}_n\times\widehat{I}_{m;n})$ 
for all $m$:
\begin{subequations}
\begin{align}
  \label{solutionq}
   \widehat{\rhob}  &\,= \,\textbf{Diag} (\Ib_0)^{-1}
  \, \overline{\qb} &\\
  \forall m \qquad \widehat{\Ib}_{m} &\,=\,
  \textbf{Diag} (\widehat{\rhob})^{-1}\, \widehat{\qb}_{m} %
  \label{solutionI}
\end{align}
\label{solution}
\end{subequations}
with $\overline{\qb} \pardef   {\textstyle\frac{1}{M}     \sum_m} \, \widehat{\qb}_{m}$. 
The solution \eqref{solution} exists as long as $I_{0;n} \neq 0$ 
and  $\widehat{\rho}_n \neq 0$, $\forall n$.  The first condition
is met if the sample is illuminated everywhere (in average), which 
is an obvious minimal requirement. For any pixel sample such that $\widehat{\rho}_{n} = 0$, 
the corresponding illumination $\widehat{I}_{m;n}$ is not defined;
this is not a problem as long as the fluorescence density $\rhob$
is the only quantity of interest.
Let us also note that the following implication holds: 
\begin{align*}
  \nonumber
  I_{0;n} \geq 0,   ~\widehat{q}_{m;n}\geq 0
  \quad \Longrightarrow \quad \widehat{I}_{m,n}\geq 0  \quad
  \text{and} \quad    \widehat{\rho}_n \geq 0.
\end{align*}
Because we are dealing with intensity patterns, the condition
$I_{0;n}\geq 0$ is always met, hence  the positivity
granted for both the density and the illumination estimates, \textit{i.e.,} the positivity constraint 
\eqref{critere1c}, is granted by \eqref{solution}. Indeed, it should be
clear that combining \eqref{critere2} and \eqref{solution} solves the original minimization 
problem \eqref{critere1}: on the one hand, the equality constraint \eqref{critere1b} 
is met since%
\footnote{Whenever $\widehat{\rho}_n=0$, the corresponding entry in the
  illumination pattern estimates \eqref{solutionI} can be set to $\widehat{I}_{m;n}=I_{0;n}/M$ 
for all $m$, hence preserving the positivity \eqref{critere1c} and the
constraint \eqref{critere1b}.}
\begin{equation}
  \label{contrainte_verifiee}
  {\textstyle \sum_m} \widehat{\Ib}_{m} = \textbf{Diag}(\widehat{\rhob})^{-1}\,
  \overline{\qb} = M \Ib_0
\end{equation}
and on the other hand, the solution \eqref{solution} minimizes the
criterion given in \eqref{critere1a} since it is built from $\{\widehat{\qb}_m \}_{m=1}^M$, 
which minimizes \eqref{critere2a}.
Finally, it is worth noting that the constrained minimization problem
\eqref{critere1} may have multiple solutions. In our reformulation,
this ambiguity issue arises in the ``minimization step''~\eqref{critere2}:
while each problem~\eqref{critere2} is convex quadratic,
and thus admits only global solutions (which in turn provide a global 
solution to problem~\eqref{critere1} when recombined according 
to~\eqref{solutionq}-\eqref{solutionI}), it may not admit \textit{unique} solutions since each criterion 
\eqref{critere2a} is not strictly convex%
\footnote{%
A constrained quadratic problem such as \eqref{critere2} is strictly convex if
and only if the matrix $\Hb$ is full rank. In our case, however, $\Hb$ is
rank deficient since its spectrum is the OTF that is strictly support-limited.} in $\qb_m$. 
Furthermore, the positivity constraint \eqref{critere2b} prevents any
direct analysis of these ambiguities. The next subsection underlines 
however the central role of this constraint in the joint
Blind-SIM strategy originally proposed in~\cite{Mudry12}.    
\begin{figure}[t]
  \centering
  \begin{tabular}{@{}l@{}c@{\kern.2cm}c}
(A)&      \begin{tabular}{c}
        \includegraphics[width=3.7cm]{
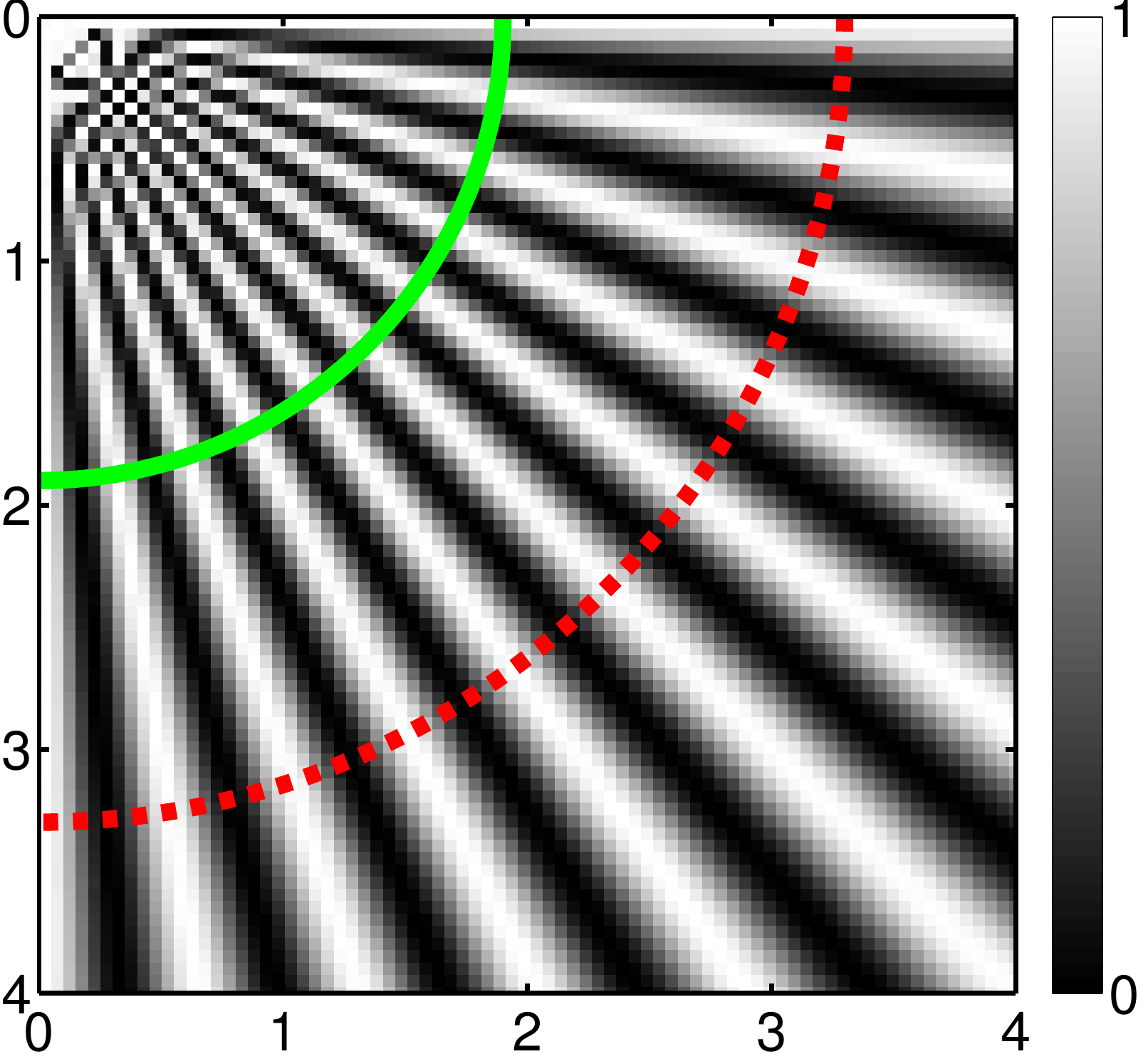}       
      \end{tabular}
    &
      \begin{tabular}{c}
        \includegraphics[width=3.8cm]{
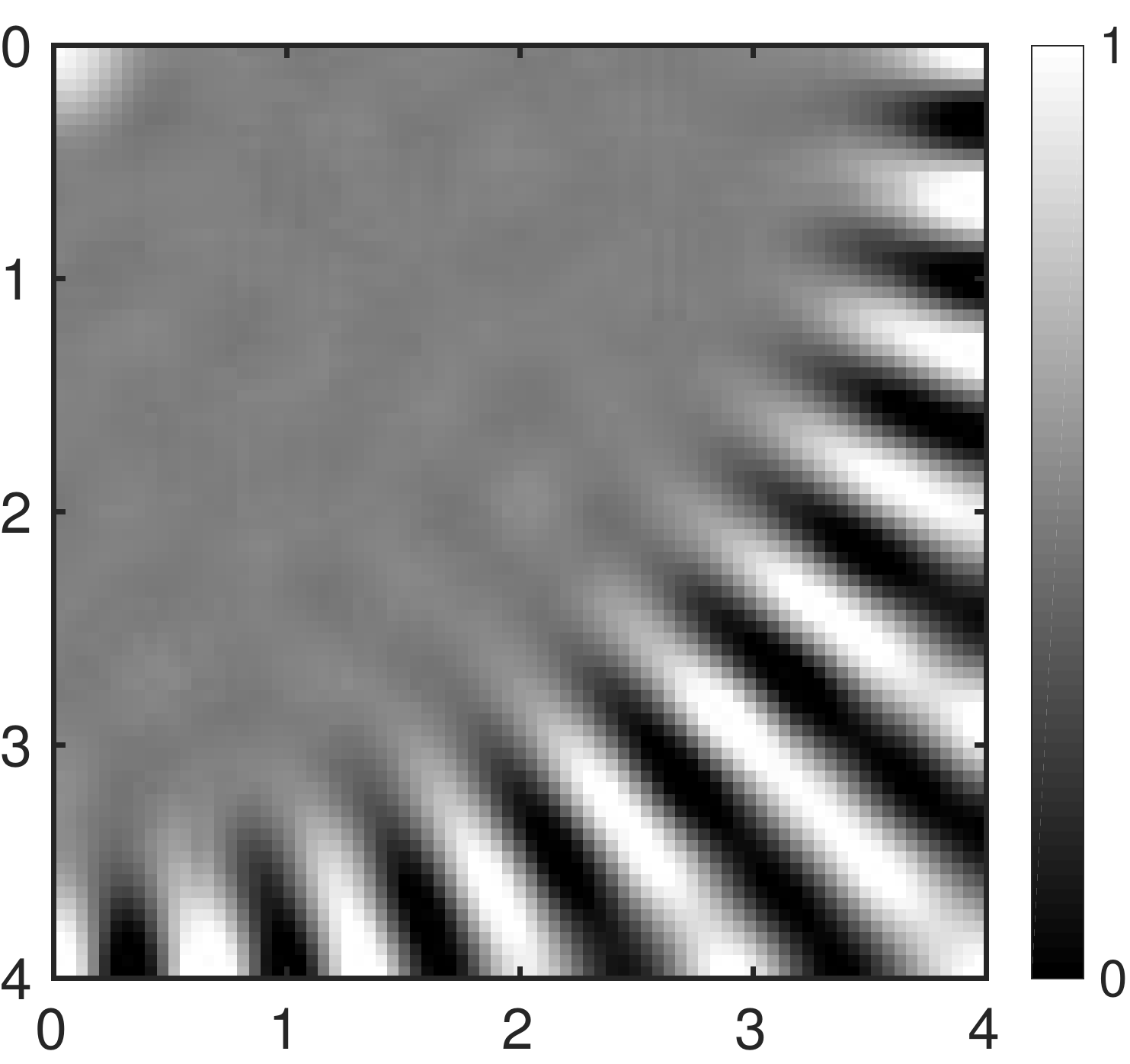}
      \end{tabular}\\[.5em]
  (B)&    \begin{tabular}{c}
        \includegraphics[width=3.8cm]{
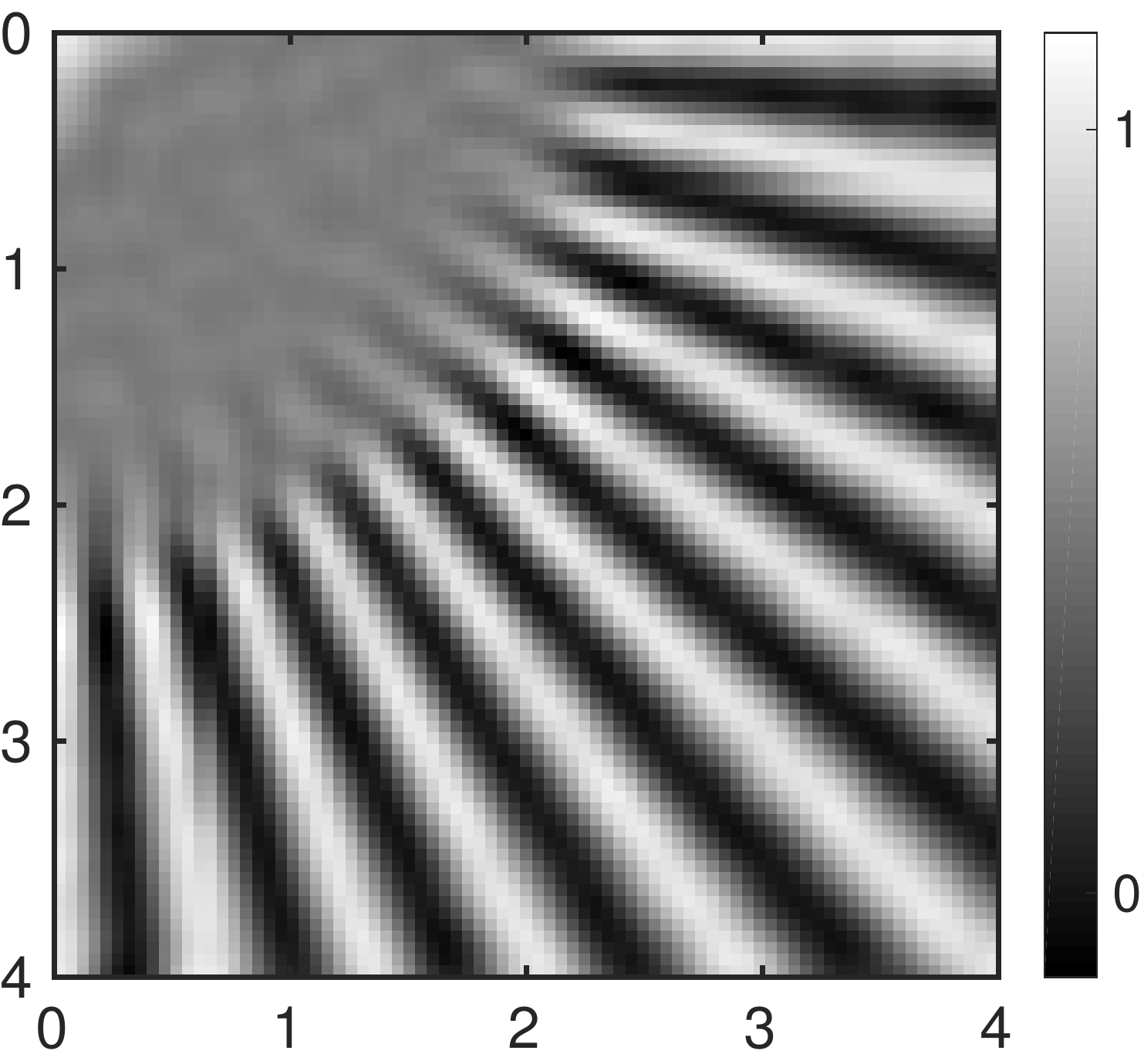}
     \end{tabular}
      &
      \begin{tabular}{c}
        \includegraphics[width=3.8cm]{
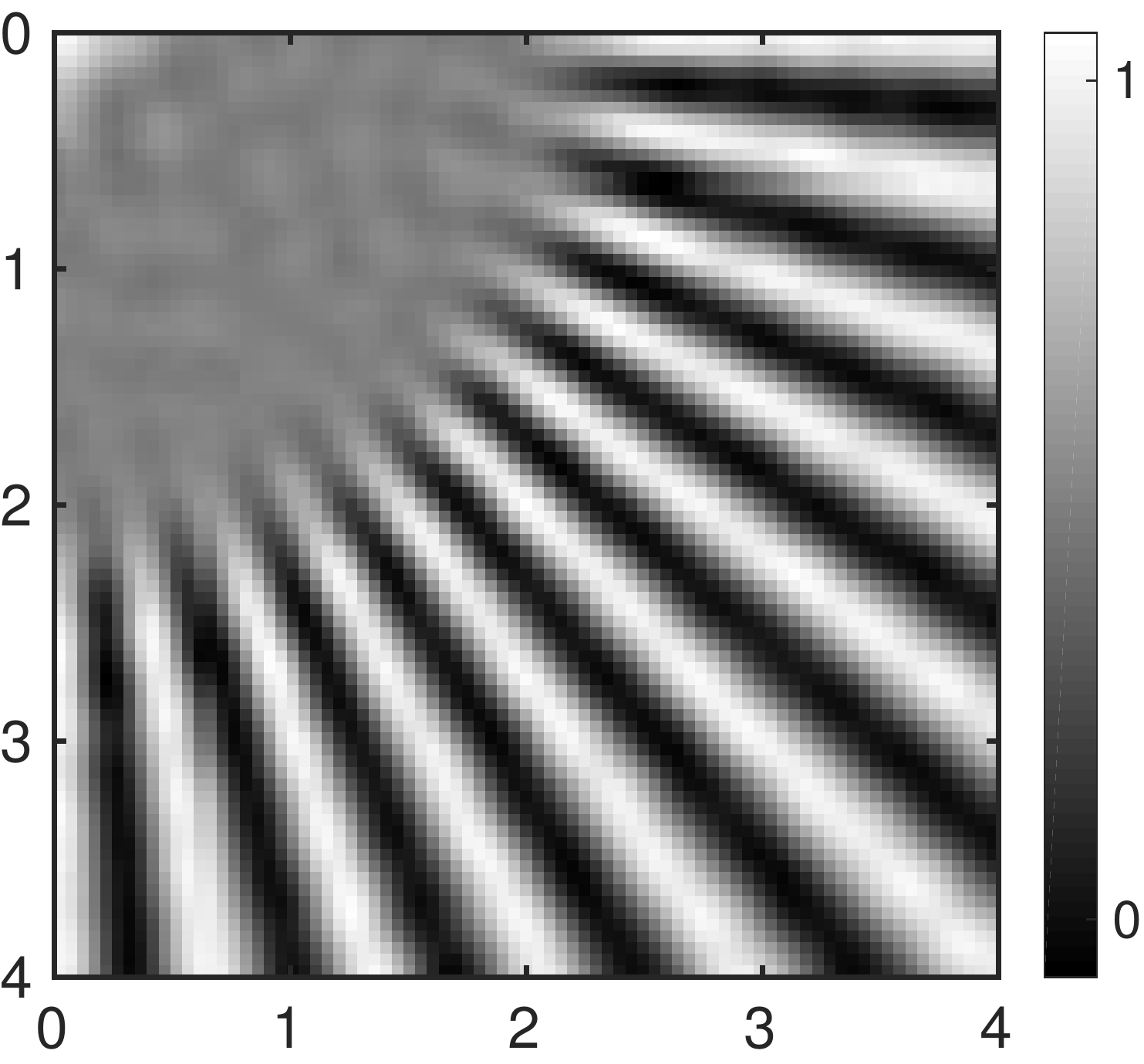}
     \end{tabular}
\end{tabular}
  \caption{[\textit{Row A}] Lower-right quarter of the (160$\times$160
    pixels) ground-truth fluorescence pattern considered in \cite{Mudry12} (left)
    and deconvolution of the corresponding wide-field image
    (right). The dashed (resp. solid) lines corresponds to the spatial 
    frequencies transmitted by the OTF support (resp. twice the OTF
    support). [\textit{Row B}] Positivity-constrained reconstruction
    from \textit{known} illumination patterns: (left) $M=9$ harmonic 
    patterns and (right) $M=200$ speckle patterns. The distance
    units along the horizontal and vertical axes are given in
    wavelength $\lambda$. }
  \label{fig:fig1}
\end{figure}

\subsection{Super-resolution unveiled}
\label{SR}

Whereas the mechanism that conveys SR with \textit{known}
structured illuminations is well understood (see~\cite{Gustafsson00}   for instance), the 
SR capacity of joint blind-SIM has not been characterized yet.
It can be made clear, however, that the positivity constraint \eqref{critere1c} 
plays a central role in this regard. 
Let $\Hb^+$ be the \textit{pseudo-inverse} of  $\Hb$  
\cite[Sec. 5.5.4]{Golub96}. Then, any solution to the 
problem \eqref{critere1a}-\eqref{critere1b}, \textit{i.e,} without positivity constraints,
 reads
\vspace{.5em}
\begin{subequations}
  \label{solutionFourier}
    \begin{align}
      \label{solutionFouriera}
      \widehat{\rhob} &=\textbf{Diag}(\Ib_0)^{-1} (\Hb^+ \overline{\yb} + \overline{\qb}^\perp ) \\[0em]
      \label{solutionFourierb}
      \widehat{\Ib}_m &=\textbf{Diag}(\widehat{\rhob})^{-1} (\Hb^+ \yb_m + \qb_m^\perp),  
    \end{align}
\end{subequations}
with 
$%
{\textstyle
\overline{\yb} = \frac{1}{M} \sum_m \yb_m,\quad  \text{and} \quad \overline{\qb}^\perp = \frac{1}{M} \sum_m }\qb_m^\perp
$
where $\qb_m^\perp$ is an arbitrary element of the kernel of 
$\Hb$, \ie with arbitrary frequency components above the OTF cutoff frequency.
Hence, the formulation \eqref{critere1a}-\eqref{critere1b} has 
no capacity to discriminate the correct high frequency components, 
which means that it has no SR capacity.
Under the positivity constraint \eqref{critere1c}, we thus expect 
that the SR mechanism rests on the fact that
each illumination pattern $\Ib_m$ activates the positivity  constraint
on $\qb_m$ in a frequent manner. 
%
%
%
%
\begin{figure}[t]
  \centering
  \begin{tabular}{@{}l@{}l@{}l@{}}
    (A)
    &
    \begin{tabular}{l}
      \includegraphics[height=0.19\textwidth]{
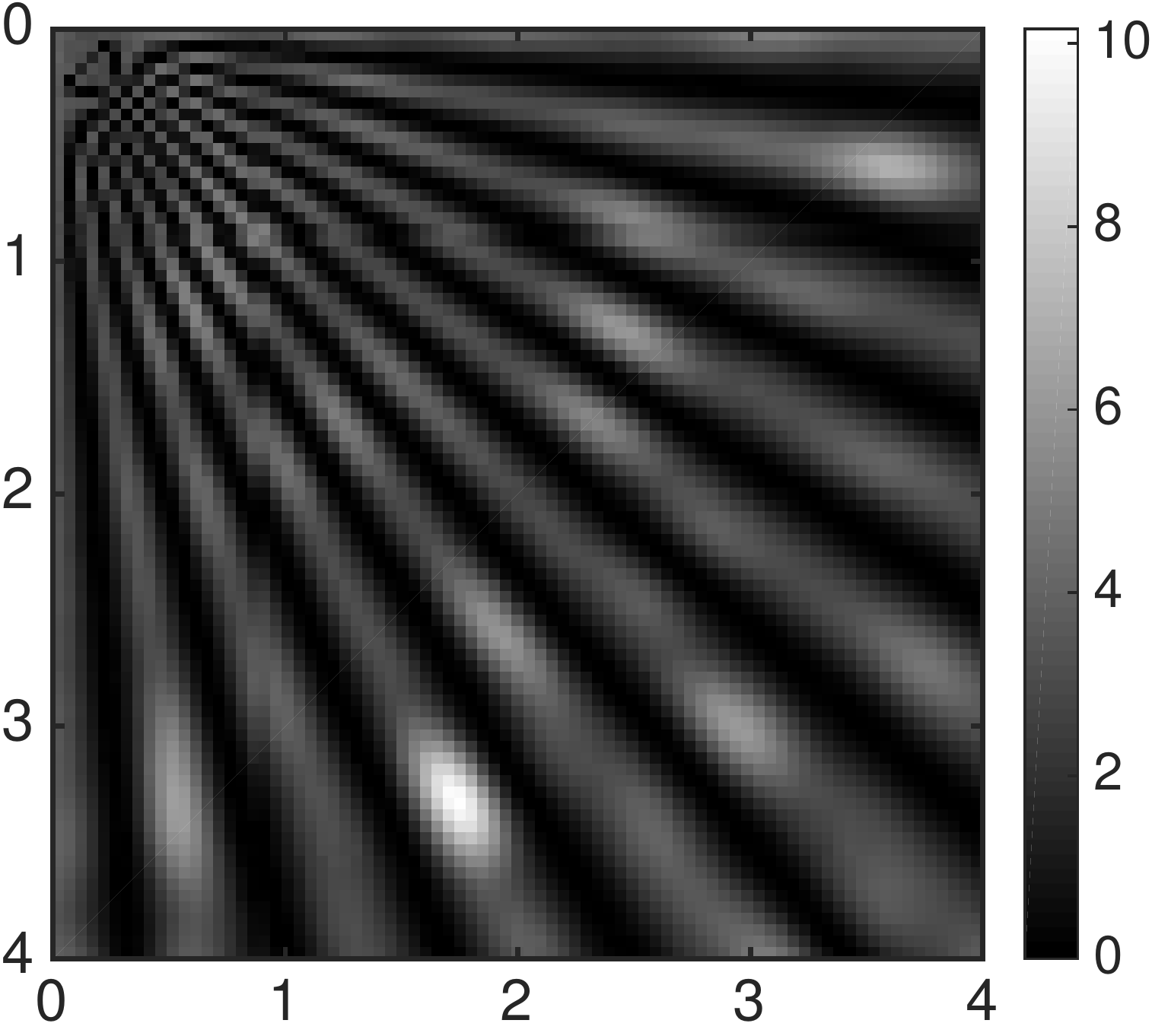}
    \end{tabular}
    &
      \begin{tabular}{l}
        \includegraphics[height=0.19\textwidth]{
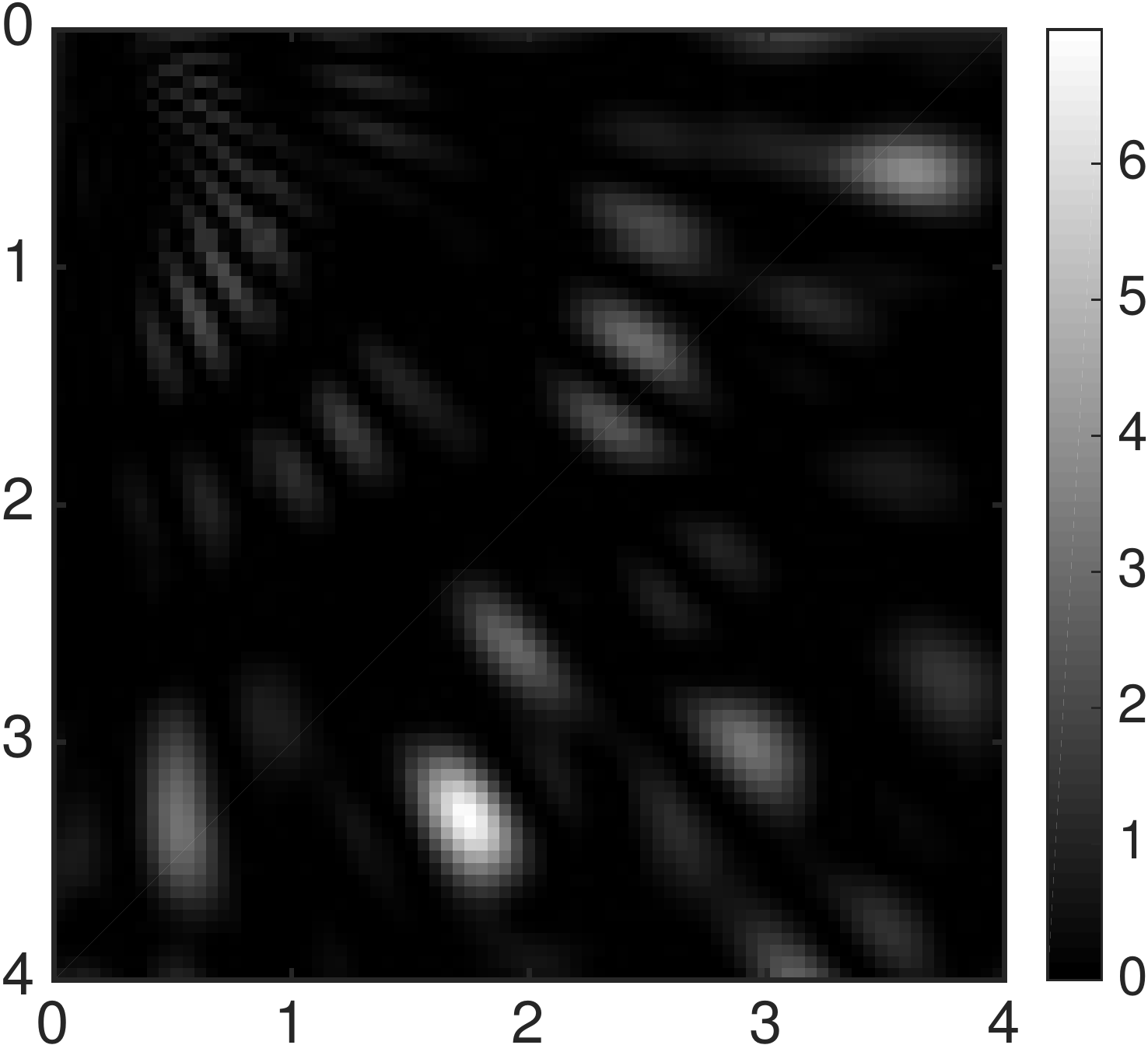}
      \end{tabular}
    \\[1em]
    (B)
    &
    \begin{tabular}{l}
      \includegraphics[height=0.19\textwidth]{
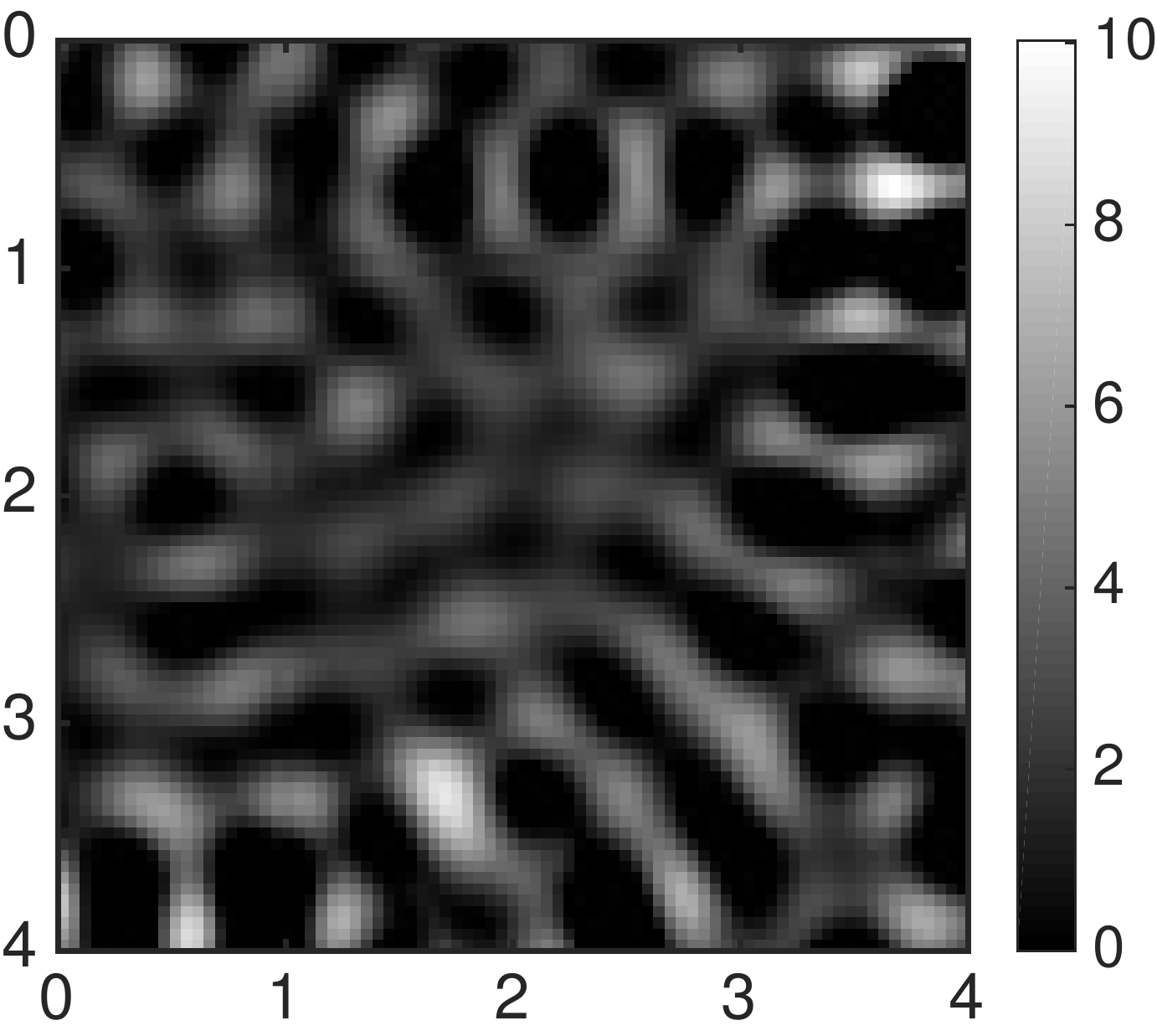}
    \end{tabular}
    &
      \begin{tabular}{l}
        \includegraphics[height=0.19\textwidth]{
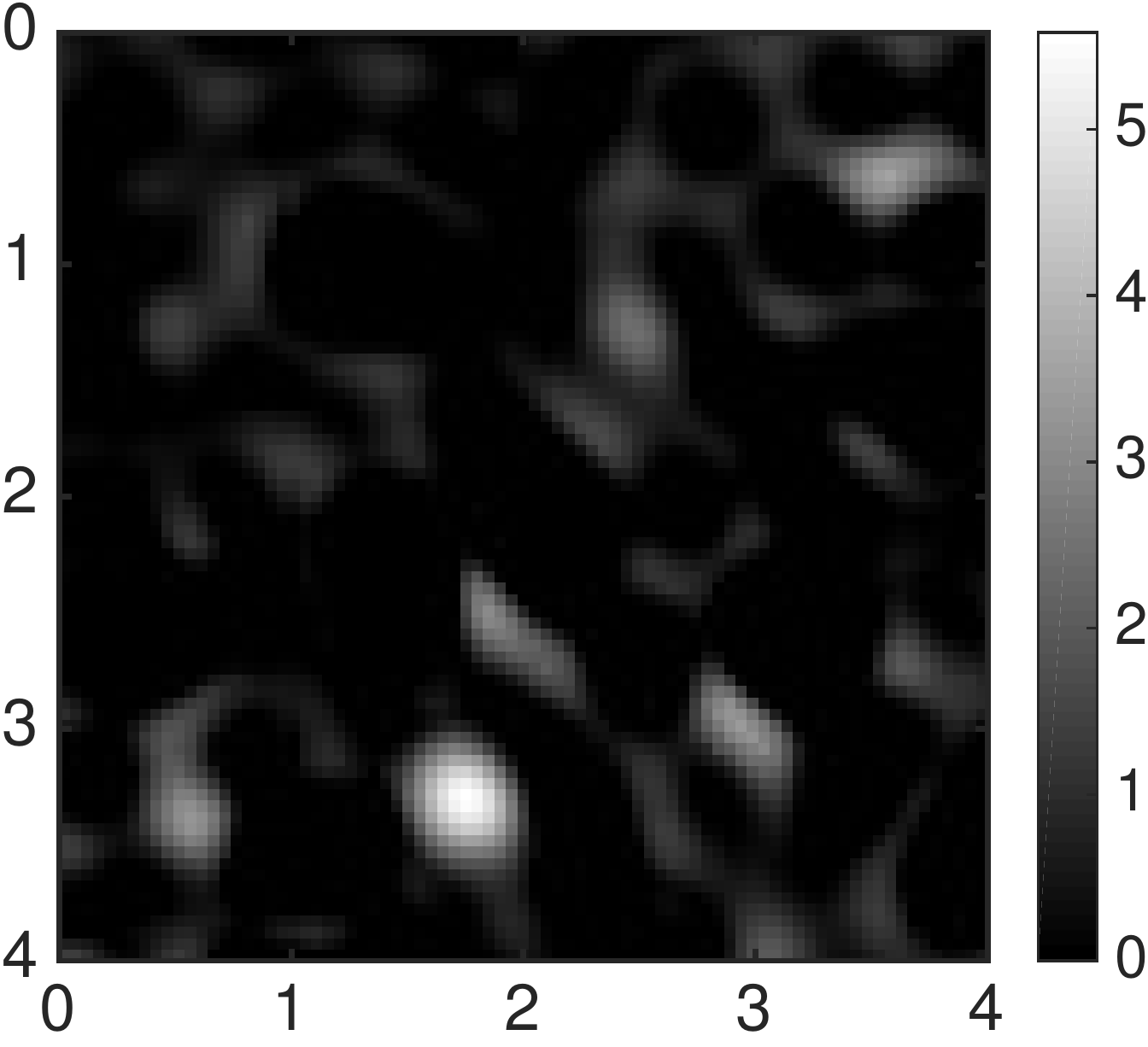}
      \end{tabular}
    \\[1em]
    (C)
    &
    \begin{tabular}{l}      
      \includegraphics[height=0.19\textwidth]{
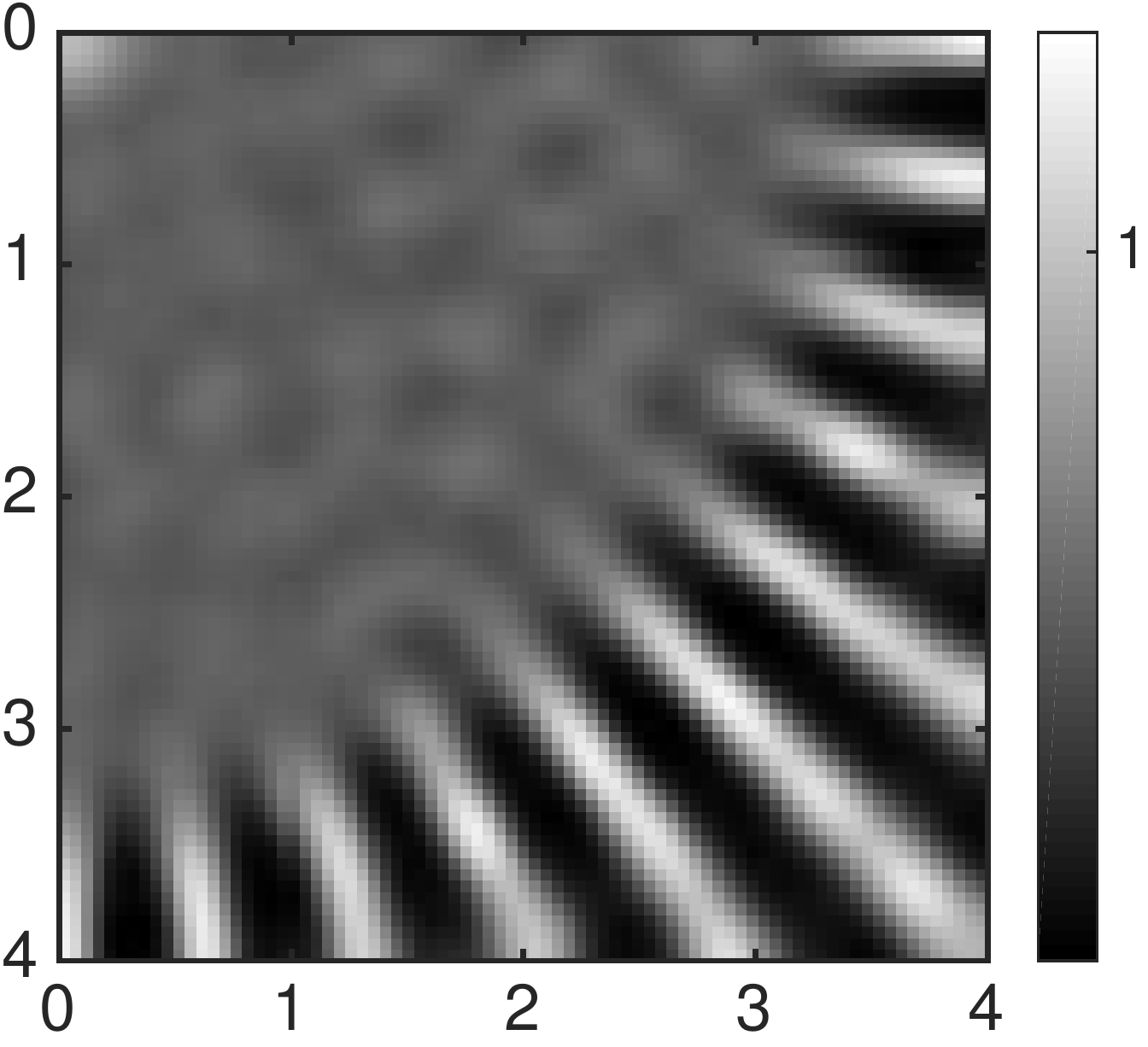}
    \end{tabular}
    &
      \begin{tabular}{l}      
        \includegraphics[height=0.19\textwidth]{
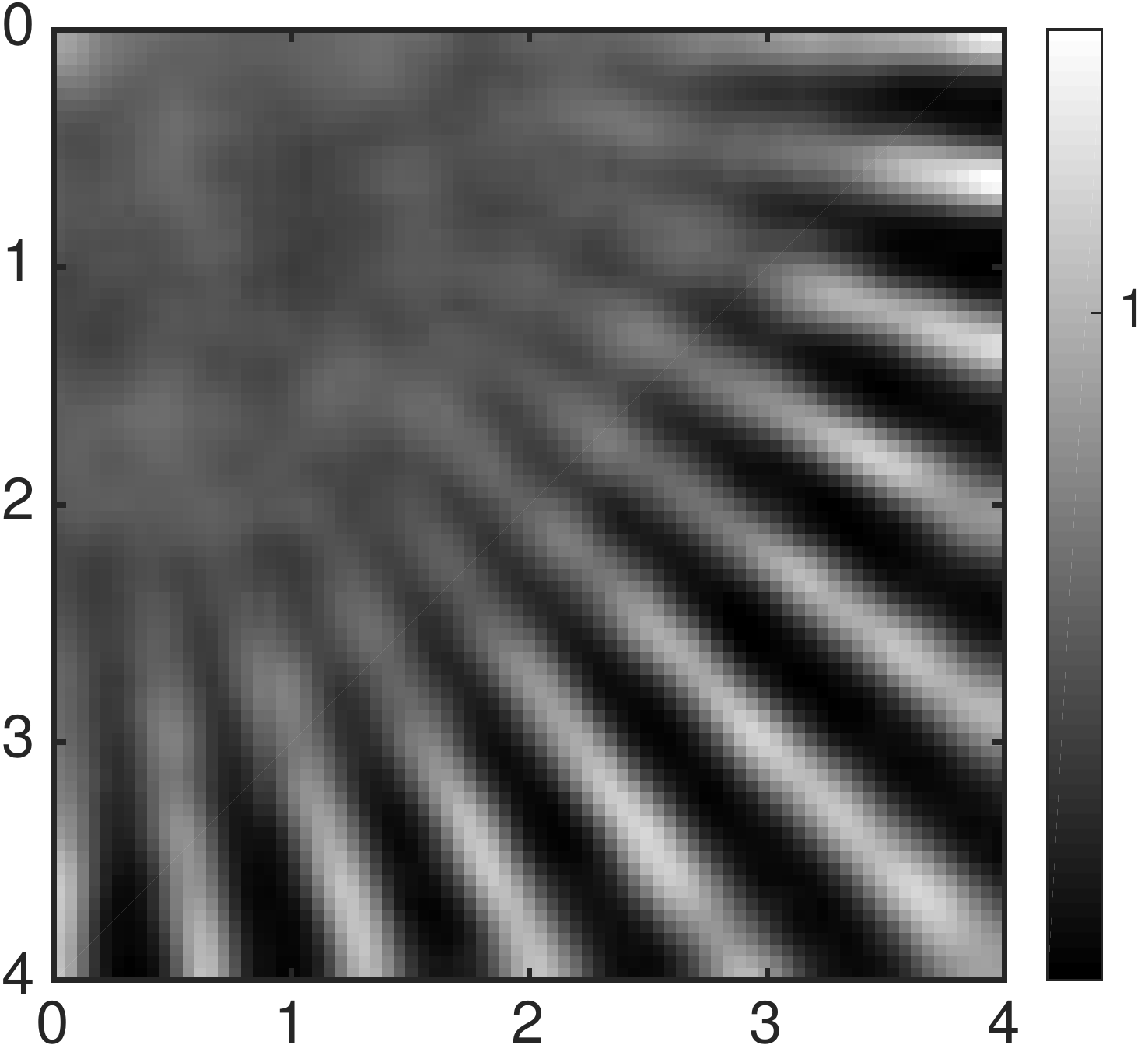}
      \end{tabular}
  \end{tabular}
  \caption{[\textit{Row} A] One product image $\qb_m=\textbf{vect}(\rho_n
    \times I_{m;n})$ built from one of  the $200$ illumination patterns
    used for generating the dataset: (left) a positive constant is
    \textit{added} to the standard speckle patterns so that the lowest value 
    is much greater that zero; (right) a positive constant is
    \textit{subtracted} to the standard speckle patterns and negative 
    values are set to zero. %
    [\textit{Row} B] Reconstruction of the product image $\qb_m$ that corresponds 
    to the one shown above. 
    [\textit{Row} C] Final reconstruction
    $\widehat{\rhob}$ achieved with the whole set of illuminations ---see 
    Subsection~\ref{SR} for details. 
     %
  }
  \label{fig:fig2}
\end{figure}


A numerical  experiment is now considered to support this assertion. 
A set of $M$ collected images are simulated 
following~\eqref{eq:observation} with the PSF $\Hc$  given by the usual
Airy pattern that reads in polar coordinates
\begin{equation}
  \label{psf}
  \Hc(r, \theta) = \frac{k_0^2}{\pi} \, \left( \frac{J_1(r\,
      k_0\,\text{NA})}{k_0\,r}\right)^2,\qquad r\geq 0,\theta\in\eR,
\end{equation}
where $J_1$ is the first order Bessel function of the first kind, NA
is the objective numerical aperture set to 1.49, and $k_0 =
2\pi/\lambda$ is the free-space wavenumber with $\lambda$ 
the emission/excitation wavelength. 
The ground truth 
is the 2D 'star-like' fluorescence pattern depicted in 
Fig.~\ref{fig:fig1}(left). The image sampling step for all the simulations 
involving the star pattern is set\footnote{%
{For an optical system modeled by \eqref{psf}, 
the sampling rate of the (diffraction-limited) acquisition 
is usually the Nyquist rate driven by the OTF cutoff 
frequency  $\nu_{\text{psf}}=2 k_0 \text{NA}$. A higher 
sampling rate is obviously needed for the super-resolved 
reconstruction, the up-sampling factor between the 
``acquisition'' and the ``processing'' rates being at 
least equal to the expected SR factor.  Here, we adopt a common sampling rate for any simulation involving the 
star-like pattern (even with diffraction-limited images), as it allows a 
direct comparison of the reconstruction results. 
}
}
 to $\lambda$/20. 
For this numerical simulation, the illumination set $\{\Ib_m\}_{m=1}^M$
consists in $M=200$ \textit{modified} speckle patterns, see Fig.~\ref{fig:fig2}(A). 
More precisely, a first set of illuminations is obtained by \textit{adding} 
a positive constant (equal to $3$) to each speckle pattern, resulting 
in illuminations that never activate the positivity constraint in~\eqref{critere2}.  
On the contrary, the second set of illuminations is built by \textit{subtracting} 
a small positive constant (equal to $0.2$) to each speckle pattern, the negative 
values being set to zero. The resulting illuminations are thus expected to activate 
the positivity constraint in~\eqref{critere2}.
For both illumination sets, low-resolution microscope images are
simulated and corrupted with Gaussian noise; in this case, 
the standard deviation was chosen so that the 
SNR of the total dataset is 40 dB. 
%
Corresponding reconstructions of the first product
image $\qb_1$ obtained \textit{via} the resolution of \eqref{critere2} is 
shown in Fig.~\ref{fig:fig2}(B), while the retrieved sample \eqref{solutionq} 
is shown in Fig.~\ref{fig:fig2}(C); for each reconstruction, the spatial mean
$\Ib_0$ in \eqref{solutionq} is set to the statistical expectation of the
corresponding illumination set.
As expected, the reconstruction with the
first illumination set is almost identical to the deconvolution 
of the wide-field image shown in Fig.~\ref{fig:fig1}(upper-right), \textit{i.e.,} there is no
SR in this case. On the contrary, the second set of 
illuminations produces a super-resolved reconstruction, hence
establishing the central role of the positivity constraint in the
original joint reconstruction problem \eqref{critere1}. 

\section{A penalized approach for joint Blind-SIM} 
\label{penalize}
As underlined in the beginning of Subsection~\ref{SR}, there is an ambiguity
issue concerning the original joint Blind-SIM reconstruction problem.
A simple way to enforce unicity is to slightly modify~\eqref{critere2}
by adding a strictly convex penalization term. We are thus led to solving
\begin{equation}
  \label{critere3}
    \min_{\qb_m \geq 0}   \left\| \yb_m - \Hb\qb_m\right\|^2 +   \varphi( \qb_m).
\end{equation}
Another advantage of such an approach is that $\varphi$ can be chosen
so that robustness to the noise is granted and/or some expected
features in the solution are enforced. In particular, the analysis
conveyed above suggests that favoring sparsity in each $\qb_m$ 
is suited since speckle or periodic illumination patterns tend to 
frequently cancel or nearly cancel the product images $\qb_m$. 
For such illuminations,  the \textit{Near-Black Object} introduced 
in Donoho's seminal paper \cite{Donoho92a} is an appropriate 
modeling and, following this line, we found that the separable 
``$\ell_1+\ell_2$'' penalty\footnote{%
{The super-resolved solution in \cite{Donoho92a} is 
obtained with a positivity constraint and a $\ell_1$ separable 
penalty. However, ambiguous solutions may exist in this case 
since the criterion to minimize is not \textit{strictly} convex.   
The $\ell_2$ penalty in~\eqref{hyperbolic} is then mostly introduced for the technical
reason that a unique solution exists for problem \eqref{critere3}.
}
}
provides super-resolved reconstructions:
\begin{equation}
  \label{hyperbolic}
  \varphi( \qb_m) \pardef     \alpha {\textstyle \sum_n |q_{m;n}|} + \beta
   ||\qb_m||^2, \qquad \alpha \geq0, \beta >0.
\end{equation}
%
%
With properly tuned $(\alpha,\beta)$, our joint Blind-SIM strategy 
is expected to bring SR if ``sparse'' illumination
patterns $\Ib_m$ are used, \textit{i.e.,} if they enforce $q_{m;n}=0$
for most (or at least many) $n$. 
More specifically, it is shown in \cite[Sec. 4]{Donoho92a} that 
SR occurs if the  number of non-zero 
$I_{m;n}$ (\textit{i.e.,} the number of non-zero components to 
retrieve in $\qb_m$) divided by $N$ is lower than $\frac{1}{2} R/N$,
with $R/N$ the \textit{incompleteness ratio} and
$R$ the rank of $\Hb$. 
In addition, the resolving power is driven by the spacing between the 
components to retrieve that, ideally, should be greater than the Rayleigh distance $\frac{\lambda}{2\,\text{NA}}$, 
see \cite[pp. 56-57]{Donoho92a}. 
%
These conditions are rather stringent and hardly met by illumination 
patterns that can be reasonably considered in practice.  These illumination patterns are usually either deterministic harmonic 
or quasi-harmonic\footnote{%
Dealing with distorted patterns is of particular practical importance 
since it allows to cope with the distortions and misalignments induced 
by the instrumental uncertainties or even by the sample itself 
\cite{Ayuk13,Jost15}.
}   
patterns, or random speckle patterns, these latter illuminations 
being much easier to generate \cite{Mudry12}. Nevertheless, in 
both cases, a SR effect is observed in joint Blind-SIM.
Moreover, one can try to maximize this effect \textit{via} the tuning 
of some experimental parameters that are left to the designer of the 
setup. 
Such parameters are mainly: the period of the light grid and the number 
of grid shifts for harmonic patterns, the spatial correlation
length and the point-wise statistics of the speckle 
patterns.  
Investigating the SR properties with respect to these 
parameters on a theoretical ground seems out of reach. However, a numerical
analysis is possible and some illustrative results are now provided
that  address this question. 
Reconstructions shown in the sequel are built from \eqref{solutionq} 
\textit{via} the numerical resolution of \eqref{critere3}-\eqref{hyperbolic}. For sake of clarity, all 
the algorithmic details concerning this minimization problem are
reported  in Sec.~\ref{algo}. 
These simulations were performed with low-resolution microscope
images corrupted by additive Gaussian noise such that the
signal-to-noise ratio (SNR) of the dataset
$\{\yb_m\}_{m=1}^M$ is 40 dB. In addition, we note that this \textit{penalized}
joint Blind-SIM strategy requires an explicit tuning of some hyper-parameters, namely
$\alpha$ and $\beta$ in the regularization function \eqref{hyperbolic}. Further
details concerning these parameters are reported in Sec.~\ref{nbom}.
\begin{figure}[t]
  \centering
  \begin{tabular}{@{\kern0pt}l@{\kern0pt}l@{\kern-2pt}l}
    (A)
    &
    \begin{tabular}{l}
      \includegraphics[height=0.19\textwidth]{
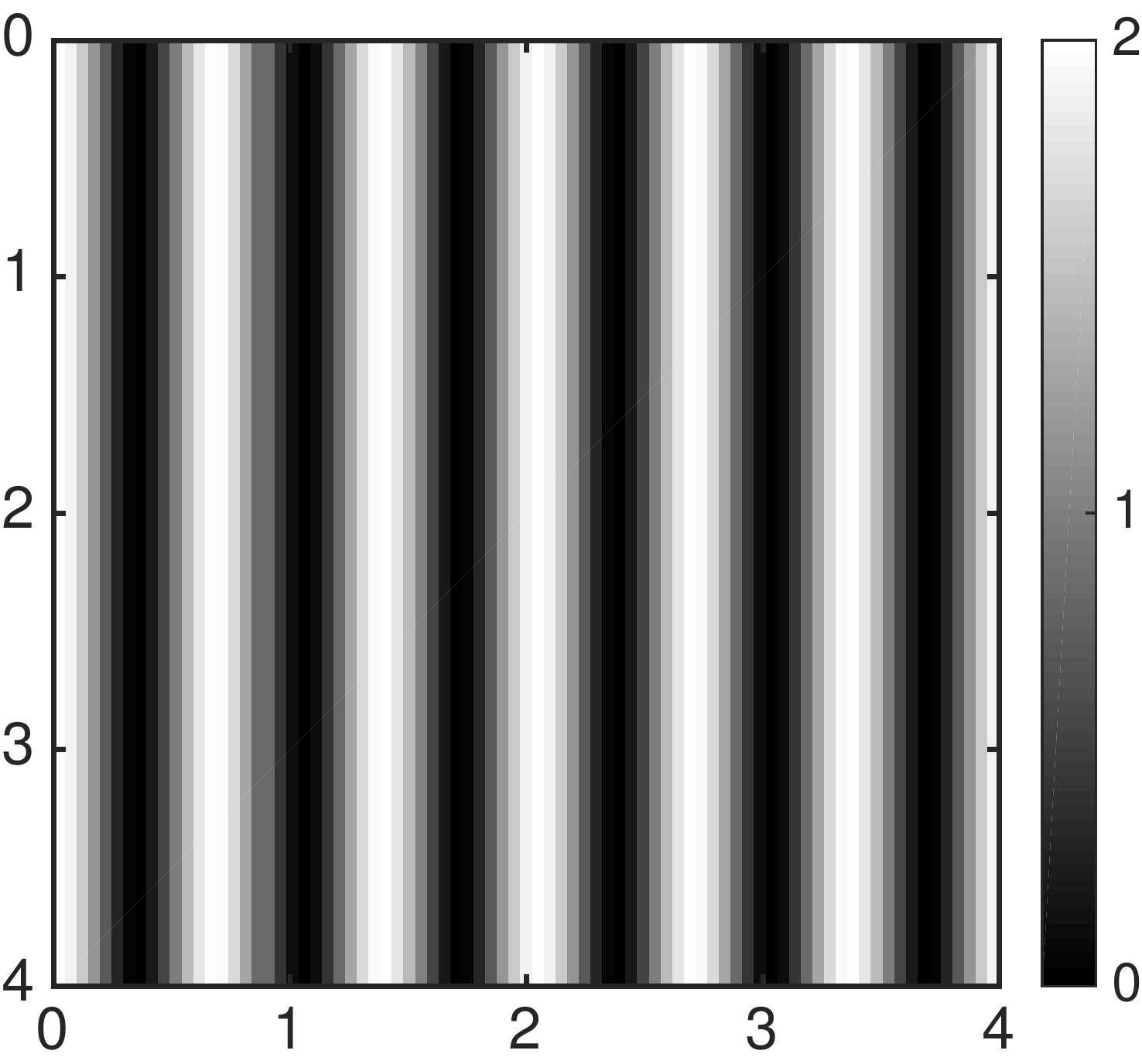}
    \end{tabular}
    &
      \begin{tabular}{l}
        \includegraphics[height=0.19\textwidth]{
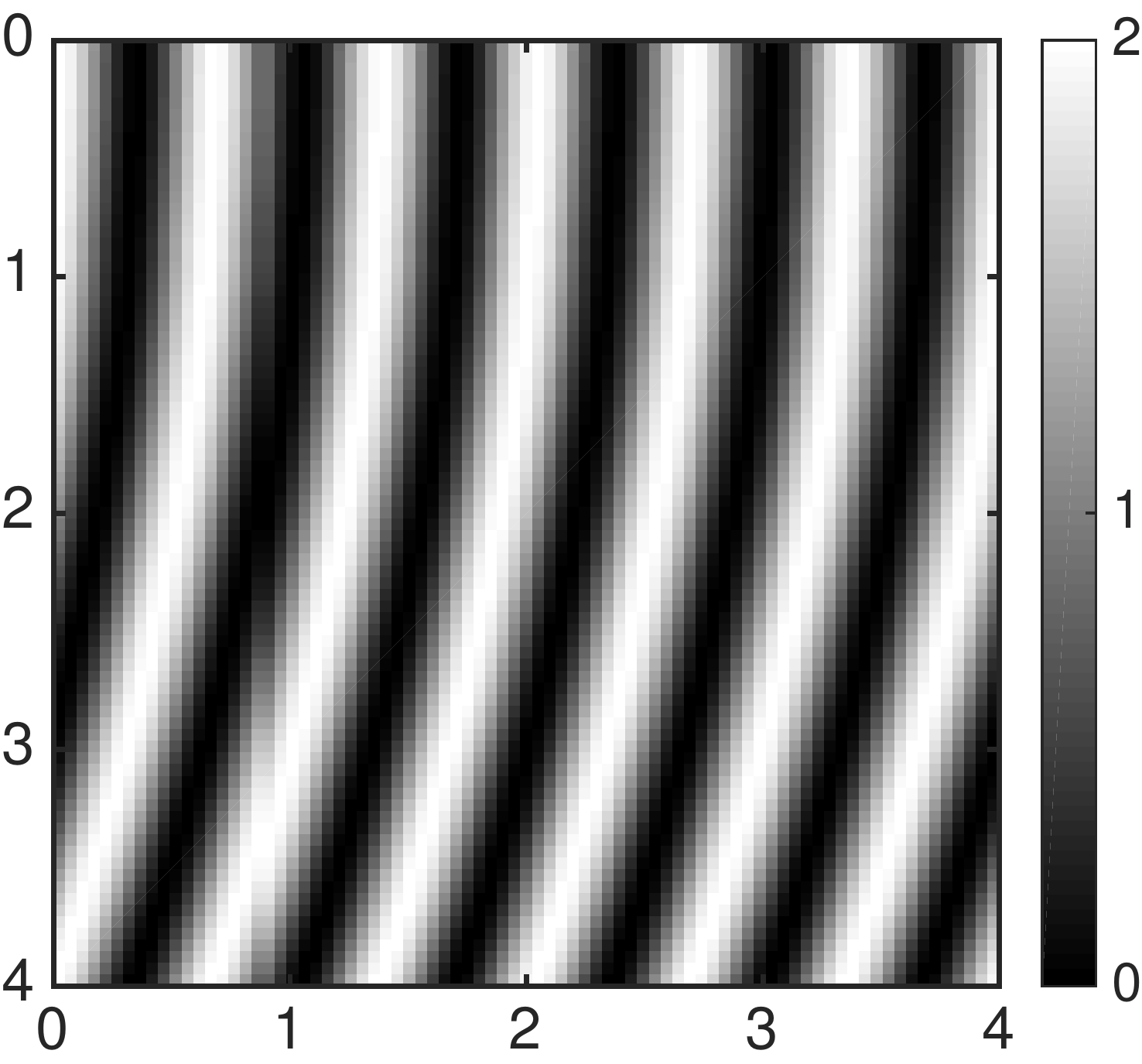}
      \end{tabular}\\[1em]
    (B)
    &
    \begin{tabular}{l}
      \includegraphics[height=0.19\textwidth]{
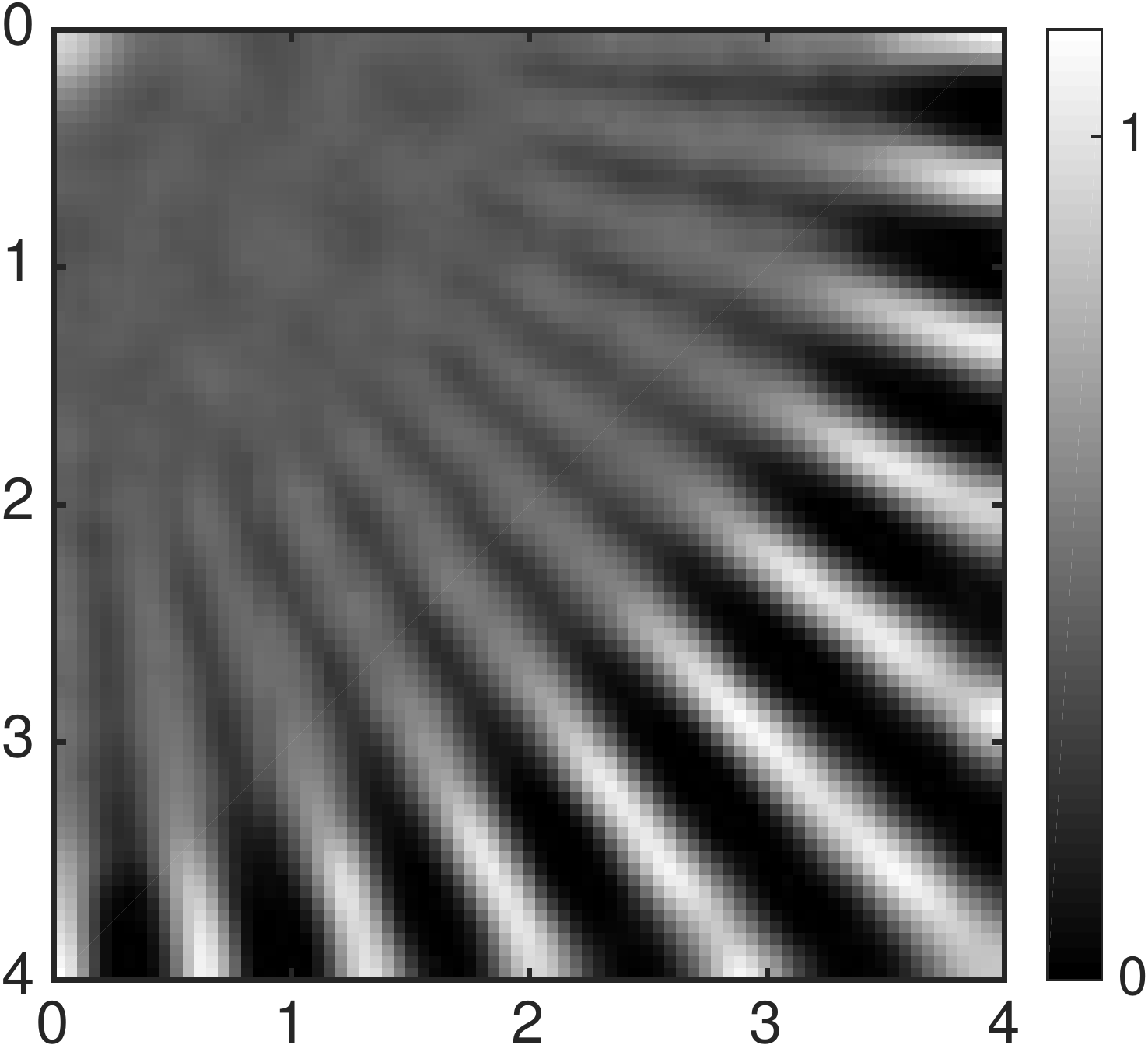} 
    \end{tabular}
    &
      \begin{tabular}{l}
        \includegraphics[height=0.19\textwidth]{
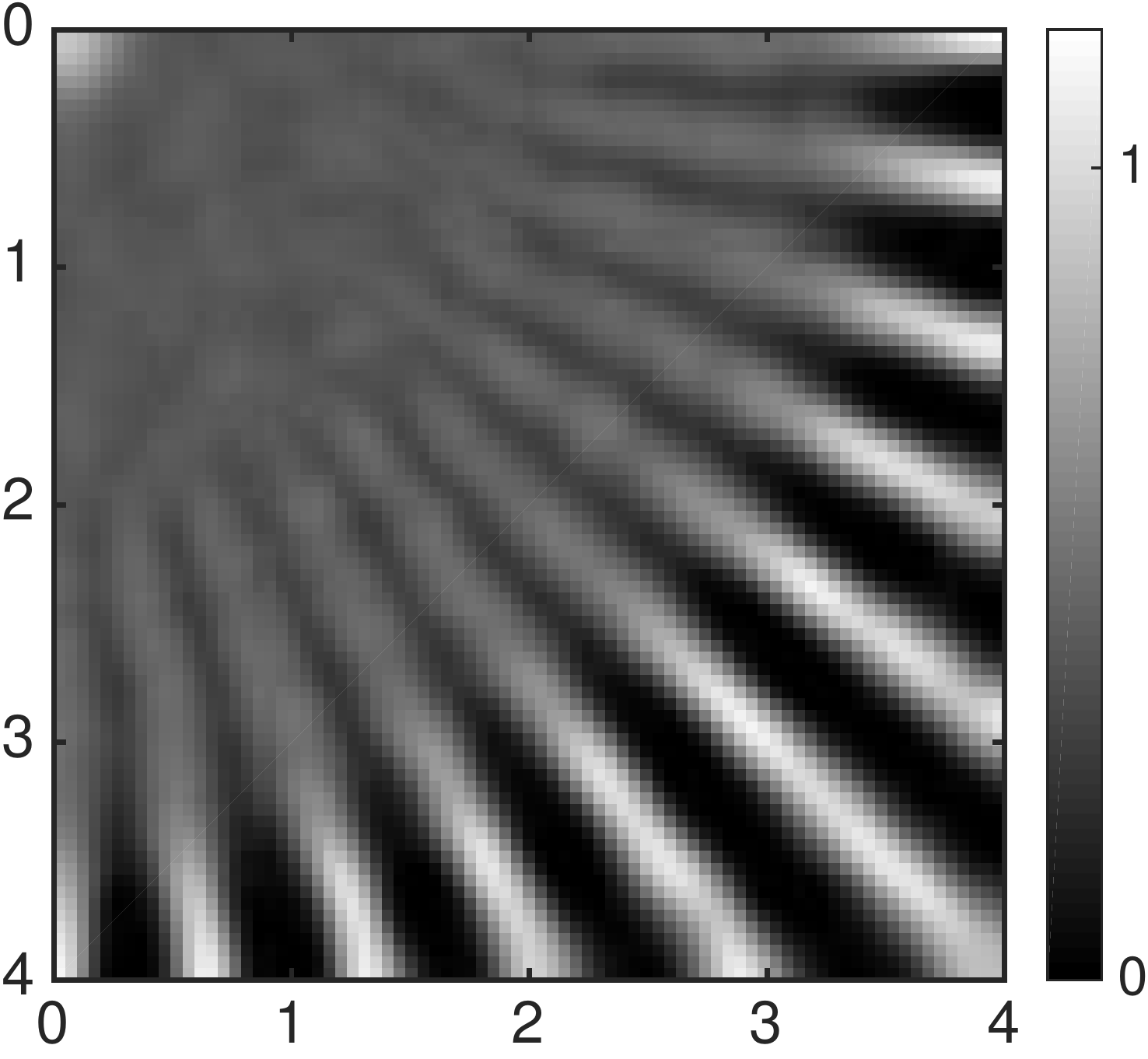}
      \end{tabular}\\[1em]
    (C)
    &
    \begin{tabular}{l}
     \includegraphics[height=0.19\textwidth]{
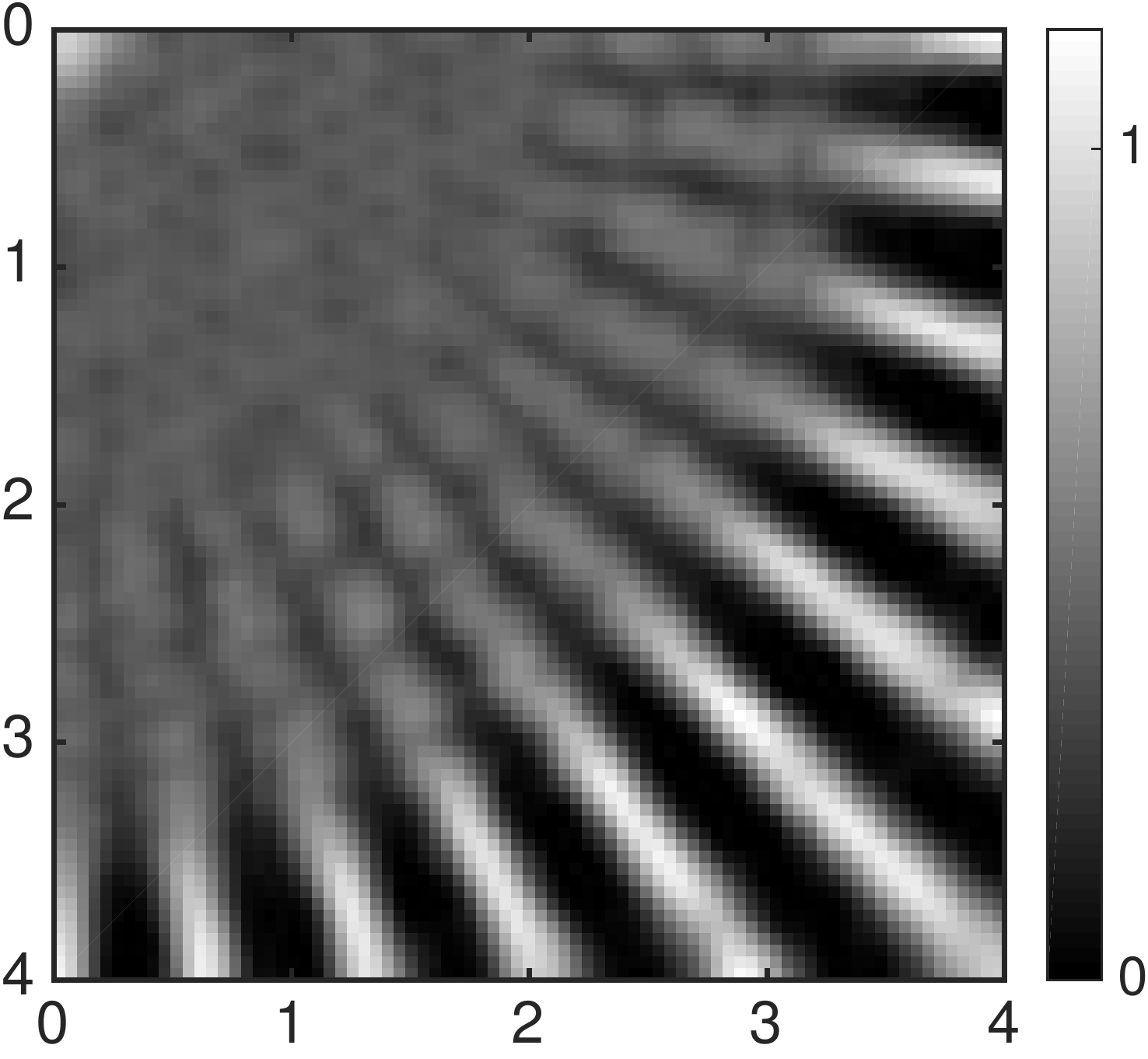}
    \end{tabular}
    &
      \begin{tabular}{l}
        \includegraphics[height=0.19\textwidth]{
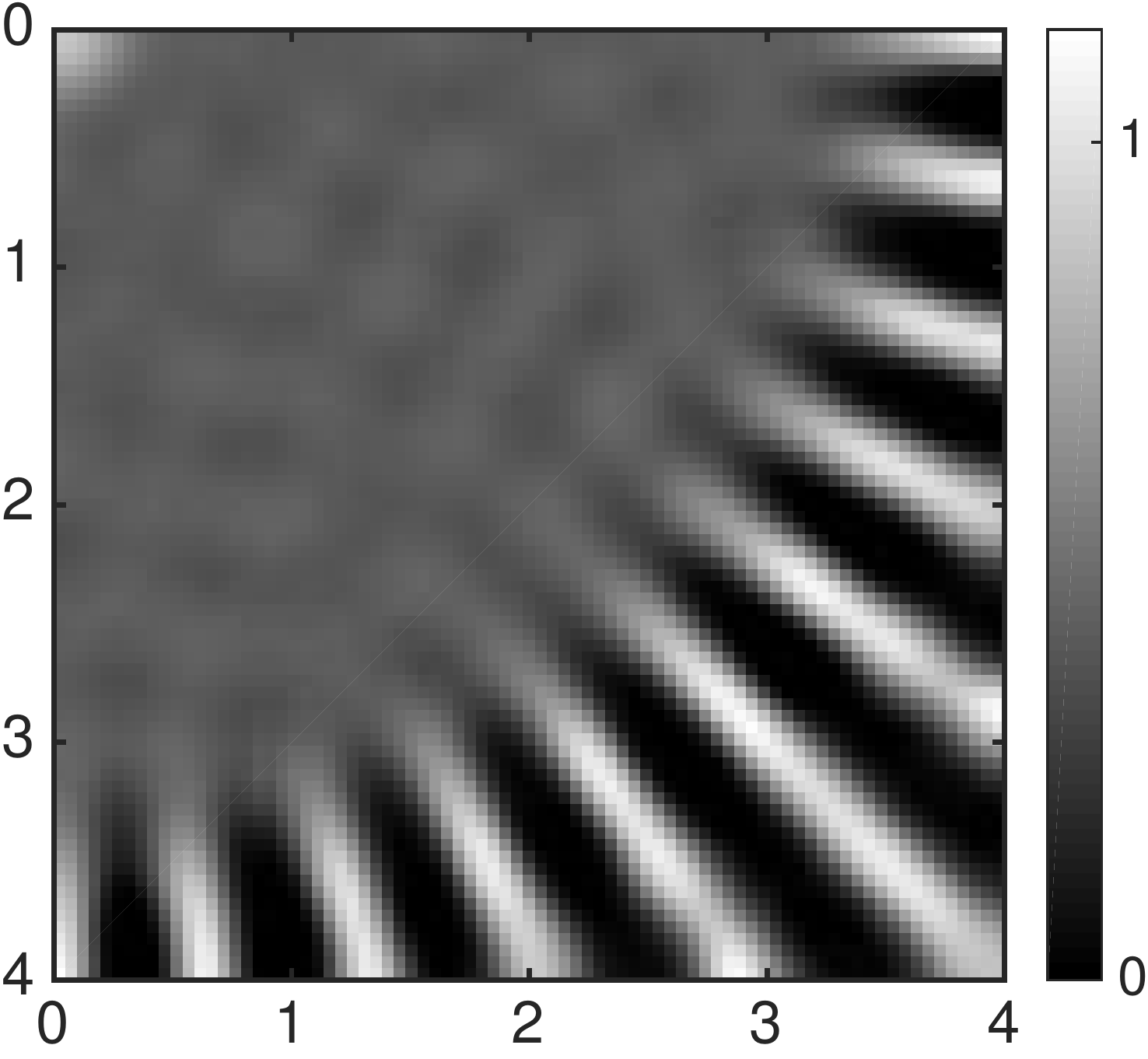}
      \end{tabular}\\[1em]
    (D)
    &
      \begin{tabular}{l}
        \includegraphics[height=0.19\textwidth]{
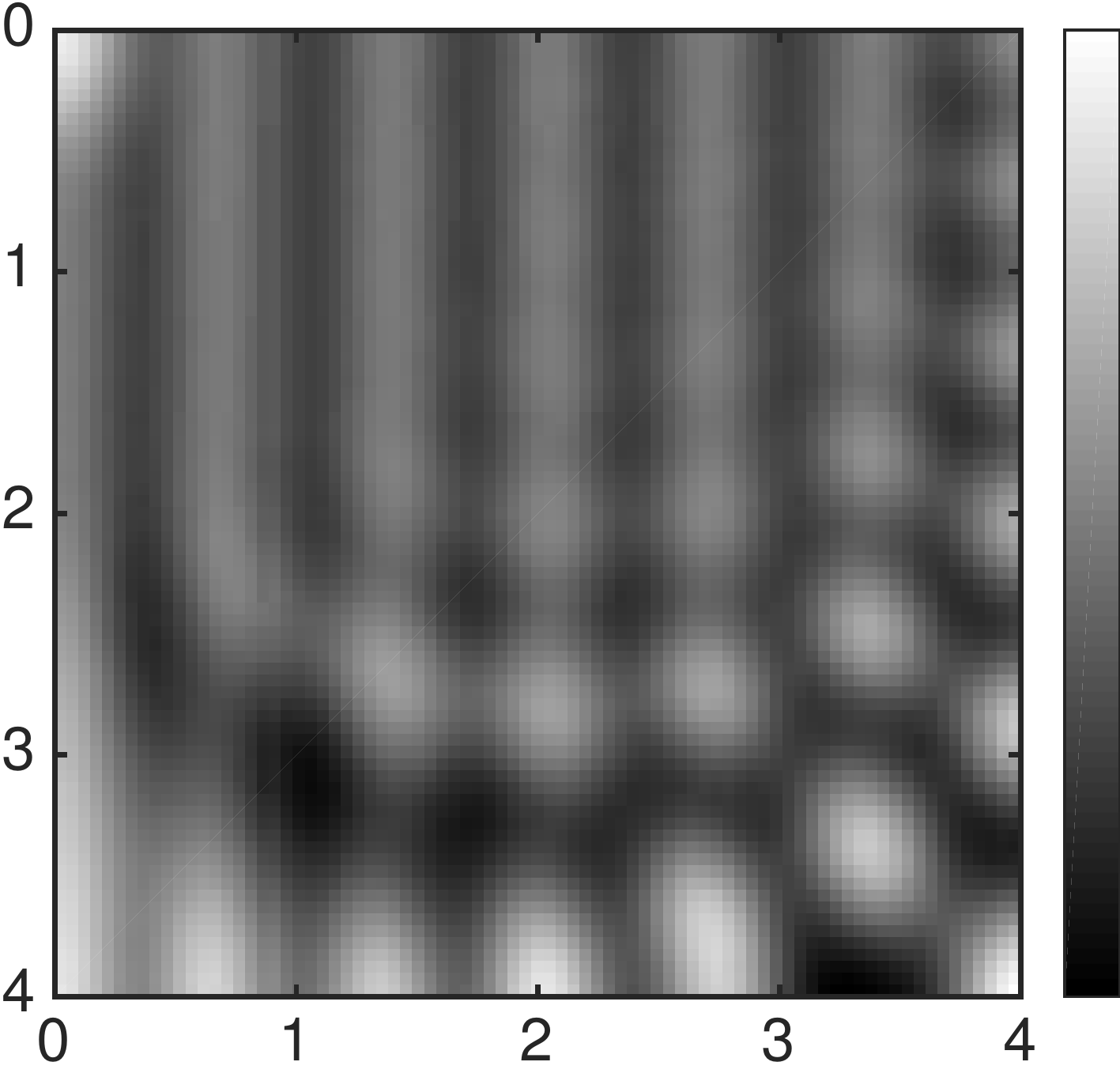}
      \end{tabular}
    &
      \begin{tabular}{l}
        \includegraphics[height=0.19\textwidth]{
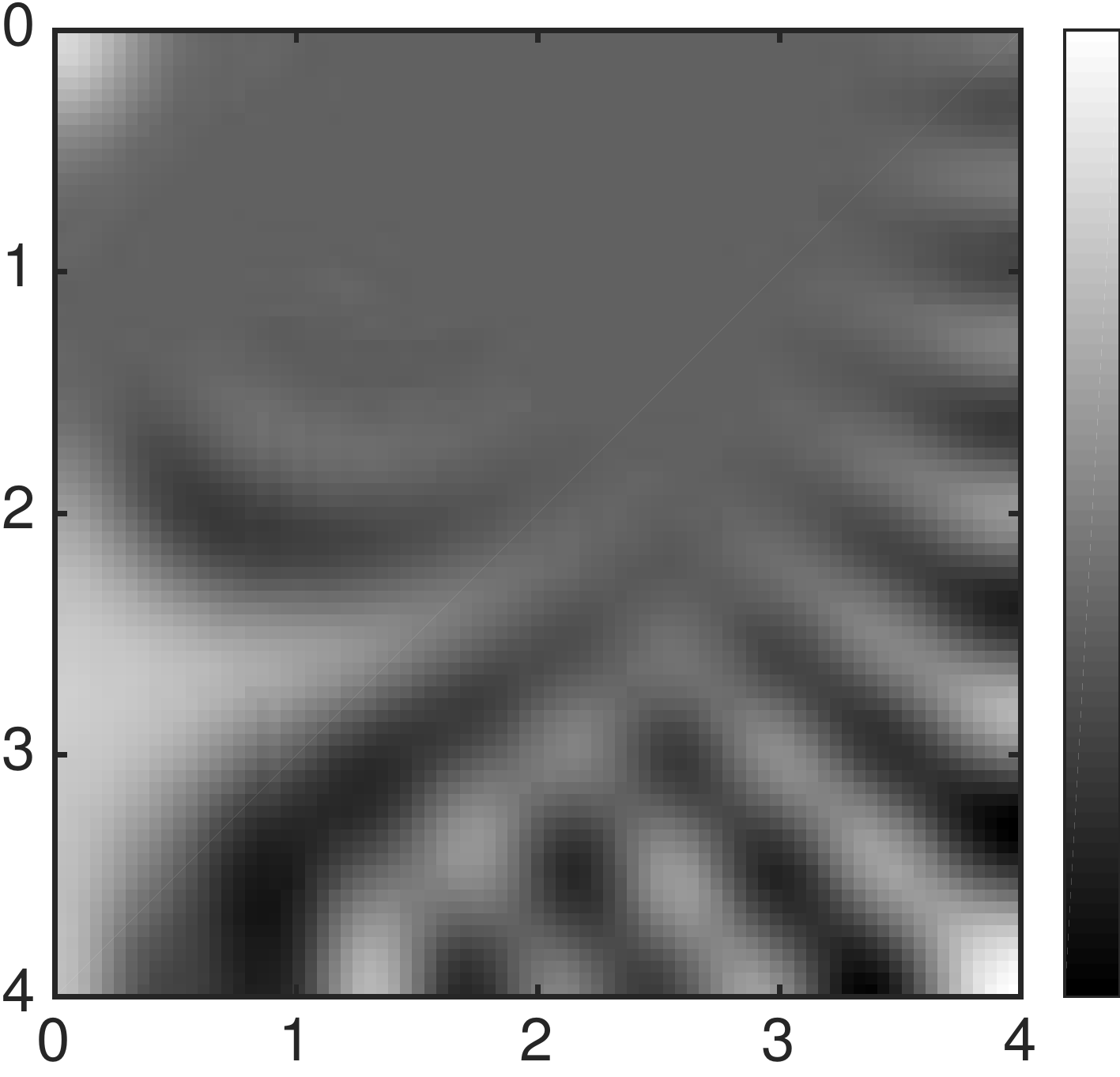}
      \end{tabular}
  \end{tabular}
  \caption{%
    \textbf{Harmonic patterns:} [\textit{Row} A] One illumination pattern $\Ib_m$ drawn from the set
    of regular (left) and distorted (right) harmonic patterns.
    [\textit{Row} B] Corresponding penalized joint Blind-SIM
    reconstructions.  
    [\textit{Row} C] (left) Decreasing the number of phase shifts 
     from 6 to 3 brings some reconstruction 
    artifacts, see  (B-left) for comparison. 
    (right)  Increasing the modulation frequency $||\nub||$
    of the harmonic patterns above the OTF cutoff frequency prevents the
    super-resolution to occur.
    [\textit{Row} D] Low-resolution image $\yb_m$ drawn 
    from the dataset for a modulation frequency $||\nub||$ 
    lying inside (left) and outside (right) the OTF domain---see 
    Sec.~\ref{Results_harmonic} for details.  
  }
  \label{fig:fig3}
\end{figure}

\subsection{Regular and distorted harmonic patterns} 
\label{Results_harmonic}
We first consider unknown harmonic patterns defining a 
``standard'' SIM experiment with $M=18$ patterns. More precisely, 
the illuminations are  harmonic patterns of the form 
$I(\rb) = 1 + \cos(2\pi \nub^t\rb + \phi)$ where $\phi$ is 
the phase shift, and with $\rb=(x,y)^t$ and $\nub=(\nu_x,\nu_y)^t$ 
the spatial coordinates and the spatial frequencies of the harmonic 
function, respectively. 
Distorted versions of these patterns (deformed by optical 
aberrations such as astigmatism and coma) were also considered. 
Three distinct orientations $\theta \pardef \tan^{-1}(\nu_y/\nu_x) 
\in \{0, 2\pi/3, 4\pi/3\}$, for each  of which 
six phase shifts of one sixth of the  period, were considered. The 
frequency of the harmonic patterns $||\nub|| := (\nu_x^2 +
\nu_y^2)^{1/2}$ is set to 80\% of the OTF cutoff
frequency, \textit{i.e.,} it lies inside the OTF support.  
\begin{figure}[t]
  \centering
  \begin{tabular}{@{\kern0pt}l@{\kern0pt}l@{\kern-2pt}l}
    (A)
    &
      \begin{tabular}{l}
        \includegraphics[height=0.19\textwidth]{
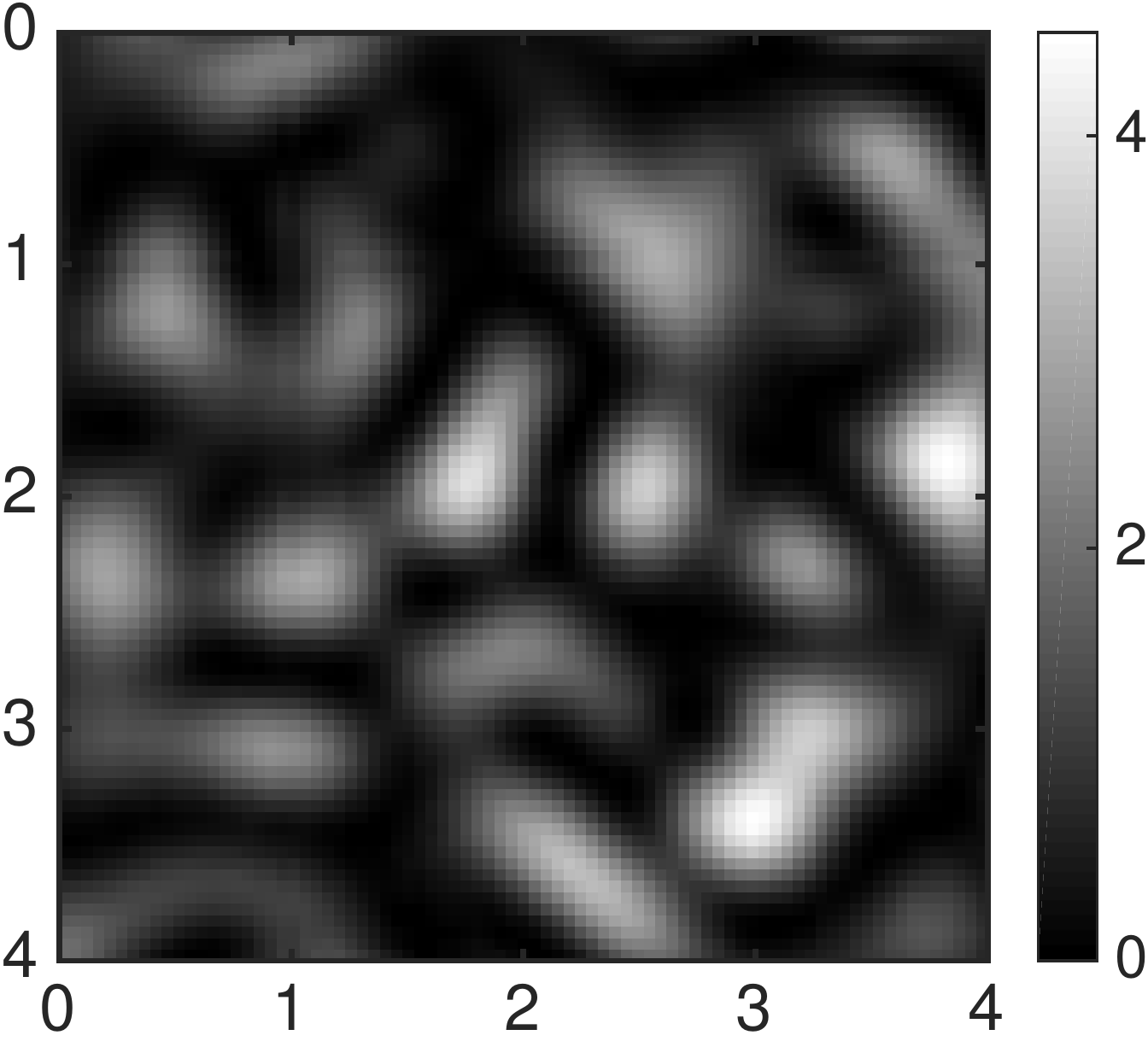}
      \end{tabular}
    &
      \begin{tabular}{l}
        \includegraphics[height=0.19\textwidth]{
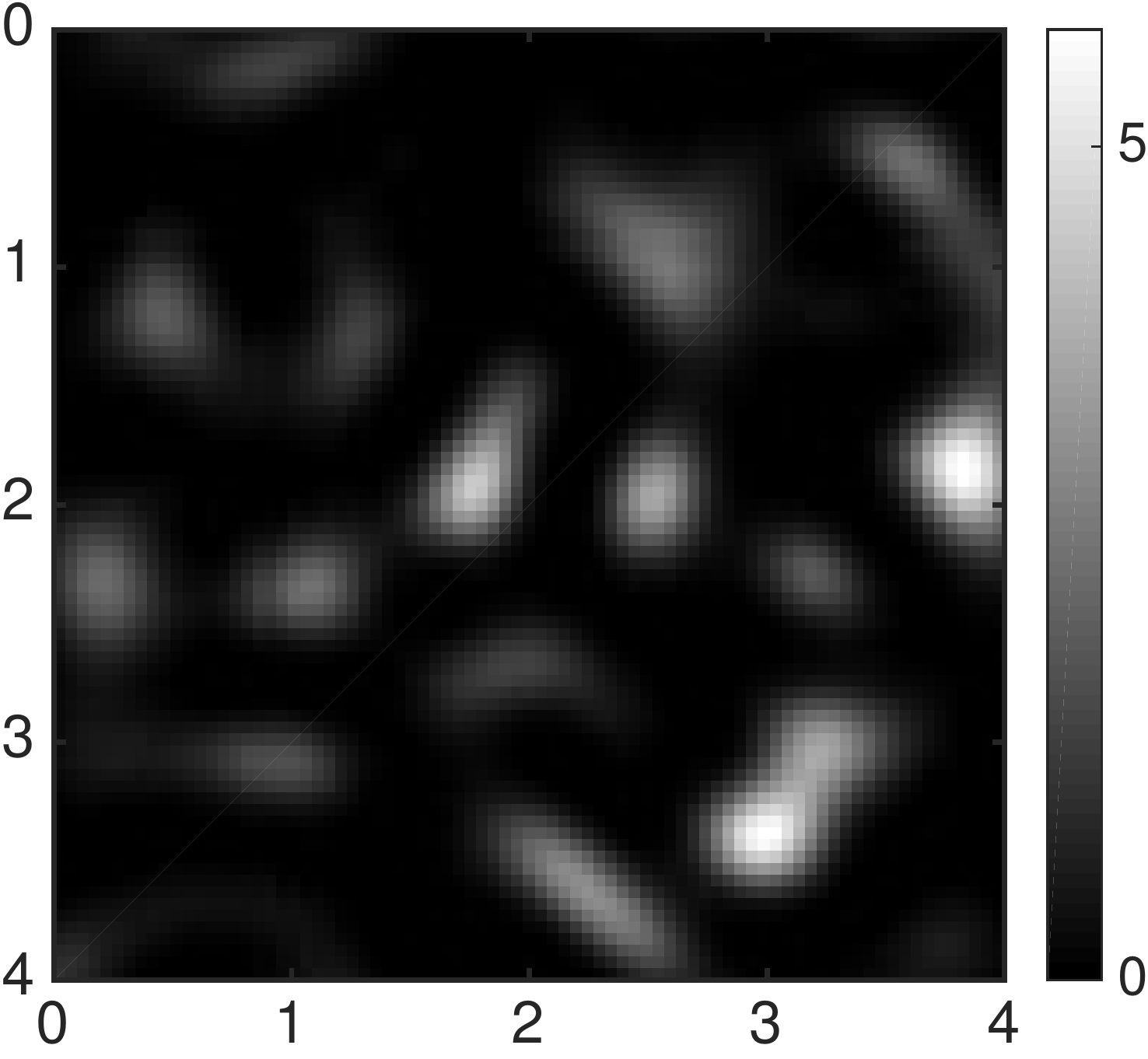}
      \end{tabular}\\[1em]
    (B)
    &
      \begin{tabular}{l}
        \includegraphics[height=0.19\textwidth]{
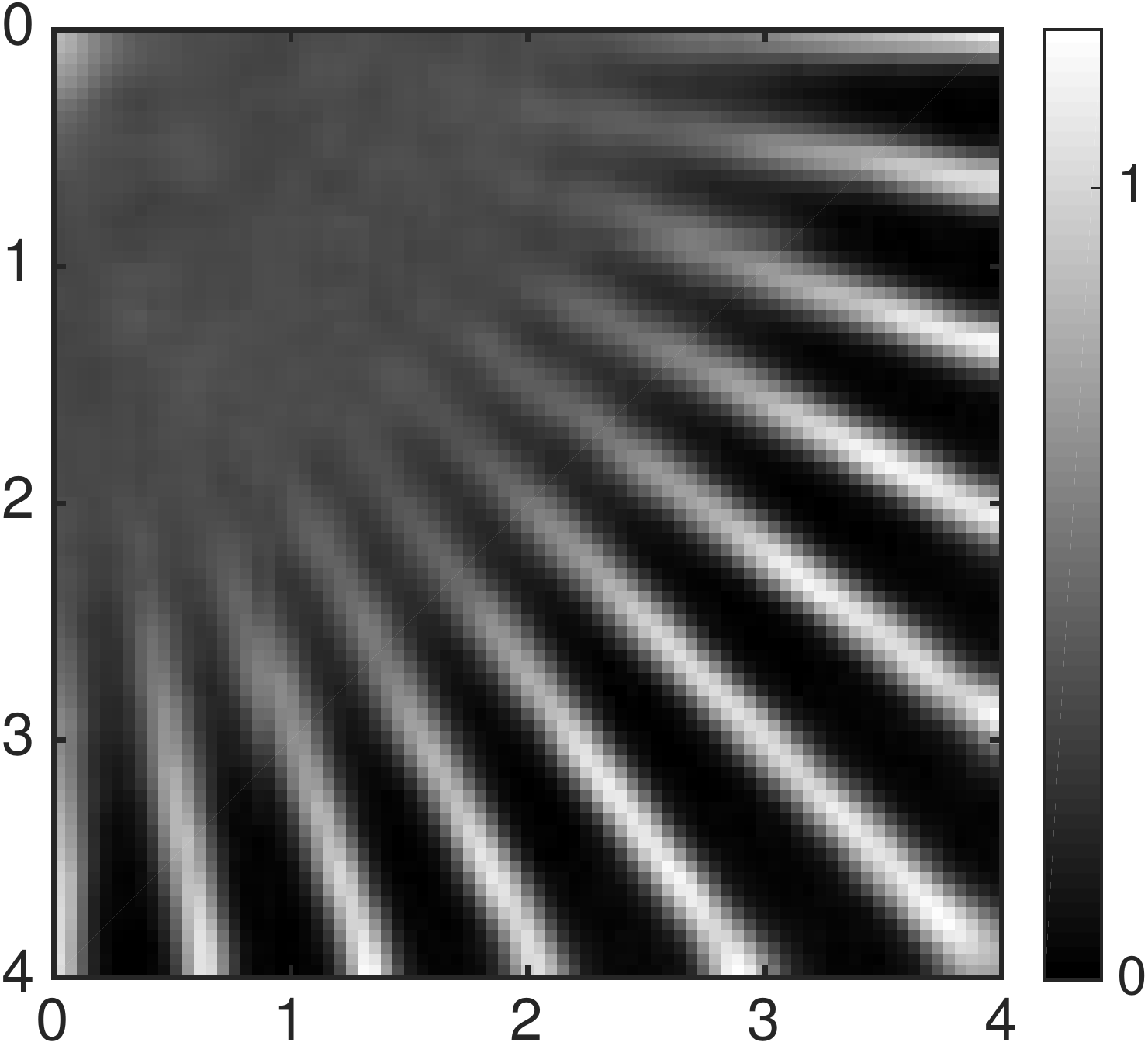}
      \end{tabular}
    &
      \begin{tabular}{l}
        \includegraphics[height=0.19\textwidth]{
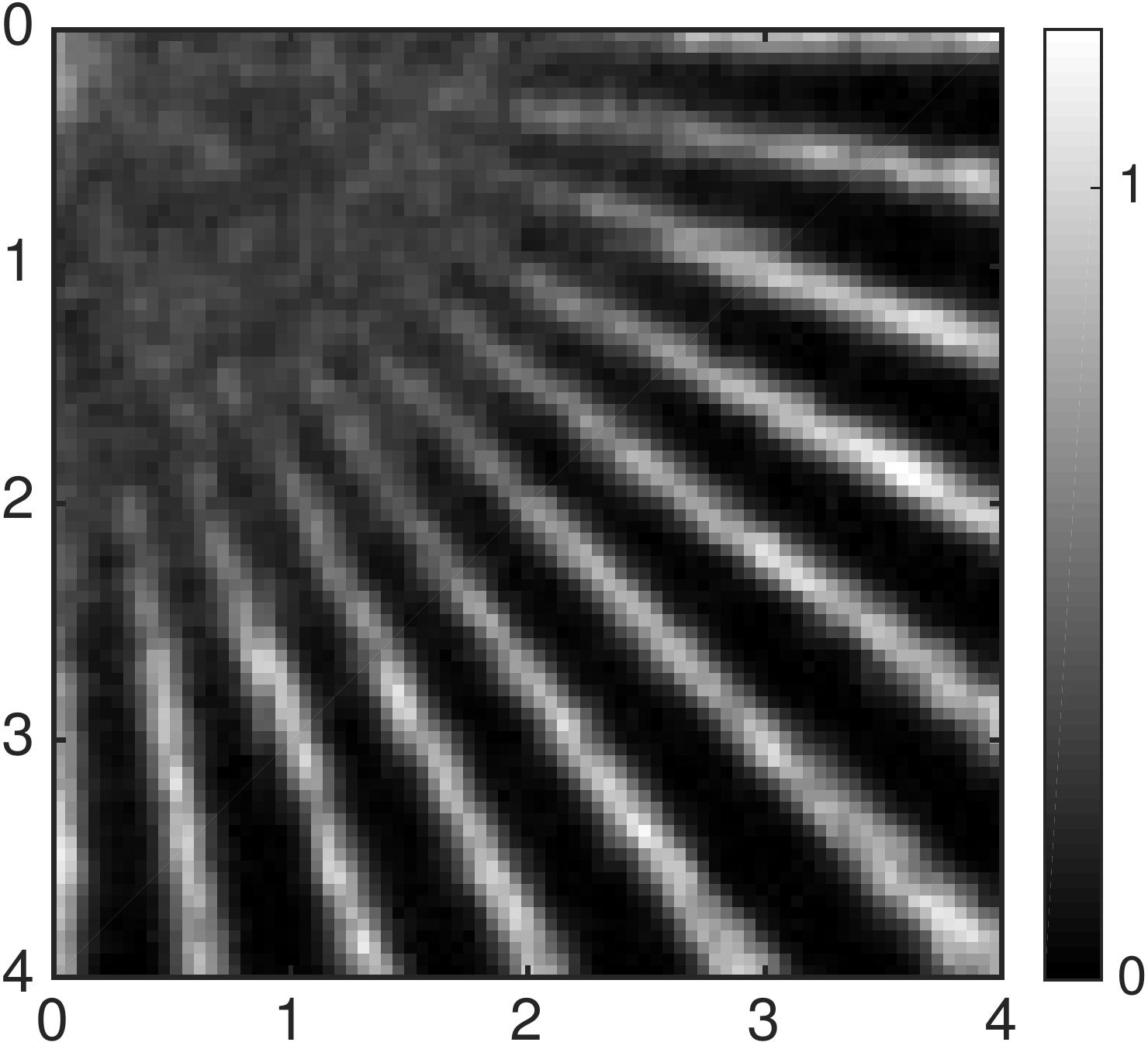}
      \end{tabular}
  \end{tabular}
  \caption{%
    \textbf{Speckle patterns:} [\textit{Row} A]  One speckle
    illumination such that $\text{NA}_{\text{ill}} = \text{NA}$  
    (left) and its ``squared'' counterpart (right).    
    [\textit{Row} B] Corresponding penalized joint Blind-SIM
    reconstructions from $M=1000$ speckle (left) and 
    ``squared'' speckle (right)  patterns 
    %
  }
  \label{fig:fig4}
\end{figure}
One regular and one distorted pattern are depicted in 
Fig.~\ref{fig:fig3}(A) and the penalized joint Blind-SIM 
reconstructions are shown in Fig.~\ref{fig:fig3}(B). For both
illumination sets, a clear SR effect occurs, which is 
similar to the one obtained with the original approach presented in
\cite{Mudry12}. As expected, however, the reconstruction quality 
achieved in this blind context is lower than what can be obtained 
with standard harmonic SIM ---for the sake of comparison, 
see Fig.~\ref{fig:fig1}(B). 
In addition, we note that some artifacts may appear if the number 
of phase shifts for each orientation is decreased, see Fig.~\ref{fig:fig3}(C-left). 
If we keep in mind that the retrieved sample $\widehat{\rhob}$ in
\eqref{solutionq} gains SR by the summation of 
(super-resolved) product images $\widehat{\qb}_m$, these artifacts 
are driven (at least partially) by the fact that fewer shifts result in illumination sets that, as a whole, misses to uniformly 
cancel the plane. In other words, the illumination patterns do 
cancel the object, but not ``that frequently'' to bring 
a uniform SR effect in the final reconstruction.  
Let us finally investigate how the modulation frequency $\nub$ used for
the generation of the patterns does impact the SR of the penalized
joint Blind-SIM reconstruction.  
The first finding is that the SR
is almost completely lost when $\norm{\nub}$ lies beyond the OTF 
cutoff frequency.  As an illustration, the penalized joint Blind-SIM reconstruction
shown in Fig.~\ref{fig:fig3}(C-right) is obtained with $||\nub||$ set to 
120\% of the OTF cutoff frequency, see also Fig.~\ref{fig:fig1}(upper right) 
for a comparison with deconvolution of the wide-field image. In this case,
each harmonic carrier $\Ib_m$ is completely filtered out from the 
low-resolution image $\yb_m$, see Fig.~\ref{fig:fig3}(D). 
As a result, the SR effect driven by each pattern $\Ib_m$ 
is lost since the sparse deconvolution of $\yb_m$ \eqref{critere3} does 
not provide any super-resolved localization of the zeros driven by the
 illumination patterns.
\begin{figure}[t]
  \centering
  \begin{tabular}{@{\kern0pt}l@{\kern0pt}l@{\kern-2pt}l}
    (A)
    &
      \begin{tabular}{l}
        \includegraphics[height=0.19\textwidth]{
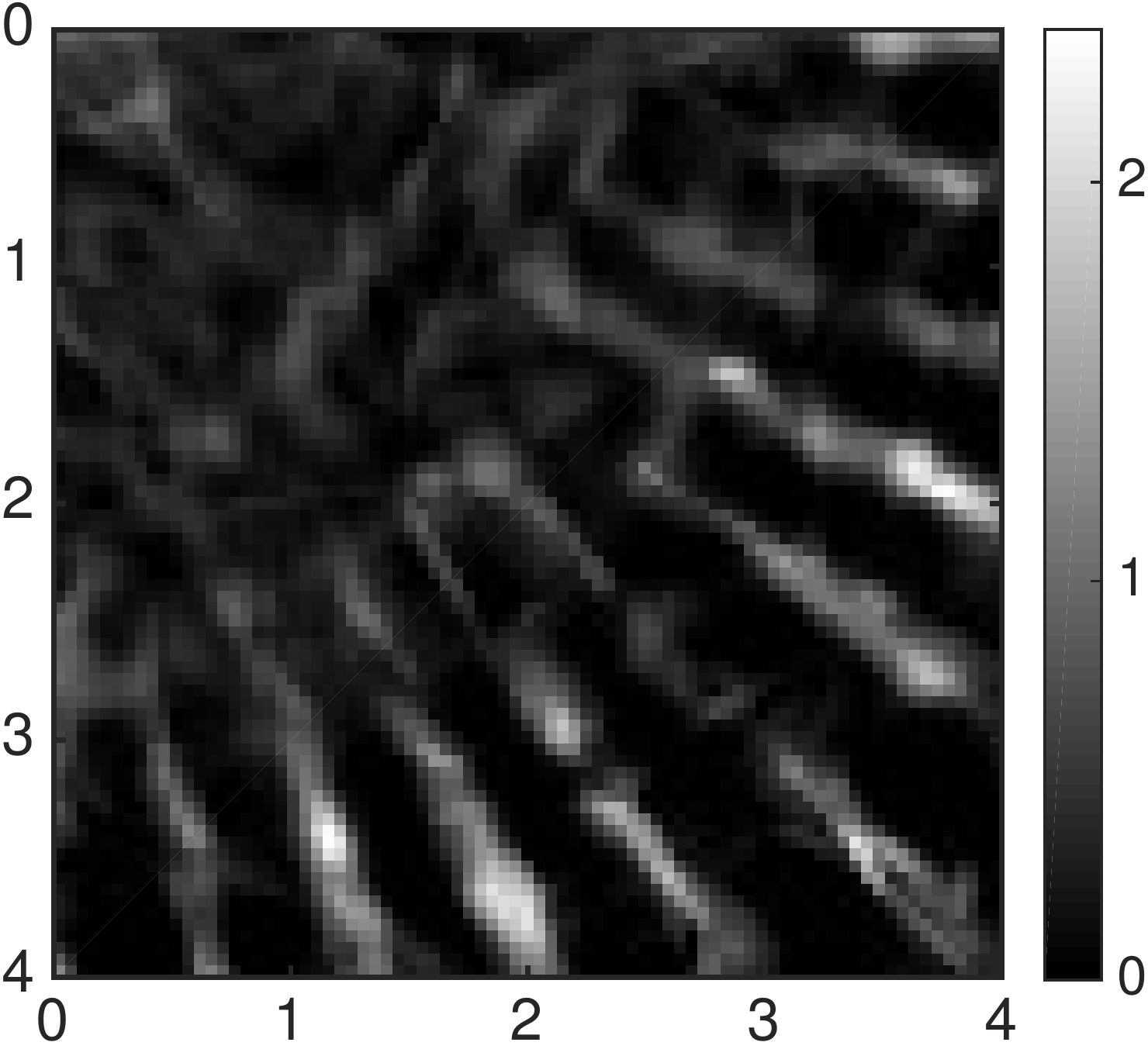}
      \end{tabular}
    &
      \begin{tabular}{l}
        \includegraphics[height=0.19\textwidth]{
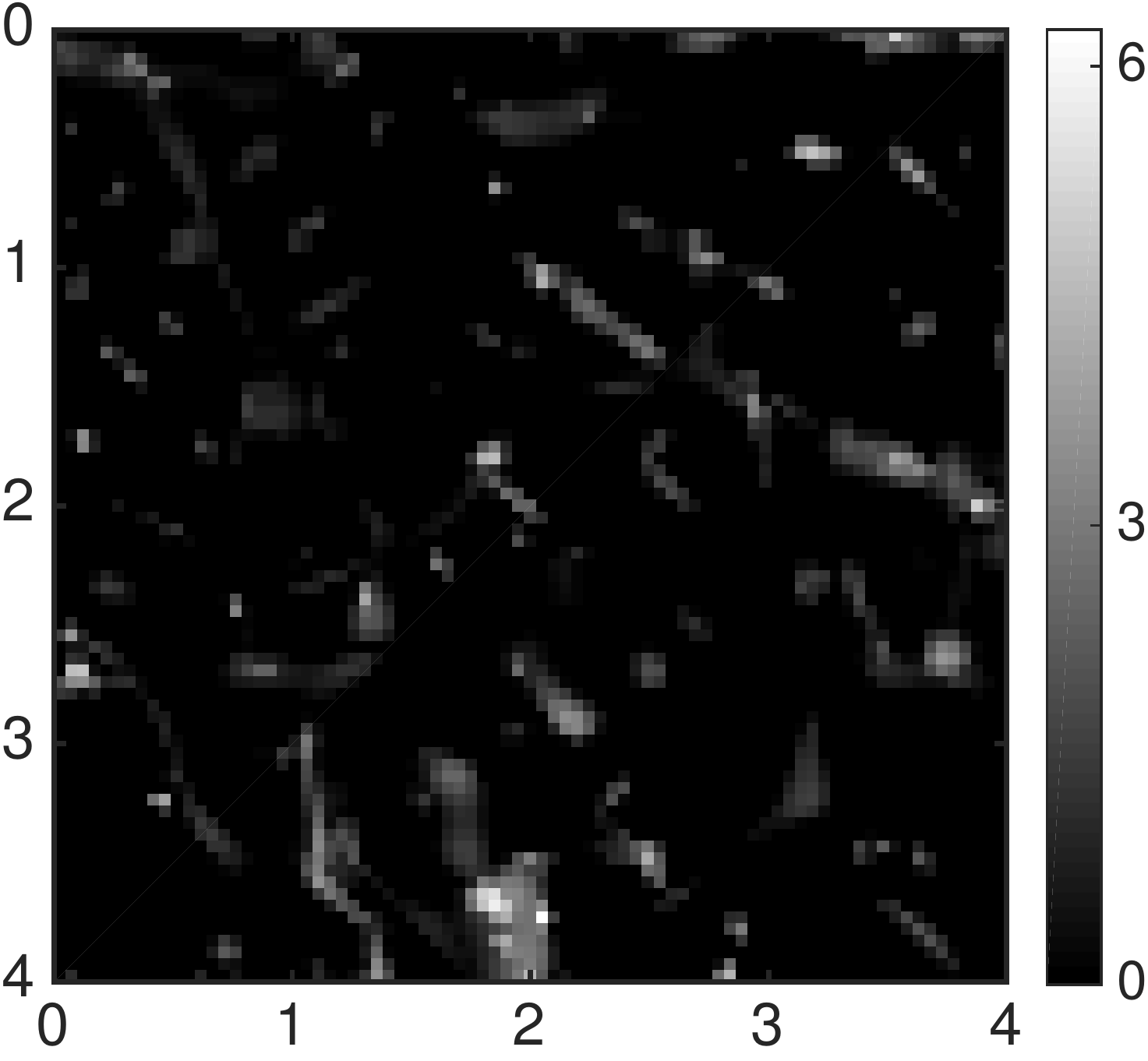}
      \end{tabular}\\[1em]
    (B)
    &
      \begin{tabular}{l}
        \includegraphics[height=0.19\textwidth]{
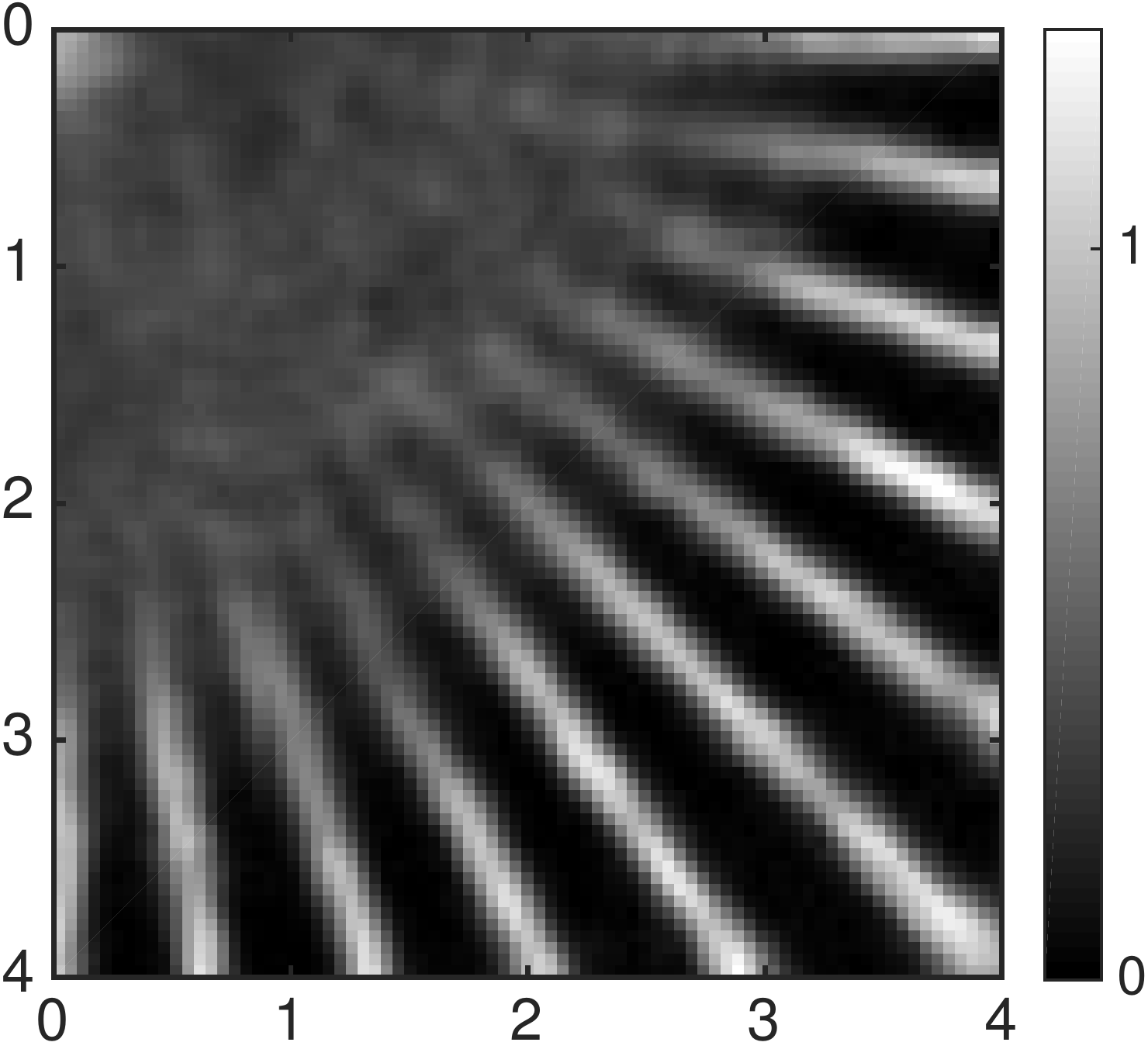}
      \end{tabular}
    &
      \begin{tabular}{l}
        \includegraphics[height=0.19\textwidth]{
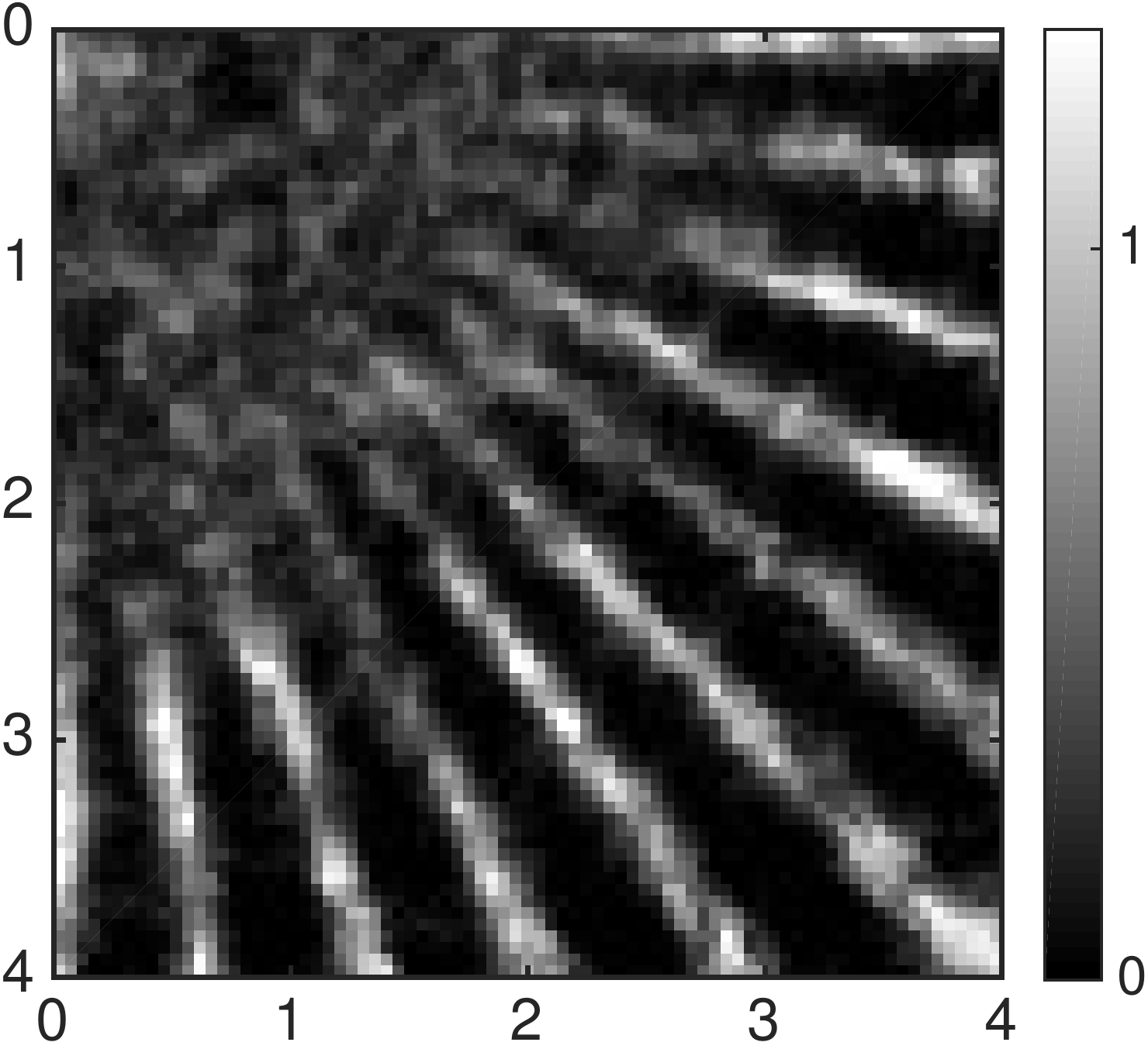}
      \end{tabular}\\[1em]
    (C)
    &
      \begin{tabular}{l}
        \includegraphics[height=0.19\textwidth]{
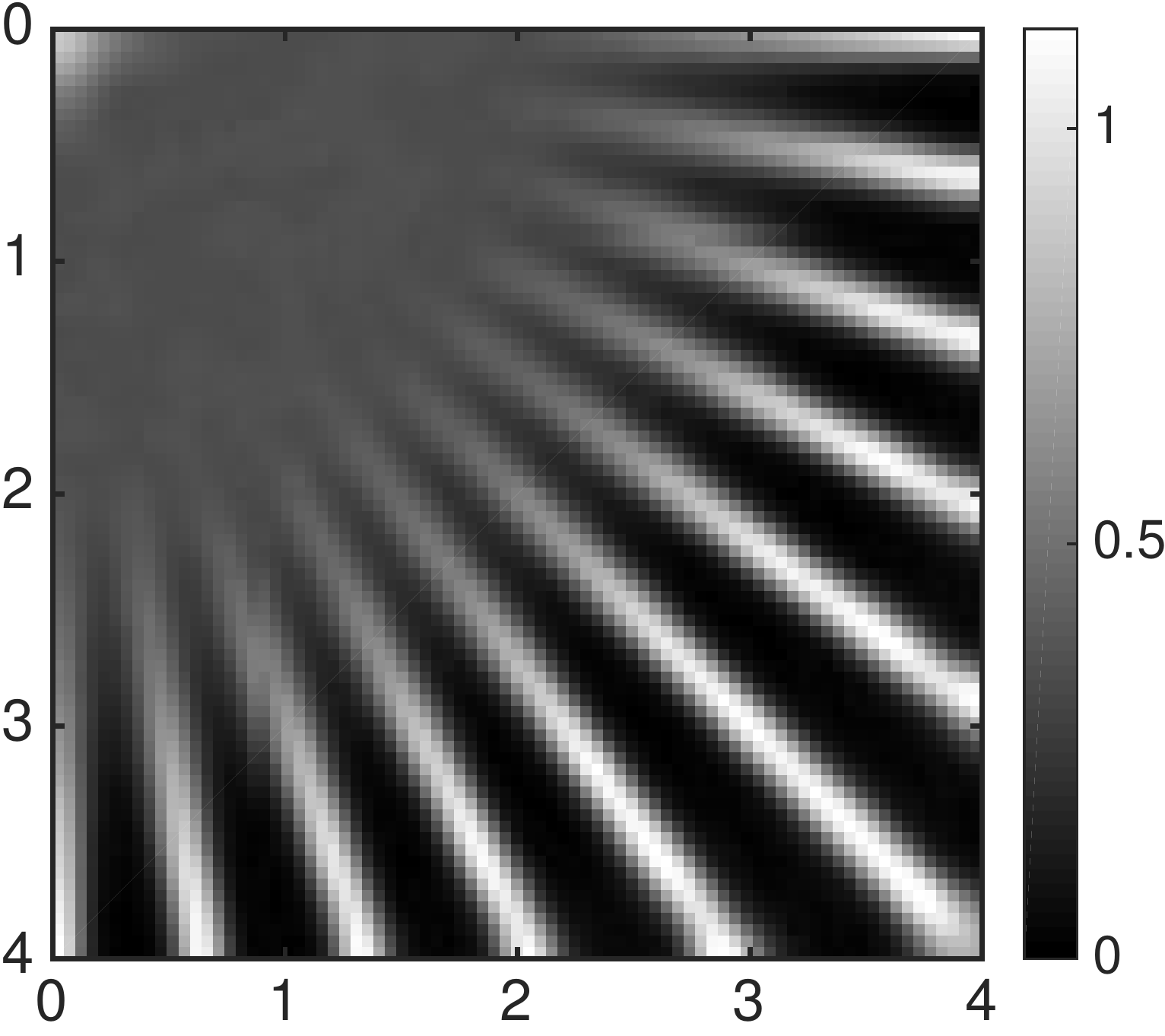}
      \end{tabular}
    &
      \begin{tabular}{l}
        \includegraphics[height=0.19\textwidth]{
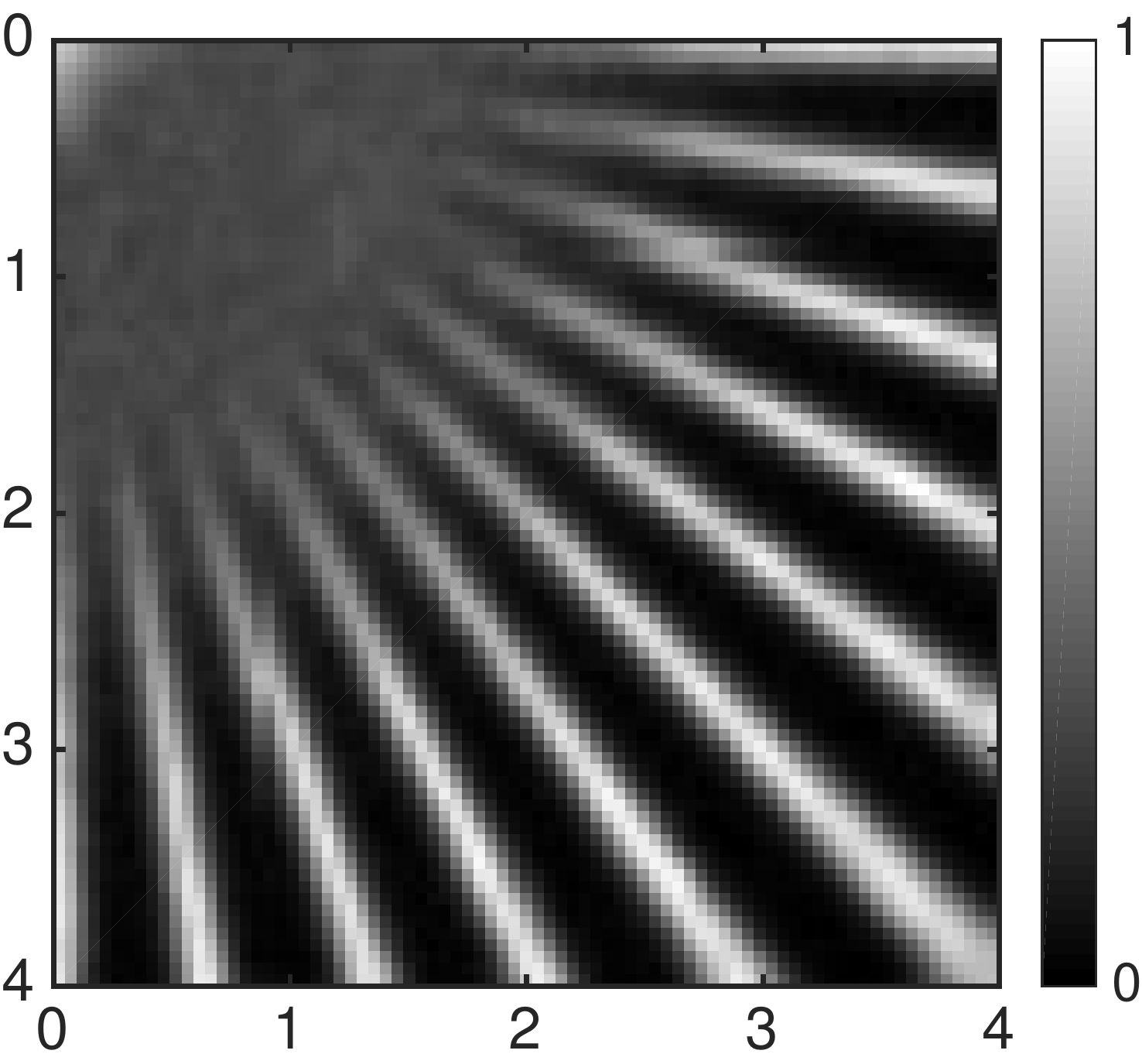}
      \end{tabular}
  \end{tabular}
  \caption{%
    \textbf{Speckle patterns (continued):} Penalized joint Blind-SIM
    reconstructions from standard speckle (left) and  ``squared'' 
    speckle (right) patterns. The number of illumination patterns
    considered  for reconstruction is $M=10$ (A),  $M=200$ (B) and 
    $M=10000$ (C).
  }
  \label{fig:fig4bis}
\end{figure}

\subsection{Speckle illumination patterns} 
\label{Results_speckle}
We now consider second-order stationary speckle illuminations 
$\Ib_m$ with known first order statistics $I_{0;n}=I_0, \forall
n$. Each one of these patterns is a fully-developed speckle 
drawn from the point-wise intensity of a correlated circular 
Gaussian random field. 
The correlation is adjusted so that the pattern $\Ib_m$ exhibits a 
spatial correlation of the form \eqref{psf} but with ``numerical 
aperture'' parameter NA$_{\text{ill}}$ that sets the correlation length 
to $\frac{\lambda}{2 \,\text{NA}_{\text{ill}}}$ within the random field. As an illustration, the speckle pattern shown in 
Fig.~\ref{fig:fig4}(A-left) was generated in the standard 
case\footnote{%
It is usually considered that $\text{NA}_{\text{ill}}=\text{NA}$ if the 
illumination and the collection of the fluorescent light are 
performed through the same optical device.} 
$\text{NA}_{\text{ill}}=\text{NA}$.
%
{From this set of regular (fully-developed) speckle
patterns, we also consider another set of random illumination 
patterns built by squaring each speckle
pattern, see Fig.~\ref{fig:fig4}(A). These ``squared'' 
patterns are considered hereafter because they give a deeper 
insight about the SR mechanism at work in joint Blind-SIM. 
Moreover, we discuss later in this subsection that these patterns 
can be generated with other microscopy techniques, hence extending 
the concept of random illumination microscope to other optical 
configurations.
From a statistical viewpoint, the 
probability distribution function (pdf) of ``standard'' 
and ``squared''  speckle patterns differ.
For instance, the pdf of the squared speckle intensity
is more concentrated around zero\footnote{%
Assuming a fully-developed speckle, the fluctuation in $I_{m;n}$  
is driven by an  exponential pdf with parameter $I_0$ whereas the
pdf of the ``squared'' point-wise intensity  $J_{m;n} := I_{m,n}^2$ 
is a Weibull distribution with \textit{shape parameter} $k=0.5$ 
and \textit{scale parameter} $\lambda=I_0^2$.} 
than the exponential pdf of the standard speckle intensity.
} 
In addition, the spatial
correlation is also changed since the power spectral density of 
the ``squared'' random field spans twice the initial support of 
its speckle counterpart \cite{denk90}.
As a result, the ``squared'' speckle grains are 
sharper, and they enjoy larger spatial separation. According 
to previous SR theoretical results \cite[p. 57]{Donoho92a} (see also the beginning of Sec.~\ref{penalize}),
these features may bring more SR in joint Blind-SIM than 
standard speckle patterns. This assumption was indeed corroborated 
by our simulations. For instance, the reconstructions in Fig.~\ref{fig:fig4}(B) 
were obtained from a single set of $M=1000$ speckle
patterns such that $\text{NA}_{\text{ill}} = \text{NA}$: in this case,  the
``squared'' illuminations (obtained by squaring the speckle 
patterns) provide a higher level of SR than the standard 
speckle illuminations. Figure~\ref{fig:fig4bis} shows how the
reconstruction quality varies with the number of illumination
patterns. 
%
{With very few  illuminations, the sample is retrieved 
in the few places that are activated by the ``hot spots''   of the 
speckle patterns.  This actually illustrates that the joint
Blind-SIM approach is also an ``activation'' strategy in the spirit 
of PALM \cite{betzig06} and STORM \cite{rust06}. With our strategy, 
the activation process is nevertheless enforced by the structured 
illumination patterns and not by the fluorescent markers staining the 
sample. 
}
This effect is more visible with the squared illumination patterns and, 
with these somehow sparser illuminations, the number of patterns
needs to be increased so that the fluctuations in $\sum_m\Ib_m$ 
is moderate, hence making the equality \eqref{critere1b} a 
legitimate constraint. 
We also stress that these simulations corroborate the empirical 
statement that $M\approx 9$ harmonic illuminations and 
$M\approx 200$ speckle illuminations produce comparable 
super-resolved reconstructions, see Fig.~\ref{fig:fig3}(C-left) and 
Fig.~\ref{fig:fig4bis}(B-left). Obviously, imaging with 
random speckle patterns remains an attractive strategy since it is 
achieved with a very simple experimental setup, see \cite{Mudry12} for details. 
\begin{figure}[t]
  \centering
   \begin{tabular}{@{\kern0pt}l@{\kern0pt}l@{\kern-2pt}l}
   (A)
    &
      \begin{tabular}{l}
        \includegraphics[height=0.19\textwidth]{
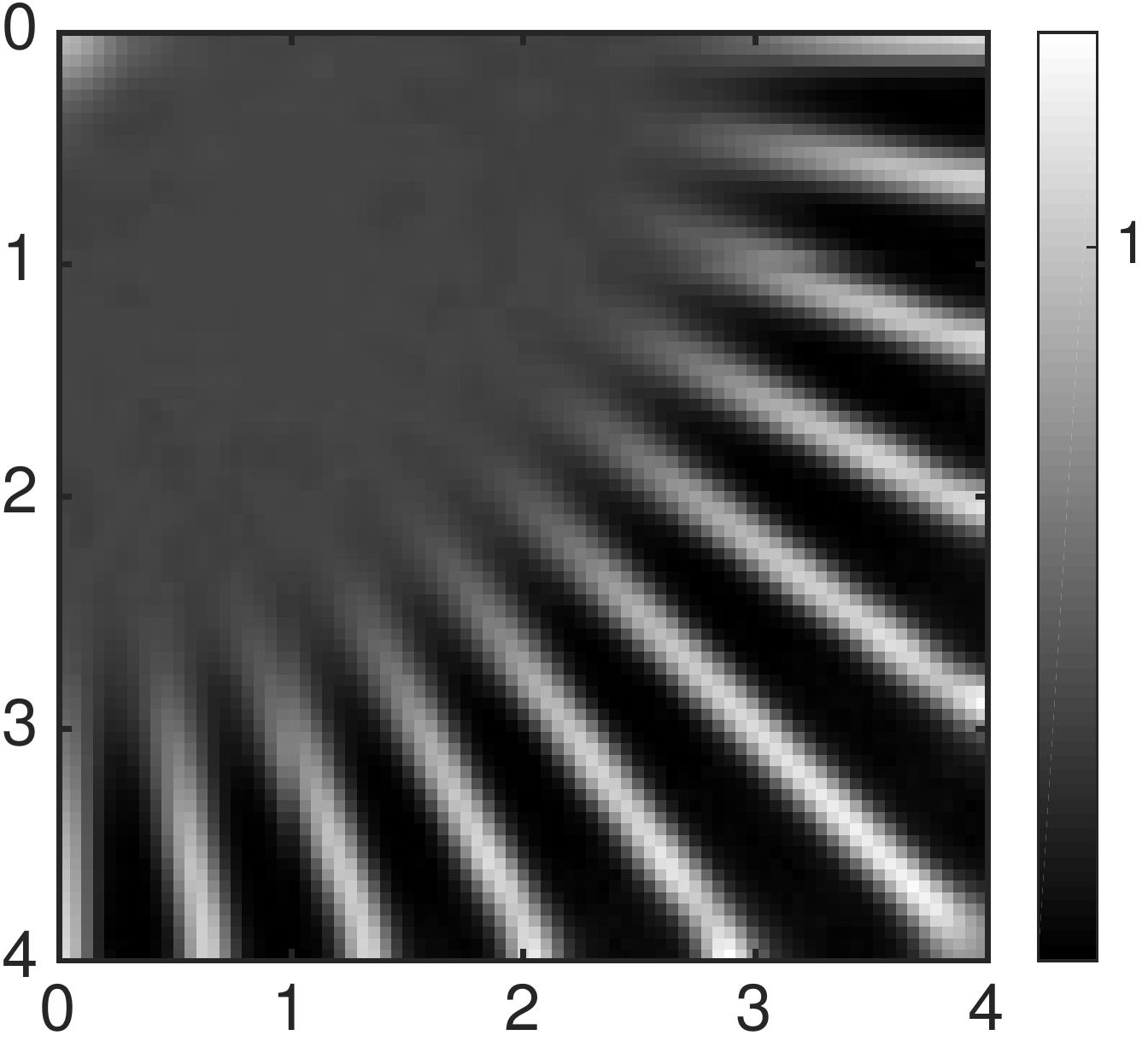}
      \end{tabular}
    &
      \begin{tabular}{l}
        \includegraphics[height=0.19\textwidth]{
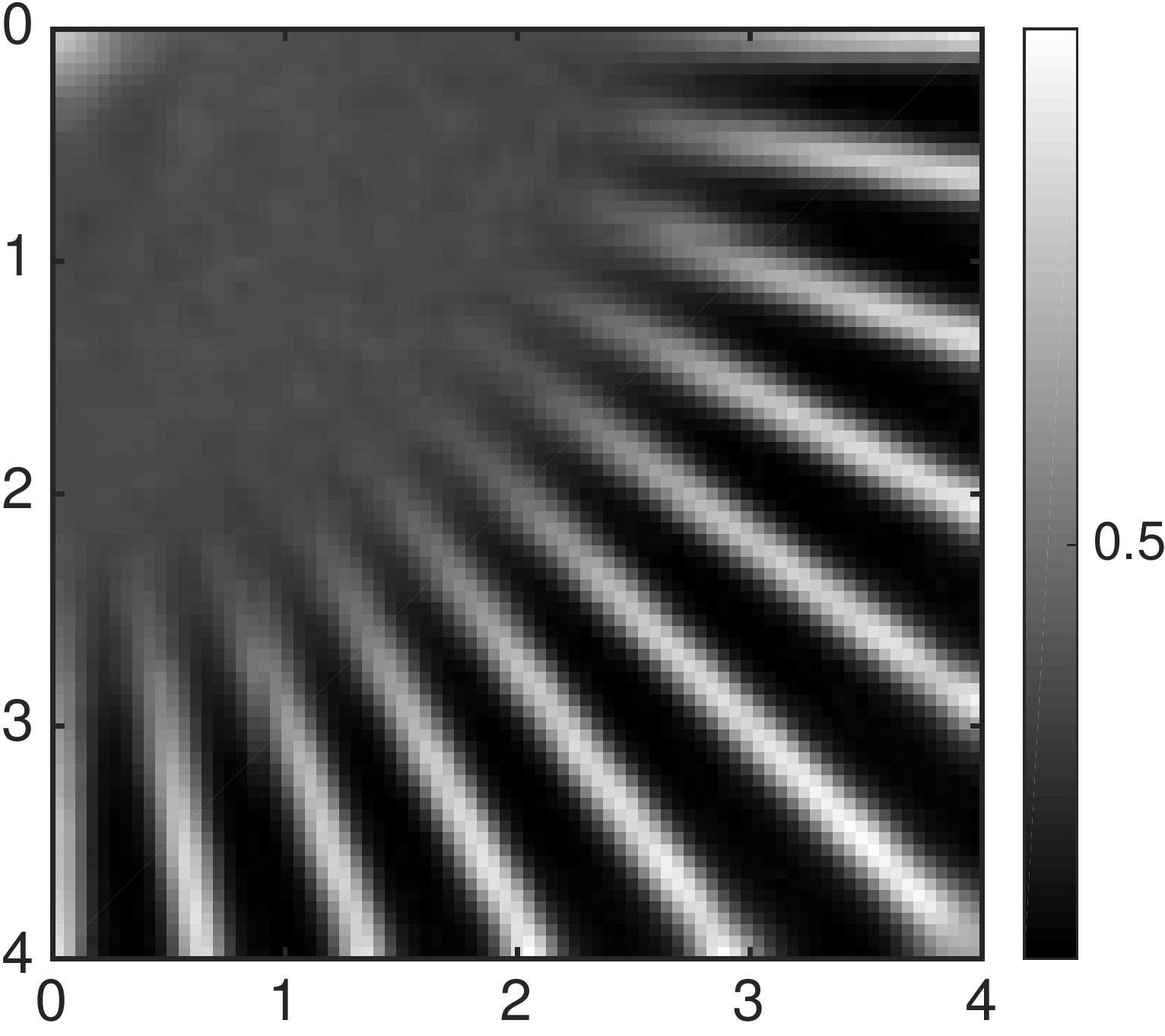}
      \end{tabular}\\[1em]
    (B)
    &
      \begin{tabular}{l}
        \includegraphics[height=0.19\textwidth]{
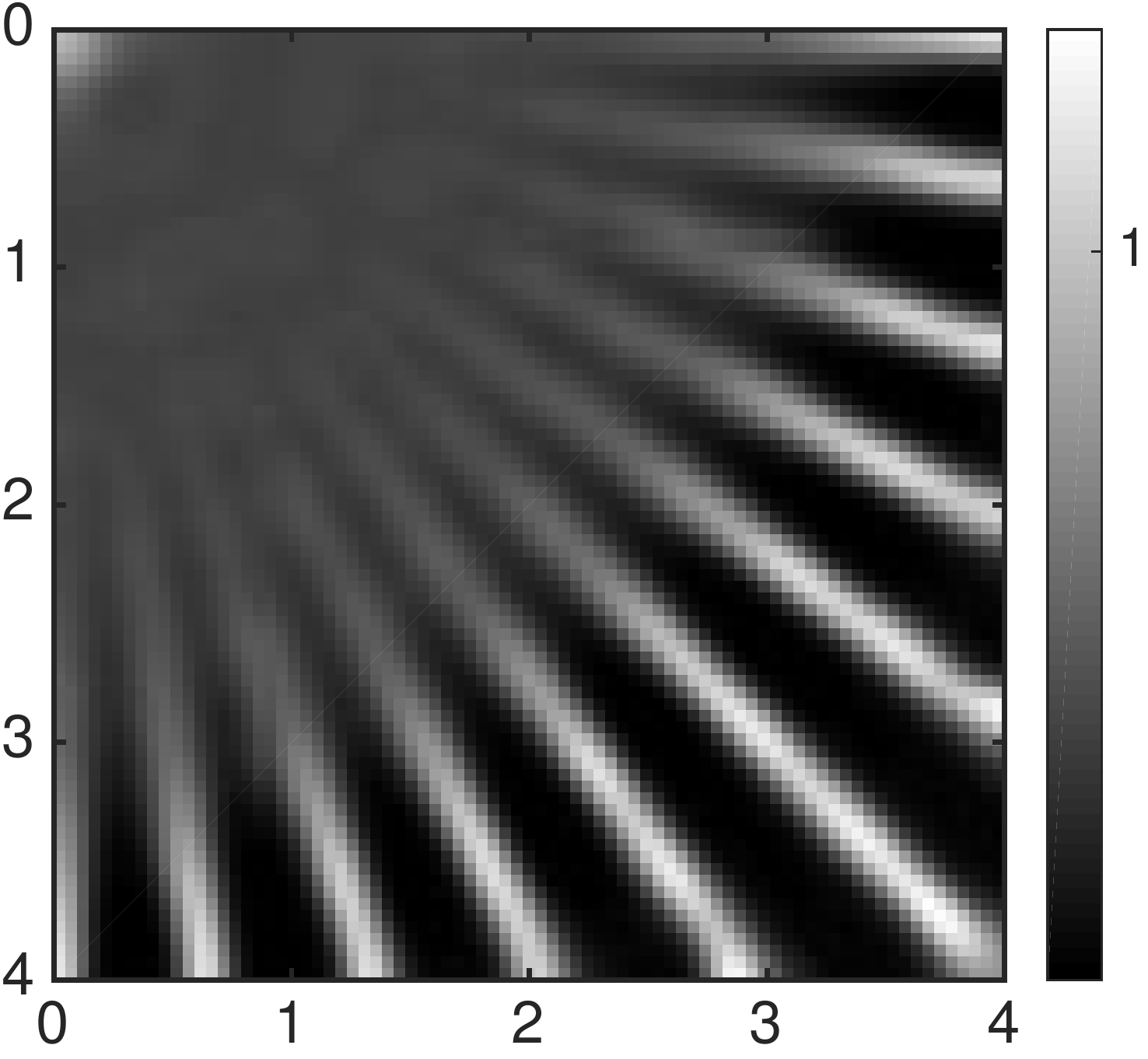}
      \end{tabular}
    &
      \begin{tabular}{l}
       \includegraphics[height=0.19\textwidth]{
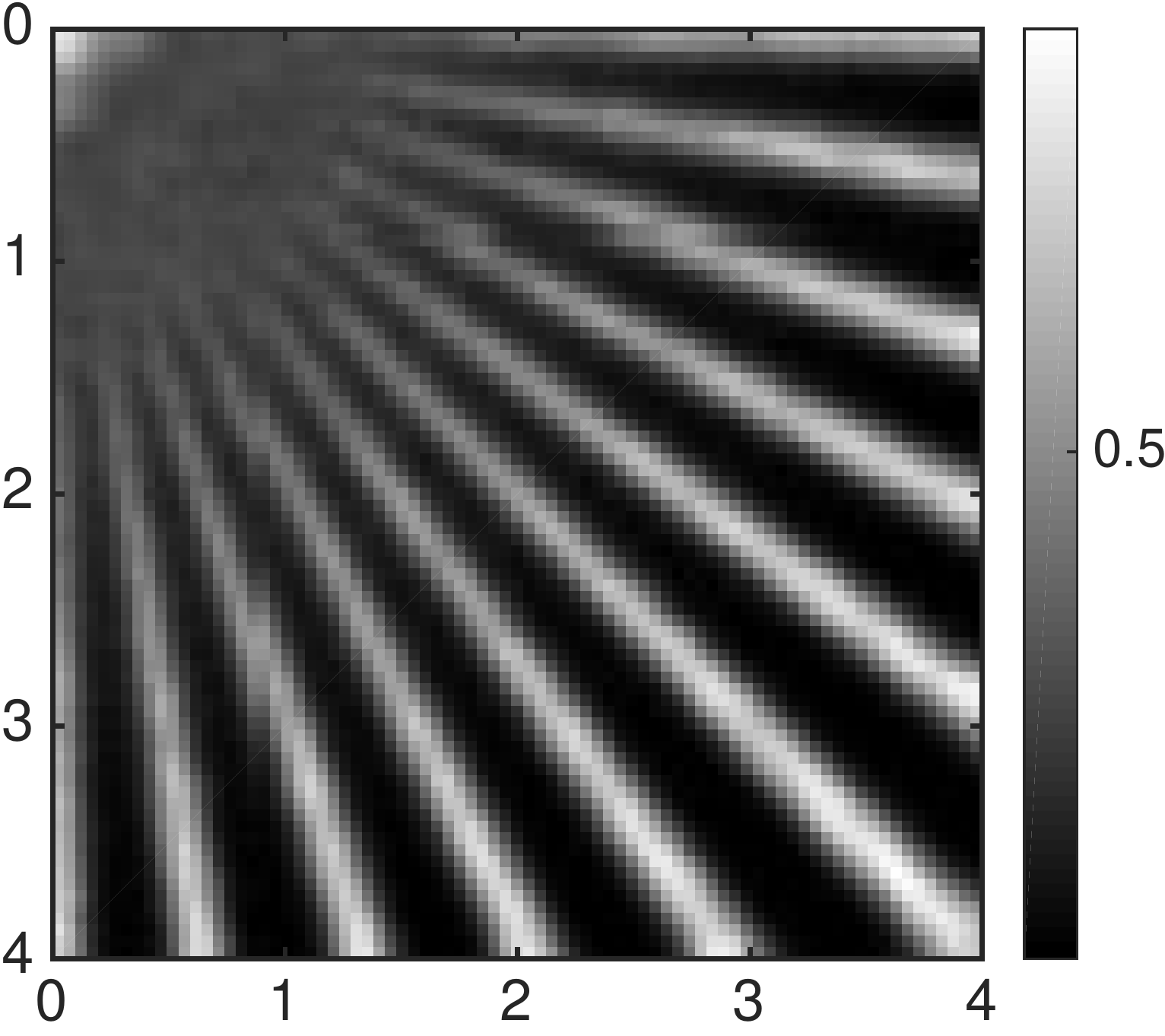}
      \end{tabular}\\[1em]
    (C)
    &
      \begin{tabular}{l}
        \includegraphics[height=0.19\textwidth]{
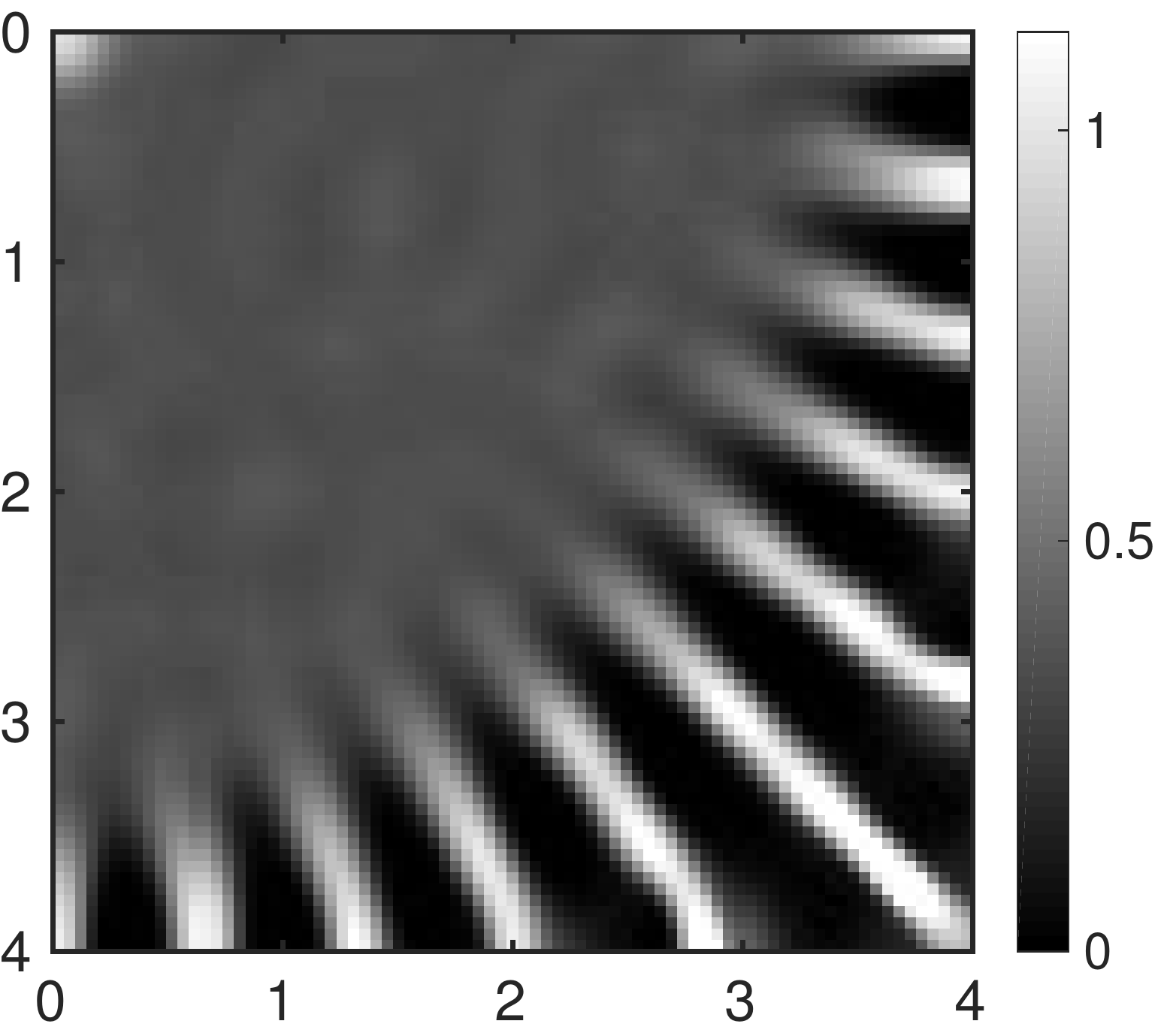}
      \end{tabular}
    &
      \begin{tabular}{l}
       \includegraphics[height=0.19\textwidth]{
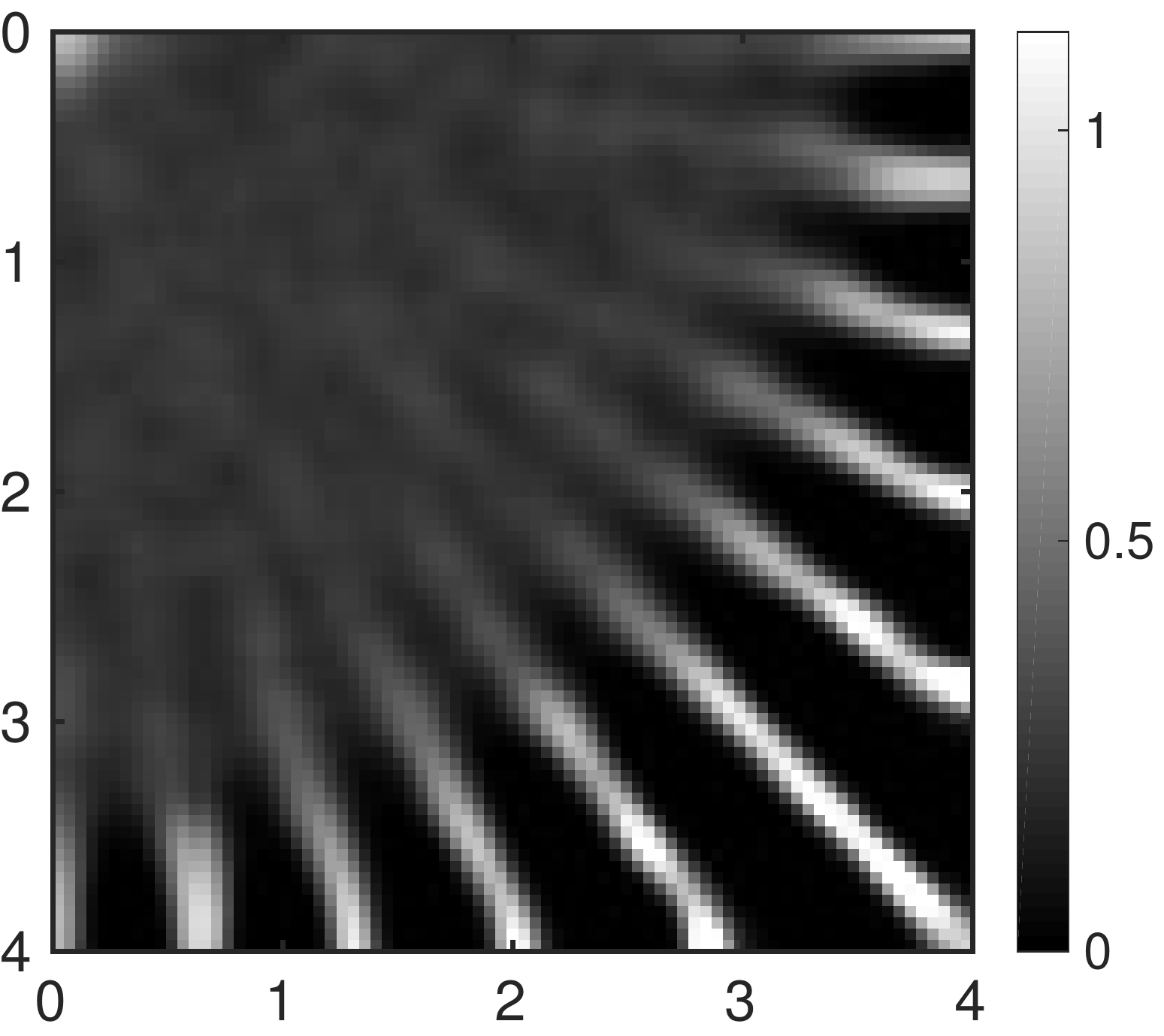}
      \end{tabular}
  \end{tabular}
  \caption{%
    \textbf{Speckle patterns (continued):} The correlation length of
    speckle and ``squared'' speckle patterns drives the level of  
    super-resolution in the penalized joint Blind-SIM reconstruction: 
    [\textit{Rows} A] reconstruction from $M=10000$ speckle patterns 
    with $\text{NA}_{\text{ill}} = 0.5\,\text{NA}$ (left) and from the 
    corresponding ``squared'' random-patterns (right). 
    [\textit{Rows} B] \textit{idem} with $\text{NA}_{\text{ill}} =    2\,\text{NA}$. 
    [\textit{Rows} C] \textit{idem} with uncorrelated patterns.  
  }
  \label{fig:fig5}
\end{figure}
For both random patterns, we also note that increasing the
correlation length  above the Rayleigh distance $\frac{\lambda}{2 \,\text{NA}}$ (\textit{i.e.}, 
setting $\text{NA}_{\text{ill}} < \text{NA}$) deteriorates the SR
whereas, conversely, taking $\text{NA}_{\text{ill}} = 2 \text{NA}$
enhances it, see Fig.~\ref{fig:fig5}-(A,B). However, the resolving power of the joint Blind-SIM estimate 
 deteriorates if the correlation length is further decreased; for
instance, uncorrelated speckle patterns are finally found to hardly 
produce any SR, see Fig.~\ref{fig:fig5}-(C). Indeed,
with arbitrary small correlation lengths, many ``hot spots'' tend 
to be generated within a single Rayleigh distance, leading to this 
loss in the resolving power. Obviously, the ``squared'' speckle 
patterns are less sensitive to this problem because they are
inherently sparser. 

{Finally, the experimental relevance of the simulations involving 
``squared'' speckle illuminations needs to
be addressed. Since a two-photon (2P) fluorescence interaction is 
sensitive to the \textit{square} of the intensity \cite{Gu95}, most of 
these simulations can actually be considered as wide-field  2P
structured illumination experiments. Unlike one-photon (\textit{i.e.}, 
fully-developed) speckle illuminations\footnote{{With one-photon
  interactions, the Stokes shift \cite{Lakowicz06} implies that the excitation 
  and the fluorescence wavelengths  are not strictly equivalent. 
  The difference is however negligible in practice (about 10\%), hence our assumption 
  that one-photon interactions occur with identical wavelengths for 
  both the excitation and the collection.}}, though, a 2P interaction requires 
an excitation wavelength $\lambda_{\text{ill}}\sim 1000$ nm
that is roughly twice the one of the collected fluorescence 
$\lambda_{\text{det}}\sim 500$ nm. The lateral 2P correlation length 
being $\frac{\lambda_{\text{ill}}}{4  \text{NA}_{\text{ill}}}$, epi-illumination 
setups with one-photon (1P) and 2P illuminations provide 
similar lateral correlation lengths. This 2P instrumental
configuration is simulated in Fig.~\ref{fig:fig5}(A-right), which 
does not show any significant SR improvement with respect to
1P epi-illumination interaction shown in Fig.~\ref{fig:fig4bis}(C-left).
The increased  SR effect driven by ``squared'' illumination patterns 
can nevertheless be obtained with 2P interactions if the excitation 
and the collection are performed with separate objectives. For
instance, the behaviors shown in Fig.~\ref{fig:fig4bis}(C-right) 
and in Fig.~\ref{fig:fig5}(B-right) can be obtained if the excitation 
NA is, respectively, twice and four times the collection NA. 
With these configurations, the 2P excitation exhibits a 
correlation length which is significantly smaller than the one 
driven by the objective PSF, and a strong SR improvement is 
observed in simulation by joint Blind-SIM.   The less spectacular 
simulation shown in Fig.~\ref{fig:fig5}(C-right) can also be considered 
as a 2P excitation, in the ``limit'' case of a very low collection 
NA. 
The 1P simulation shown in Fig.~\ref{fig:fig5}(C-left) rather mock a 
photo-acoustic imaging experiment\cite{Chaigne16}, an imaging 
technique for which the illumination lateral correlation length is 
negligible with respect to the PSF width.  
}

{As a final remark, we stress that 2P interactions 
are not the only way to generate sparse illumination patterns 
for the joint Blind-SIM. In particular, super-Rayleigh speckle 
patterns \cite{Bromberg14} are promising candidates for that 
purpose. 
}

\begin{figure*}[t]
  \centering
  \begin{tabular}{l@{\kern15pt}l@{\kern15pt}l@{\kern15pt}l}
    (A)
    &
       \begin{tabular}{l}
	  \includegraphics[height=0.27\textwidth] {
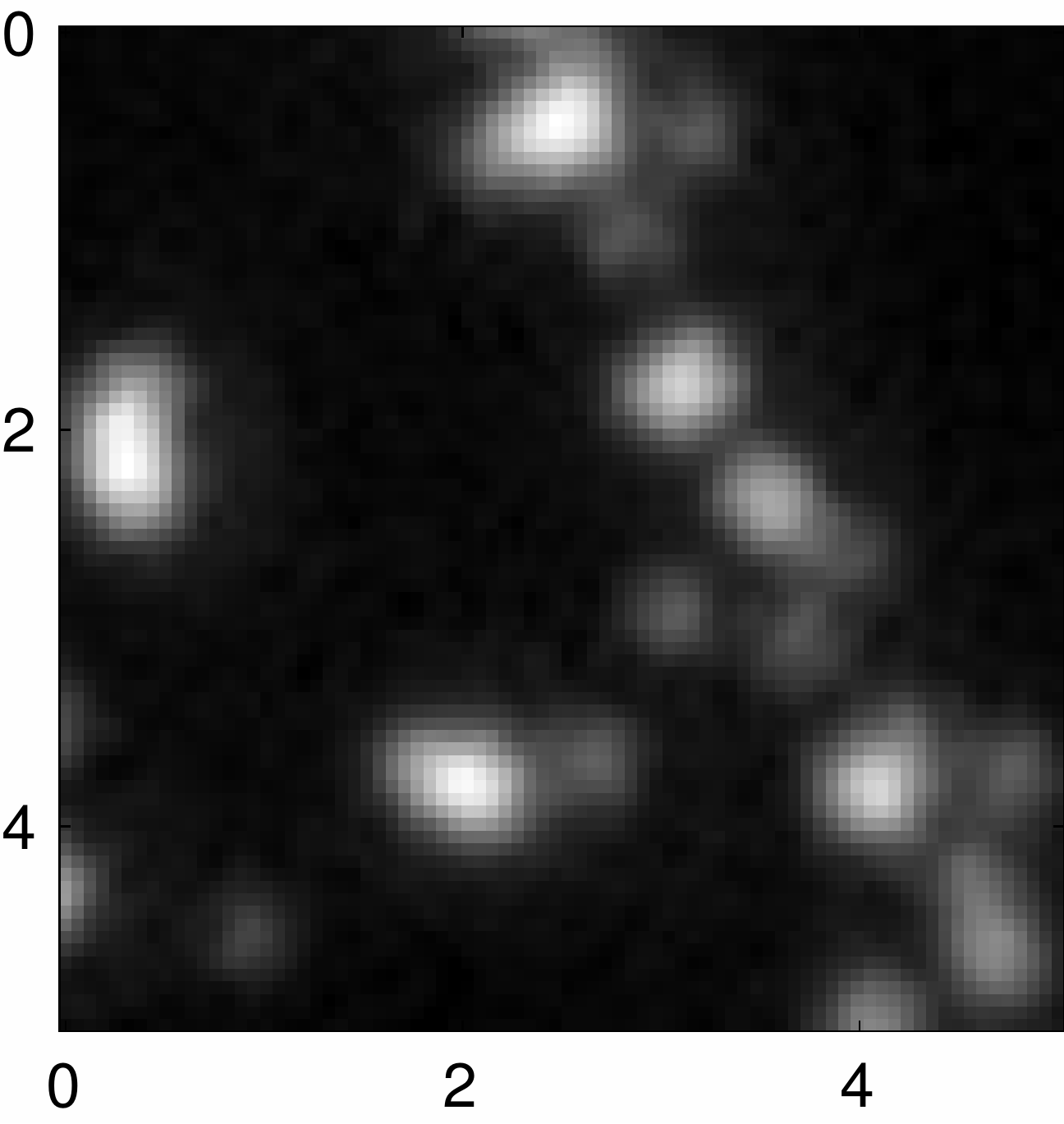}
      \end{tabular}
    &
      \begin{tabular}{l}
	  \includegraphics[height=0.27\textwidth] {
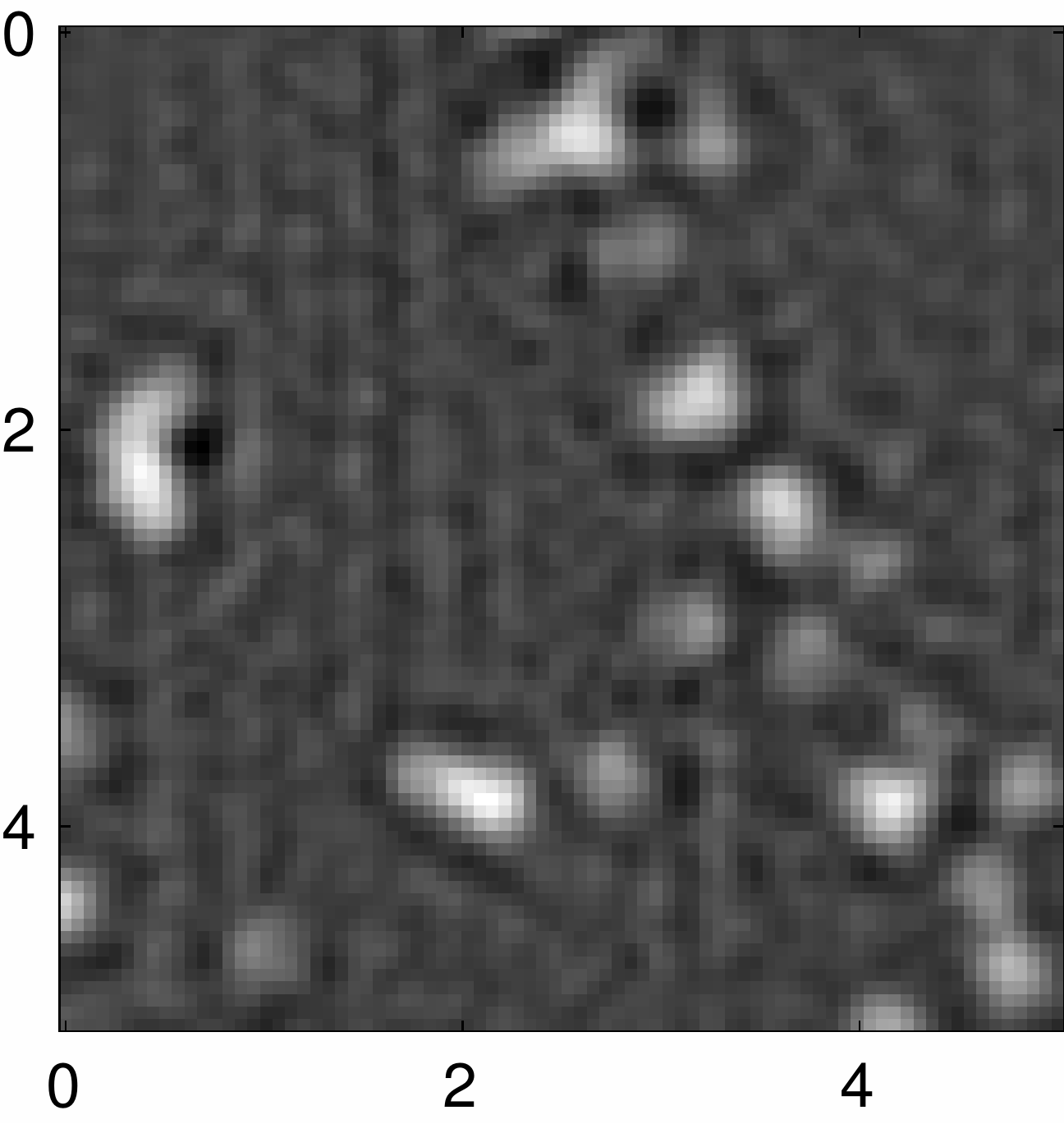}
      \end{tabular}
    &
      \begin{tabular}{l}
	  \includegraphics[height=0.27\textwidth]{
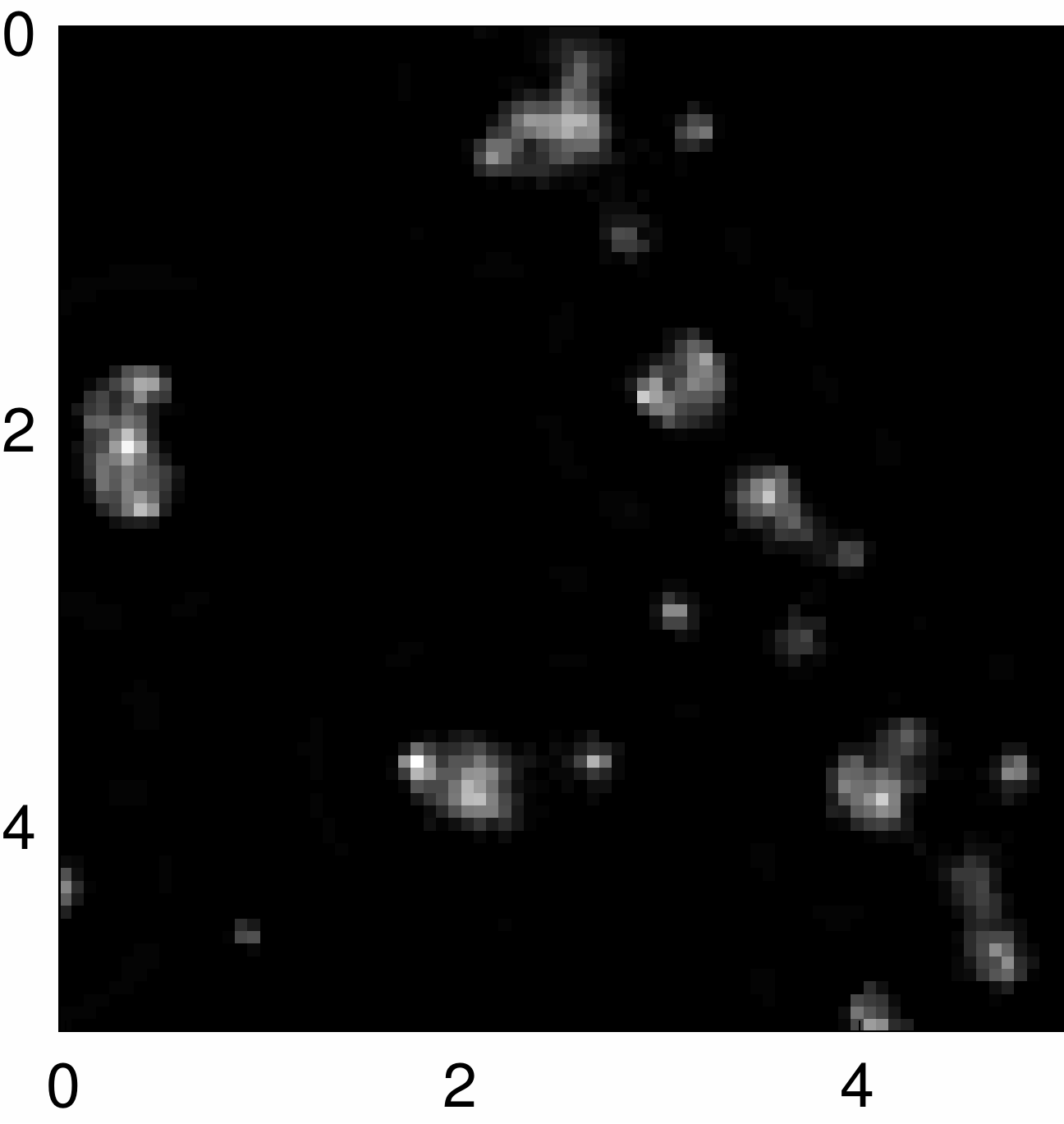}
      \end{tabular}\\
    (B)
    &
      \begin{tabular}{l}
	  \includegraphics[height=0.27\textwidth] {
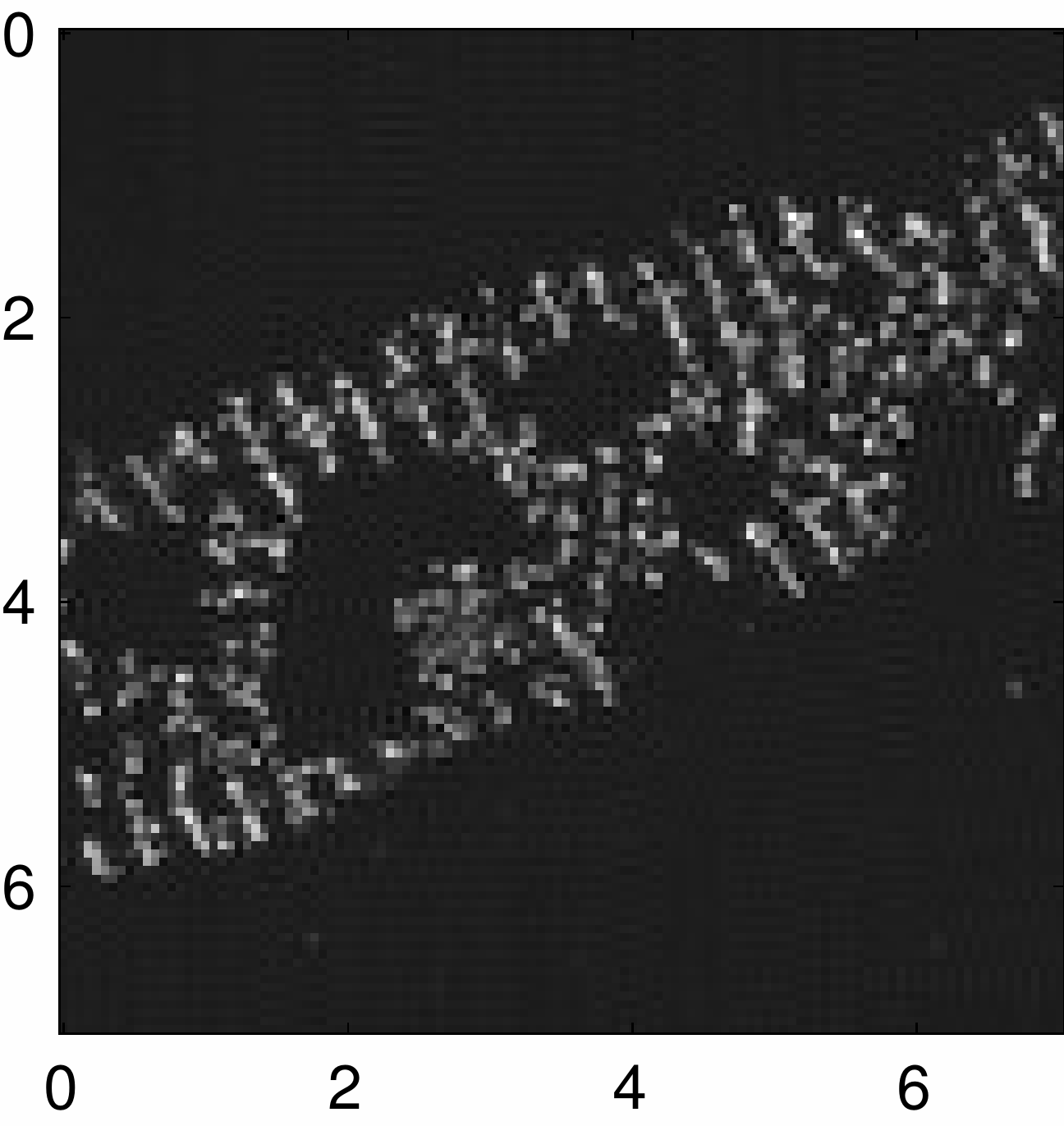}
      \end{tabular}
    &
      \begin{tabular}{l}
	  \includegraphics[height=0.27\textwidth] {
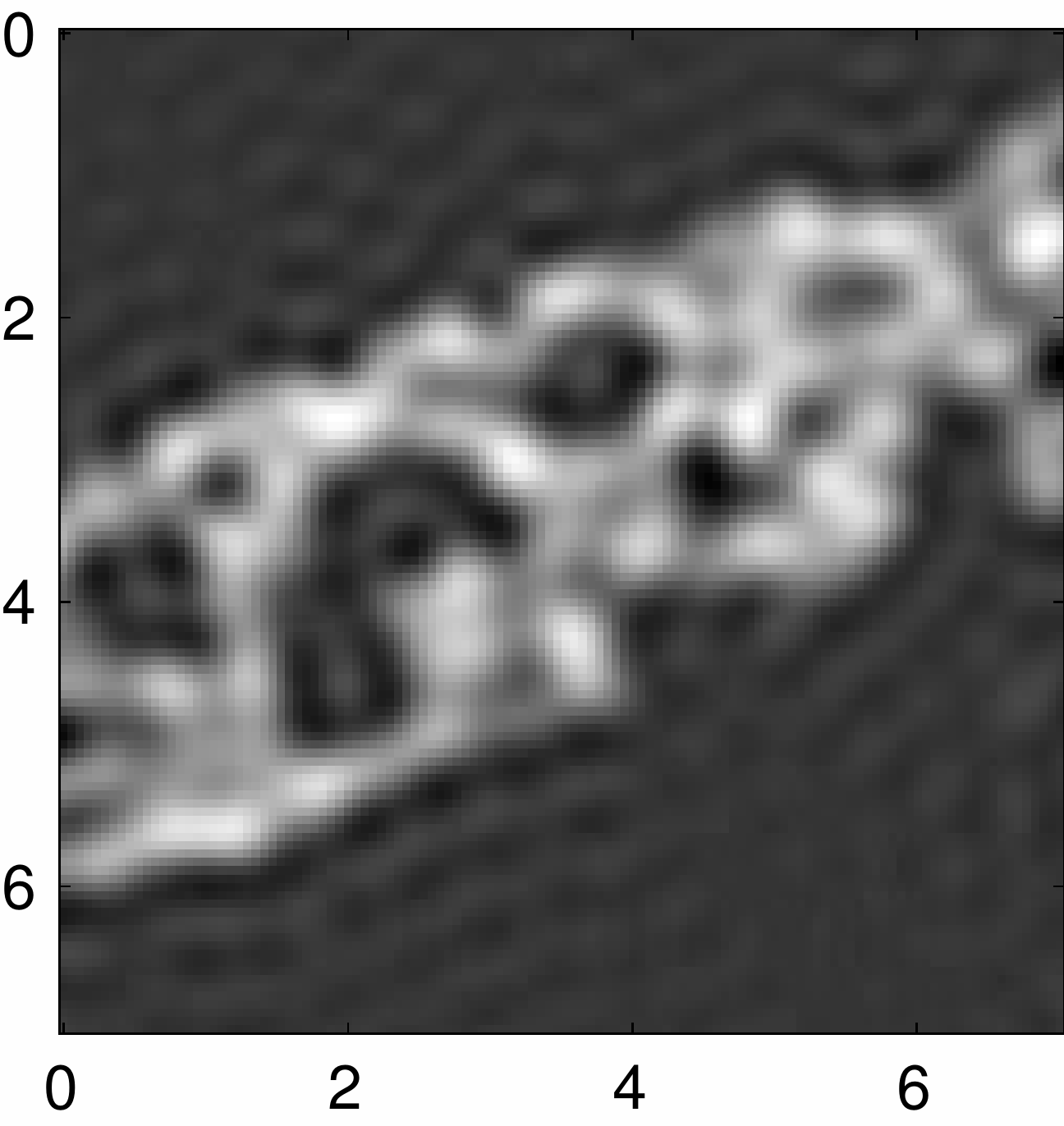}
      \end{tabular}
    &
      \begin{tabular}{l}
	  \includegraphics[height=0.27\textwidth]{
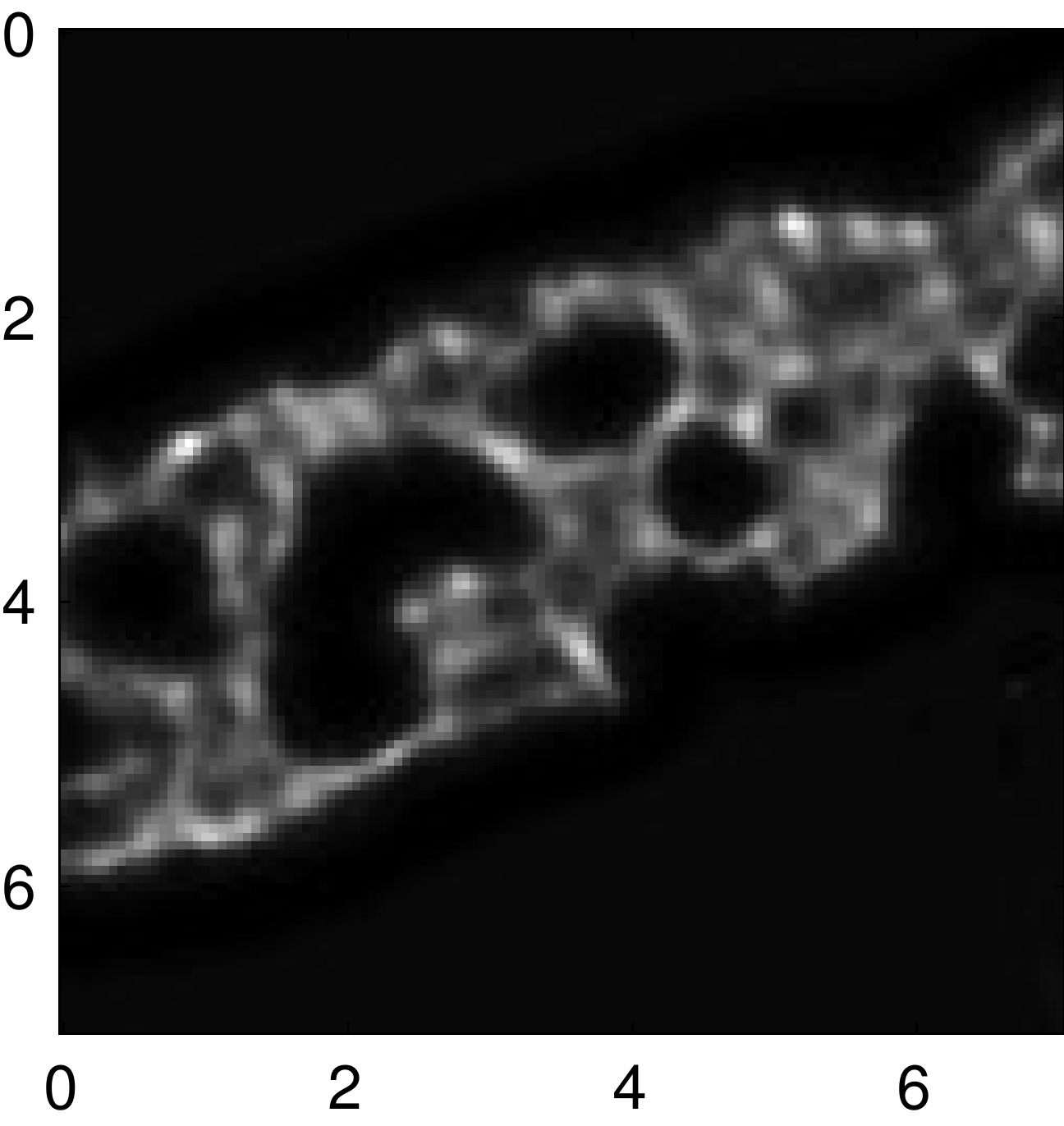}
      \end{tabular}
  \end{tabular}
  \caption{{%
{\bf Processing of real and mock data:} [\textit{Row} A] Fluorescent beads with
    diameters of 100 nm are illuminated by 100 fully-developed
    (\textit{i.e.,} one-photon) speckle   patterns through an
    illumination/collection objective ($\text{NA}=1.49$). The sum of the
    acquisitions of the  fluorescent light (left) and  its Wiener
    deconvolution (middle) provide diffraction limited images of 
    the beads.  The joint Blind-SIM reconstruction performed 
    with the hyper-parameters set to $\beta=5\times 10^{-5}$ and 
    $\alpha=0.4$ is significantly more resolved (right). The sampling rate used in these images is
    32.5 nm, corresponding to an up-sampling factor of two with
    respect to the camera sampling. 
    [\textit{Row} B] STORM reconstruction 
    of a marked rat neuron showing a lattice structure with a 190-nm 
  periodicity (left).   
  Deconvolution of the simulated wide-field image (middle). 
  Joint Blind-SIM reconstruction of the 
  sample obtained from 300 (one-photon) speckle patterns; 
  the hyper-parameters are set to $\beta=2\times 10^{-5}$ and 
    $\alpha=1.5$ (right). The sampling rate of the STORM 
    ground-truth image is 11.4 nm. The sampling rate of the 
    joint Blind-SIM reconstruction is 28.5 nm, corresponding 
    to an up-sampling factor of four with respect to the camera 
    sampling. 
		The distance units along the 
    horizontal and vertical axes are given in wavelength
    $\lambda_{\text{coll}}$, \textit{i.e.,} 520 nm in row A and 488 nm
    in row B.}}
  \label{fig:realdata}
\end{figure*}

\subsection{{Some reconstructions from real and mock data}}
{%
The star test-pattern used so far is a simple and
legitimate mean to evaluate the resolving power of our strategy 
\cite{Horstmeyer16}, but it hardly provides a convincing illustration 
of what can be expected with real data. Therefore, we 
now consider the processing of more realistic datasets with joint 
Blind-SIM. In this section, the microscope acquisitions are all designed 
so that the spatial sampling rate is equal or slightly above the Nyquist 
rate $\frac{\lambda}{4\text{NA}}$.  As a consequence, a preliminary 
up-sampling step of the camera acquisitions is performed so that their 
sampling rate reaches that of the super-resolved reconstruction.
} 

{As a first illustration, we consider a real dataset resulting from a
  test sample composed of fluorescent beads  with diameters 
  of 100~nm. A set of $100$ one-photon speckle patterns is 
  generated by  a spatial light modulator and a  laser source operating at  $\lambda_{\text{ill}} 
  = 488$~nm. The fluorescent light at  $\lambda_{\text{coll}} = 520$
  nm is collected through an objective with $\text{NA}= 1.49$ and 
  recorded by a camera. The excitation and the collection 
  of the emitted light are performed through the same objective, 
  \textit{i.e.,}  the setup is in epi-illumination mode. The total
  number of photons per camera pixels is about $65\,000$.
  In the perspective of further processing, this set of  camera 
  acquisitions is first up-sampled  with a factor of two.
  Figure~\ref{fig:realdata}(A-left) shows the sum of these (up-sampled) 
  acquisitions,  which is similar  to  a wide-field image. Wiener
  deconvolution of this image can be 
  performed  so that all spatial frequencies transmitted by the 
  OTF are equivalently contributing in a diffraction-limited image 
  of the beads, see Figure~\ref{fig:realdata}(A-middle).  The processing of 
  the dataset by the joint  Blind-SIM strategy shown  in 
  Figure~\ref{fig:realdata}(A-right) reveals several beads that are 
  otherwise unresolved on the diffraction-limited images, hence 
  demonstrating a  clear SR effect. 
  In this case, the distance between the closest 
  pair of resolved beads provides an upper bound for the final 
  resolution, that is $\lambda_{\text{coll}}/5$. 
}

{The experimental demonstration above does not 
  involve any biological sample, and we now consider a   simulation designed to be close to a “real-world” biological
  experiment. More specifically, the STORM reconstruction 
  of a marked neuron\footnote{%
    A rat hippocampal neuron in culture labelled with an
    anti-$\beta$IV-spectrin primary and a donkey anti-rabbit 
    Alexa Fluor 647 secondary antibodies, imaged by STORM 
    and processed similarly to \cite{Leterrier15}.
  }  is used  
as a  ground truth to simulate a series of microscope 
  acquisitions generated from one-photon speckle illuminations. 
  Our simulation considers 300 illuminations and acquisitions, 
  both  performed through the same objective,  at  $\lambda = 488$ nm 
  and with  NA $= 1$.  Each
  low-resolution acquisition  is finally plagued with Poisson noise,  
  the total photon budget being equal to $50\,000$ so that it fits to 
  the one of a standard fluorescence wide-field image. 
  The sample (ground truth) shown in Figure~\ref{fig:realdata}(B-left) 
  interestingly exhibits a lattice structure
  with a 190 nm  periodicity (in average) that is not resolved by the 
  diffracted-limited
  image shown in Figure~\ref{fig:realdata}(B-middle). The  
  joint  Blind-SIM reconstruction in
  Figure~\ref{fig:realdata}(A-right) shows a significant 
  improvement of the resolution, which 
  reveals some parts of the underlying structure.
}

\subsection{Tuning the regularization parameters}
\label{nbom}
\begin{figure}[t]
  \centering
  \begin{tabular}{ll}
	  \includegraphics[height=0.19\textwidth] {
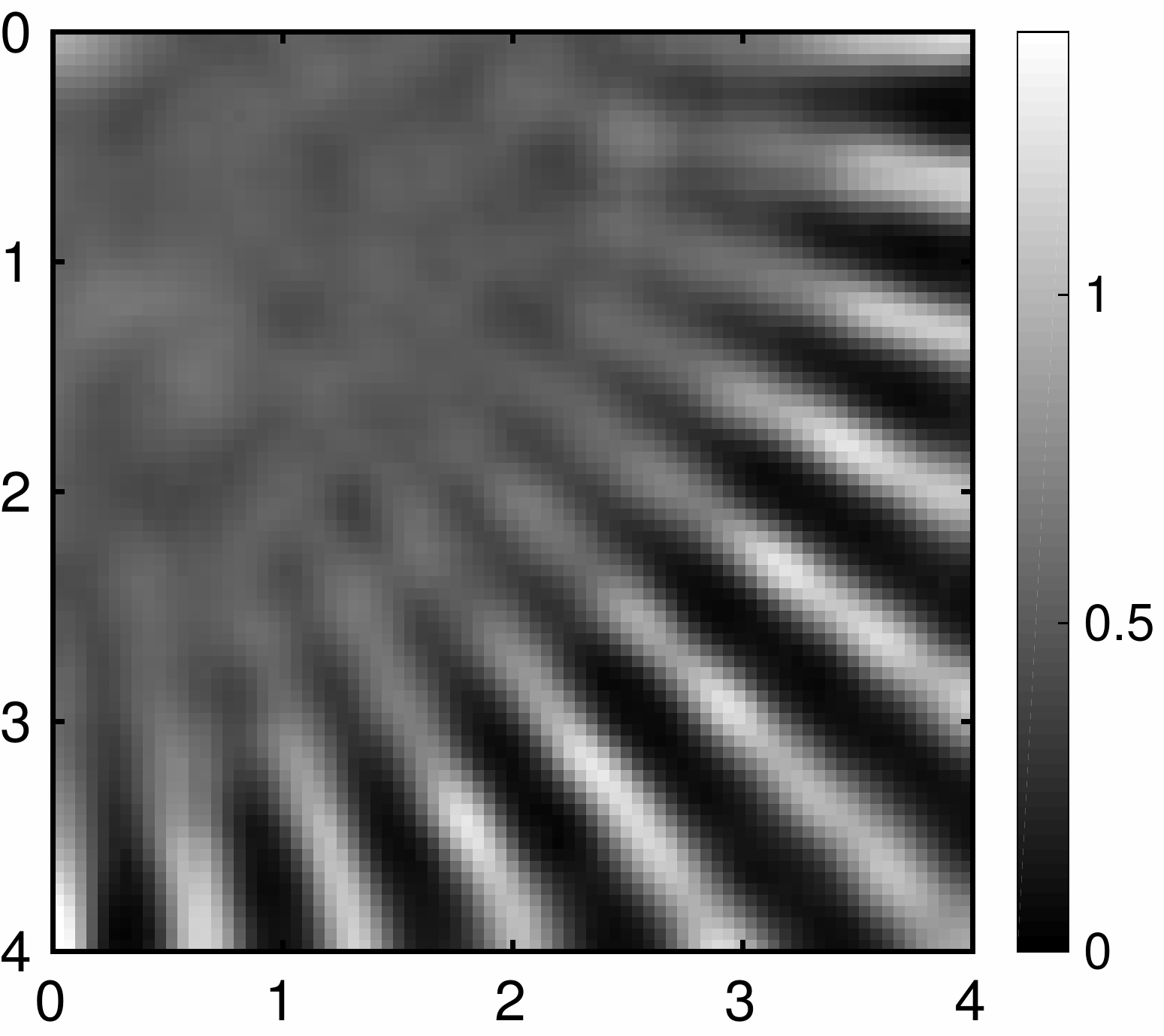}
    &
	  \includegraphics[height=0.19\textwidth]{
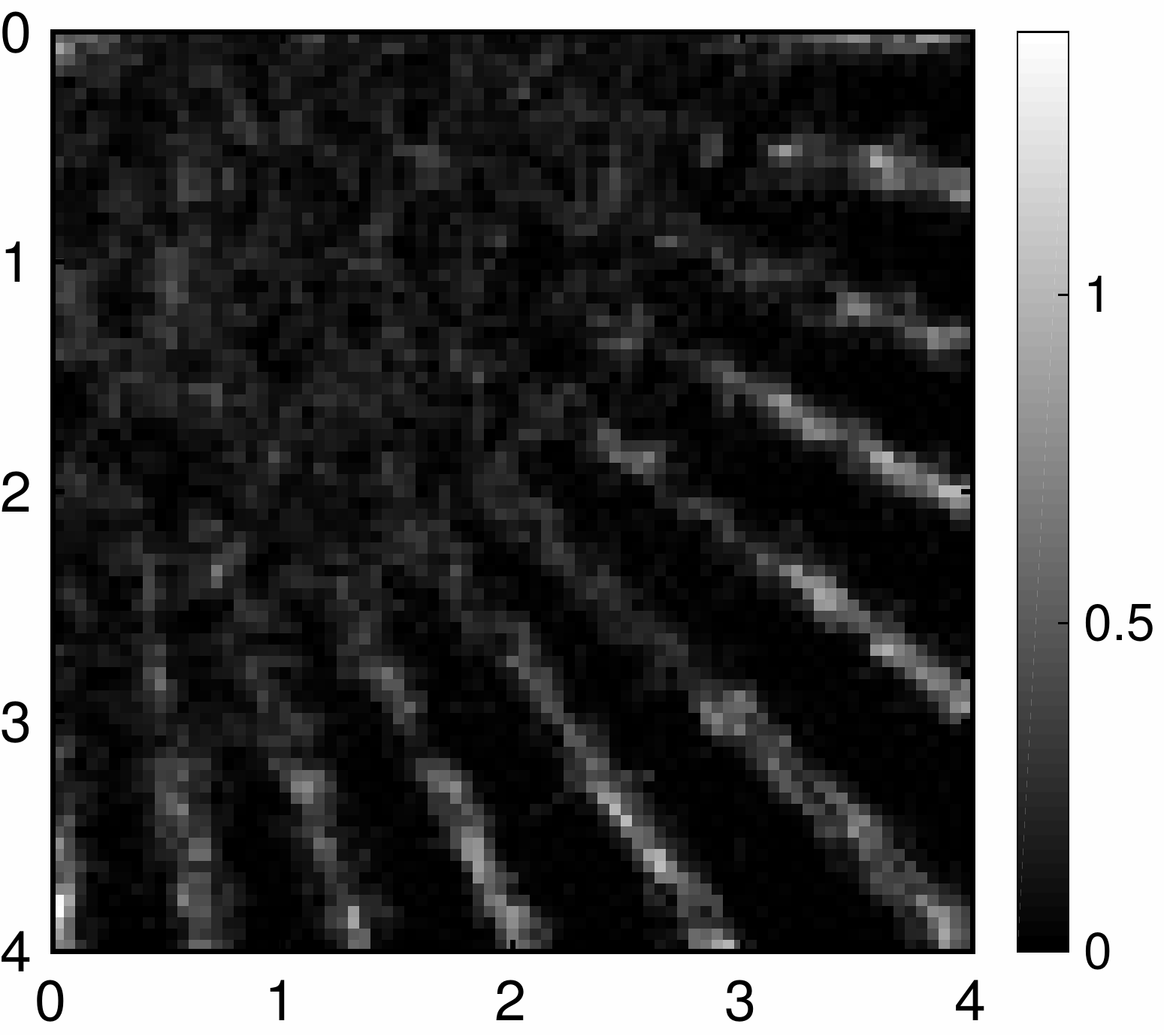}\\
  \end{tabular}
  \caption{%
    Penalized Blind-SIM reconstructions from the dataset used to generate
    the super-resolved reconstruction shown in
    Fig.~\ref{fig:fig4}(B-left). The hyper-parameter $\beta$ was set
    to $10^{-6}$ in any case, and $\alpha$  was set to
    $10^{-3}$ (left) and $0.9$ (right). For the
    sake of comparison, our tuning for the reconstruction shown in 
    Fig.~\ref{fig:fig4}(B-left) is $\beta=10^{-6}$ and $\alpha=0.3$.}
  \label{fig:surreg_downreg}
\end{figure}

The tuning of parameters $\alpha$ 
and $\beta$ in \eqref{hyperbolic} is a pivotal issue since
inappropriate values result in deteriorated reconstructions.
On the one hand, the quadratic penalty in \eqref{hyperbolic} 
was mostly introduced to ensure that the minimizer defined 
by \eqref{critere3} is unique (\textit{via} strict convexity of the 
criterion). 
{However, because high-frequency components in $\widehat{\qb}_m$ 
are progressively damped as $\beta$ increases, the latter parameter
can also be adjusted in order to prevent an over-amplification of the 
instrumental noise. A trade-off should nevertheless be sought since large values of $\beta$ prevent super-resolution to 
occur. For a given SNR, 
$\beta$ is then maintained  to a fixed (usually small) value.
For instance, we chose $\beta=10^{-6}$ for all the simulations 
involving the star pattern in this paper since they were performed
with a rather favorable SNR.
%
%
On the other hand, the
quality of reconstruction crucially depends on parameter $\alpha$. 
More precisely, larger values of $\alpha$ will provide sparser 
solutions $\wh{\qb}_m$, and thus a sparser reconstructed 
object $\wh{\rhob}$. Fig.~\ref{fig:surreg_downreg} shows an	 
example of under-regularized and over-regularized solutions, 
respectively corresponding to a too small and a too large value of $\alpha$.
The prediction of the appropriate level of sparsity to seek for 
each $\qb_m$, or equivalently the tuning of the regularization 
parameter $\alpha$, is not an easy task.
Two main approaches can be considered. One relies on automatic
tuning. For instance, a simple method called Morozov's discrepancy principle 
considers that the least-squares terms $\norm{\yb_m - \Hb\wh{\qb}_m}^2$
should be chosen in proportion with the variance of the additive
noise, the latter being assumed known \cite{Morozov84}. Other possibilities 
seek a trade-off between $\norm{\yb_m - \Hb\wh{\qb}_m}^2$ and 
$\varphi(\wh{\qb}_m)$. This is the case with the L-curve
\cite{Hansen92}, but also with the recent contribution \cite{Song16}, 
which deals with a situation comparable to ours.  Another option relies 
on a Bayesian interpretation of $\wh{\qb}_m$ as a maximum \emph{a
  posteriori} solution, which opens the way to the estimation of 
$\alpha$ marginally of $\qb_m$. In this setting,  Markov Chain 
Monte Carlo sampling \cite{Lucka12} or variational Bayes 
methods \cite{Babacan09} could be employed. 
}
{An alternate approach to automatic tuning consists in relying on 
a calibration step. It amounts to consider that similar acquisition 
conditions, applied to  a given type of biological samples, lead 
to similar ranges of admissible values for the tuning of $\alpha$. 
The validation of such a principle is however outside the 
scope  of this article as it requires various experimental
acquisitions from biological samples with known structures 
(or, at least, with some calibrated test patterns). Concerning the 
examples proposed in the present section, the much simpler strategy consisted in selecting the reconstruction 
which is visually the ``best''  among the reconstructed images 
with varying $\alpha$.}


%

\section{A new preconditioned proximal iteration}
\label{algo}

We now consider the algorithmic issues involved in the constrained 
optimization problem \eqref{critere3}-\eqref{hyperbolic}. For sake of simplicity, the 
subscript $m$  in $\yb_m$ and $\qb_m$ will be dropped. The reader 
should however keep in mind that the algorithms presented below 
only aim at solving one of the $M$ sub-problems involved in the 
final joint Blind-SIM reconstruction. Moreover, we stress that all 
simulations presented in this article are performed with a convolution 
matrix $\Hb$ with a block-circulant with circulant-block (BCCB) 
structure. The more general case of block-Toeplitz with Toeplitz-block
(BTTB) structure is shortly addressed at the end of 
Subsection~\ref{blind-SIM_PPDS}.

At first, let us note that \eqref{critere3}-\eqref{hyperbolic} is an
instance of the more general problem 
%
\begin{equation}
  \label{critereProx}
  \min_{\qb\in\eR^N} \left[ f(\qb) \pardef g(\qb) + h(\qb) \right] 
\end{equation}
where $g$  and $h$ are closed-convex functions that may not share 
the same regularity assumptions: $g$ is supposed to be a smooth 
function with a $L$-Lipschitz continuous gradient $\boldsymbol{\nabla}
g$, but $h$ does not need to be smooth. Such a \textit{splitting} aims at solving constrained 
non-smooth optimization problems by proximal (or forward-backward) 
iterations. The next subsection presents the basic proximal algorithm
and the well-known FISTA  that usually improves the convergence speed.

\subsection{Basic proximal and FISTA iterations} 
\label{standardProx}
We first present the resolution of \eqref{critereProx} in a general
setting, then  the penalized joint Blind-SIM problem  \eqref{critere3}
is addressed as our particular case of interest.
\medskip

\subsubsection{General setting} Let $\qb^{(0)}$ be an arbitrary initial guess, the basic proximal 
update $k\rightarrow k+1$ for minimizing the convex criterion $f$ is \cite{combettes2005,Beck10,combettes11a}
\begin{equation}
  \label{algoGP}
  \qb^{(k+1)} ~\longleftarrow~ \prox{\gamma h} \left( \qb^{(k)} -  \gamma  \boldsymbol{\nabla} g(\qb^{(k)})\right)
\end{equation}
where $\prox{\gamma h}$ is the proximity operator 
(or \textit{Moreau envelope}) of the function $\gamma h$ \cite[p.339]{rockafellar70}
\begin{equation}
  \label{eq:prox1}
  \prox{\gamma h}  (\qb) \pardef \argmin_{\xb \in \eR^N}
  \left[ h(\xb) + \frac{1}{2\gamma}|| \xb - \qb ||^2 \right].
\end{equation}
Although this operator defines the update implicitly, an explicit
form is actually available for many of the functions met in signal 
and image processing applications, see for instance 
\cite[Table 10.2]{combettes11a}. The Lipschitz constant $L$ granted to $\boldsymbol{\nabla} g$  
plays an important role in the convergence of iterations \eqref{algoGP}. 
In particular, global convergence toward a solution of
\eqref{critereProx} occurs as long as the step size $\gamma$ is chosen such 
that $0 <\gamma < 2/L $.  However, the convergence speed 
is usually very low and the following accelerated
version named FISTA~\cite{Beck09}  
is usually preferred 
 \begin{subequations}
  \label{algoFISTA}
\begin{align}
  \label{algo2}
&  \qb^{(k+1)} ~\longleftarrow~ \prox{\gamma h} \left( \omegab^{(k)} -  \gamma  \boldsymbol{\nabla} g(\omegab^{(k)})\right)\\[.5em]
\label{algo3}  
&  \omegab^{(k+1)} ~\longleftarrow~\qb^{(k+1)} + {\textstyle \frac{k-1}{k+2}}  \big(\qb^{(k+1)} - \qb^{(k)}\big).
  \end{align}
\end{subequations}
The convergence speed toward $\min_\qb f(\qb)$ achieved by \eqref{algoFISTA}  
is $O(1/k^2)$, which is  often considered as a substantial gain
compared to the $O(1/k)$ rate of the basic proximal
iteration. It should be noted however that this ``accelerated'' form may not 
always provide a faster convergence speed with respect to its standard 
counterpart, see for instance \cite[Fig. 10.2]{combettes11a}. 
FISTA was nevertheless found to be faster for solving the constrained minimization problem involved in joint Blind-SIM, see Fig.~\ref{fig:precondEffect}. We finally stress that convergence of 
\eqref{algoFISTA} is granted for $0<\gamma < 1/L$~\cite{Beck09}.
\begin{figure}[t]
  \centering
  \begin{tabular}{c@{ }c@{\kern1.cm}c@{ }c}
    \ylabel{\footnotesize   (a) FISTA 10    it.}&\figc[height=0.19\textwidth]{
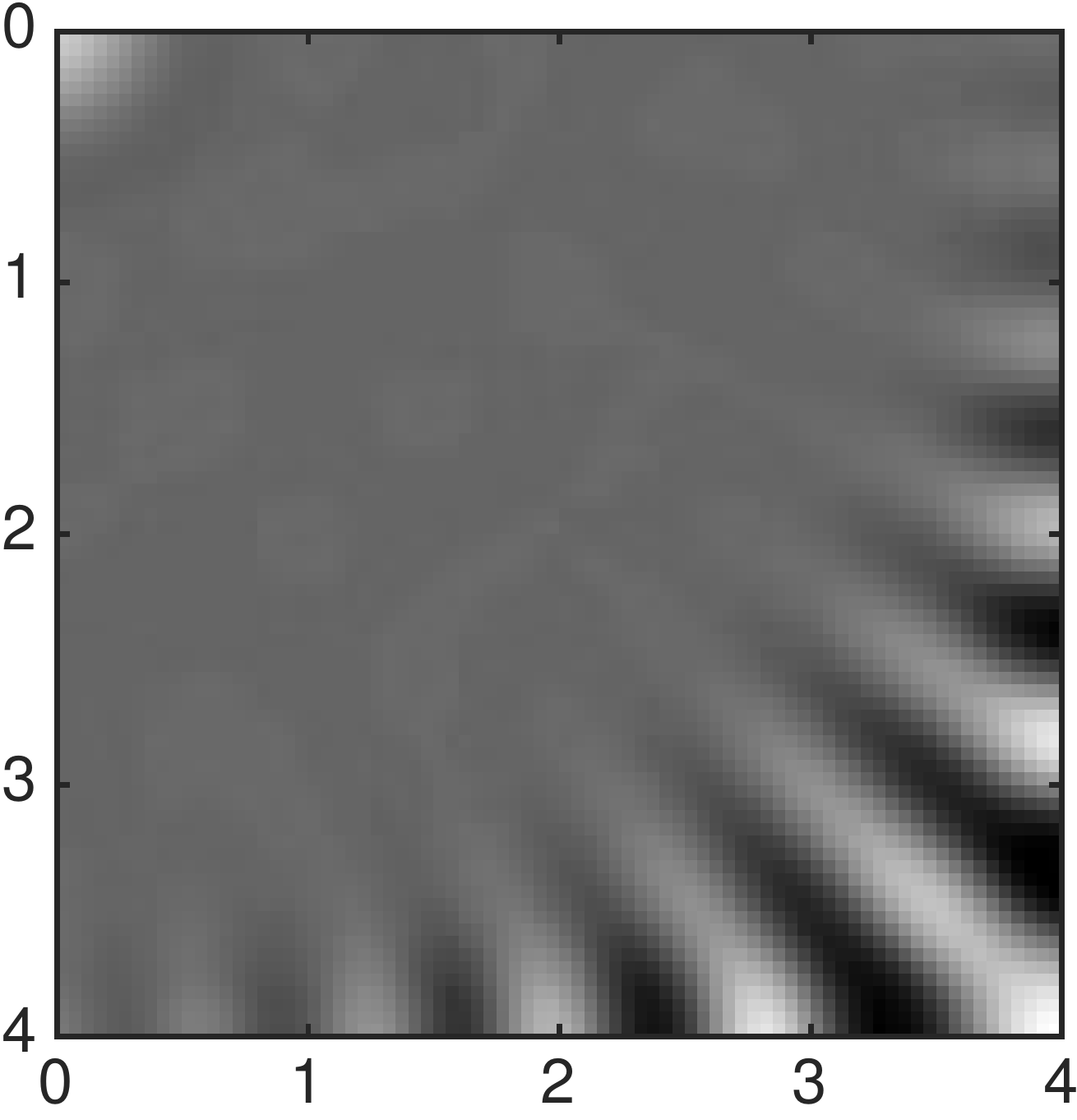}
    &
      \ylabel{\footnotesize   (d) PPDS 10      it.}&\figc[height=0.19\textwidth]{
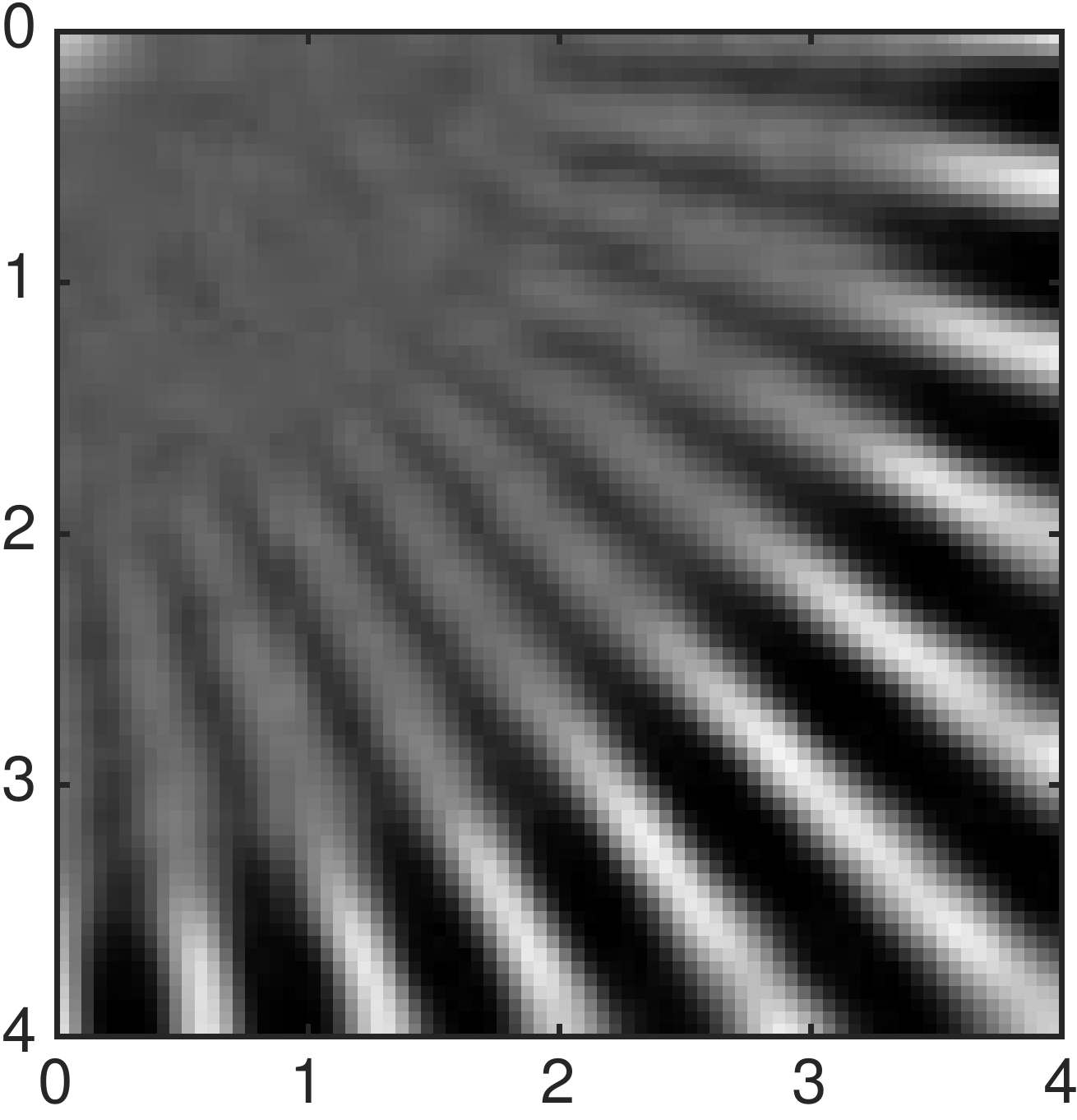}
    \\[.5em]
    \ylabel{\footnotesize   (b) FISTA 50    it.}&\figc[height=0.19\textwidth]{
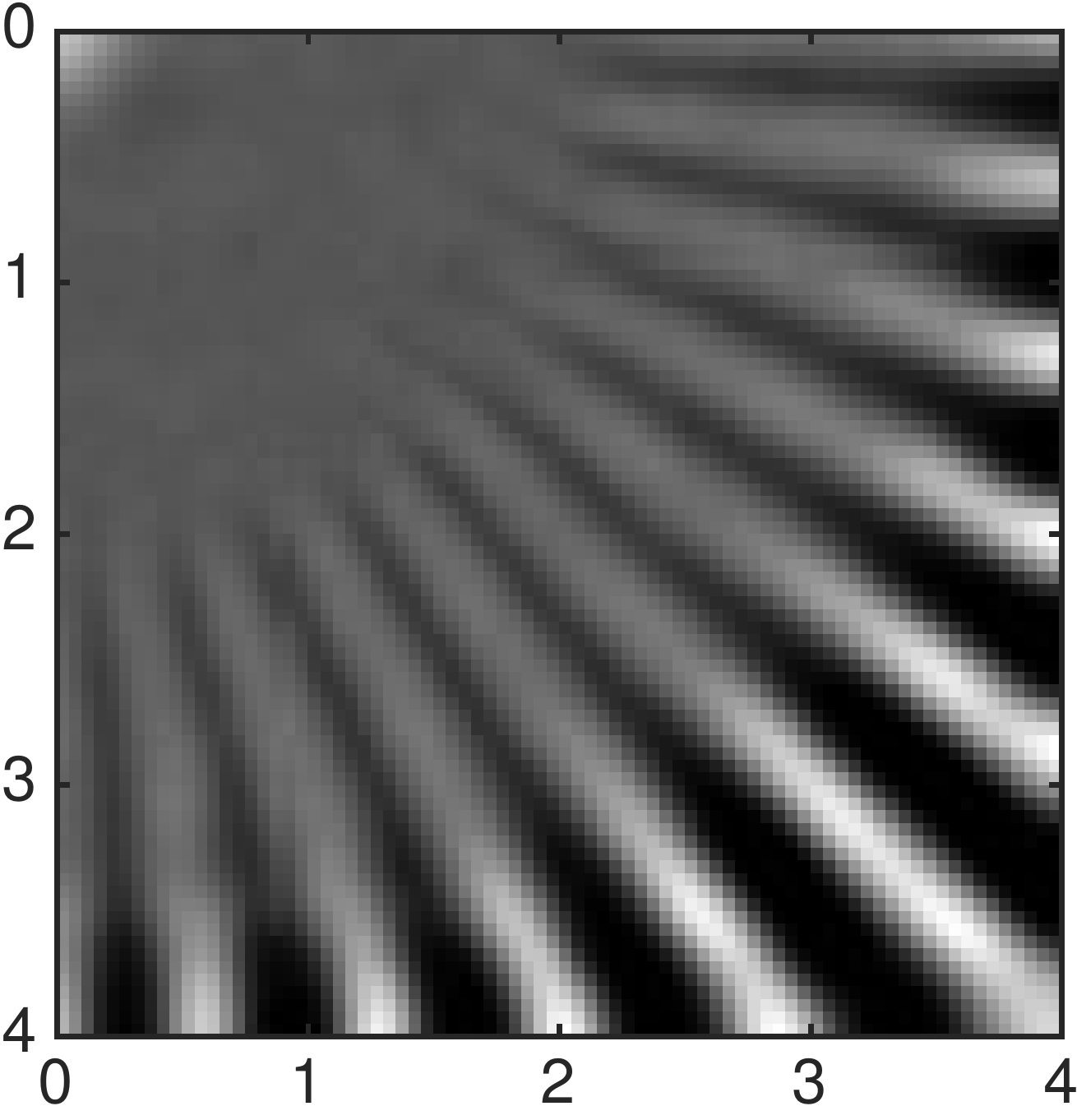}
    &
      \ylabel{\footnotesize   (e) PPDS 50      it.}&\figc[height=0.19\textwidth]{
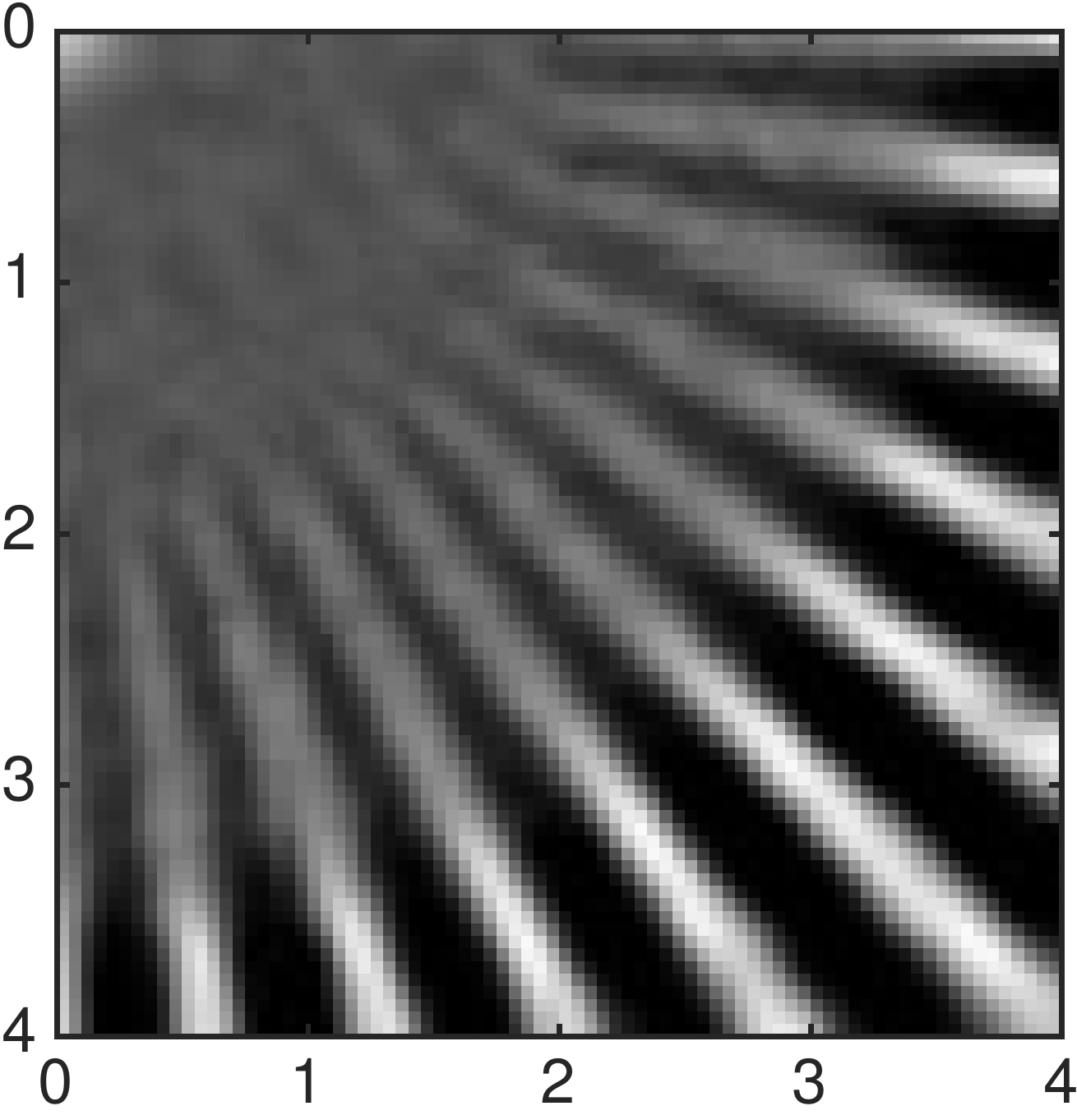}
    \\[.5em]
    \ylabel{\footnotesize   (c) FISTA 1000    it.}&\figc[height=0.19\textwidth]{
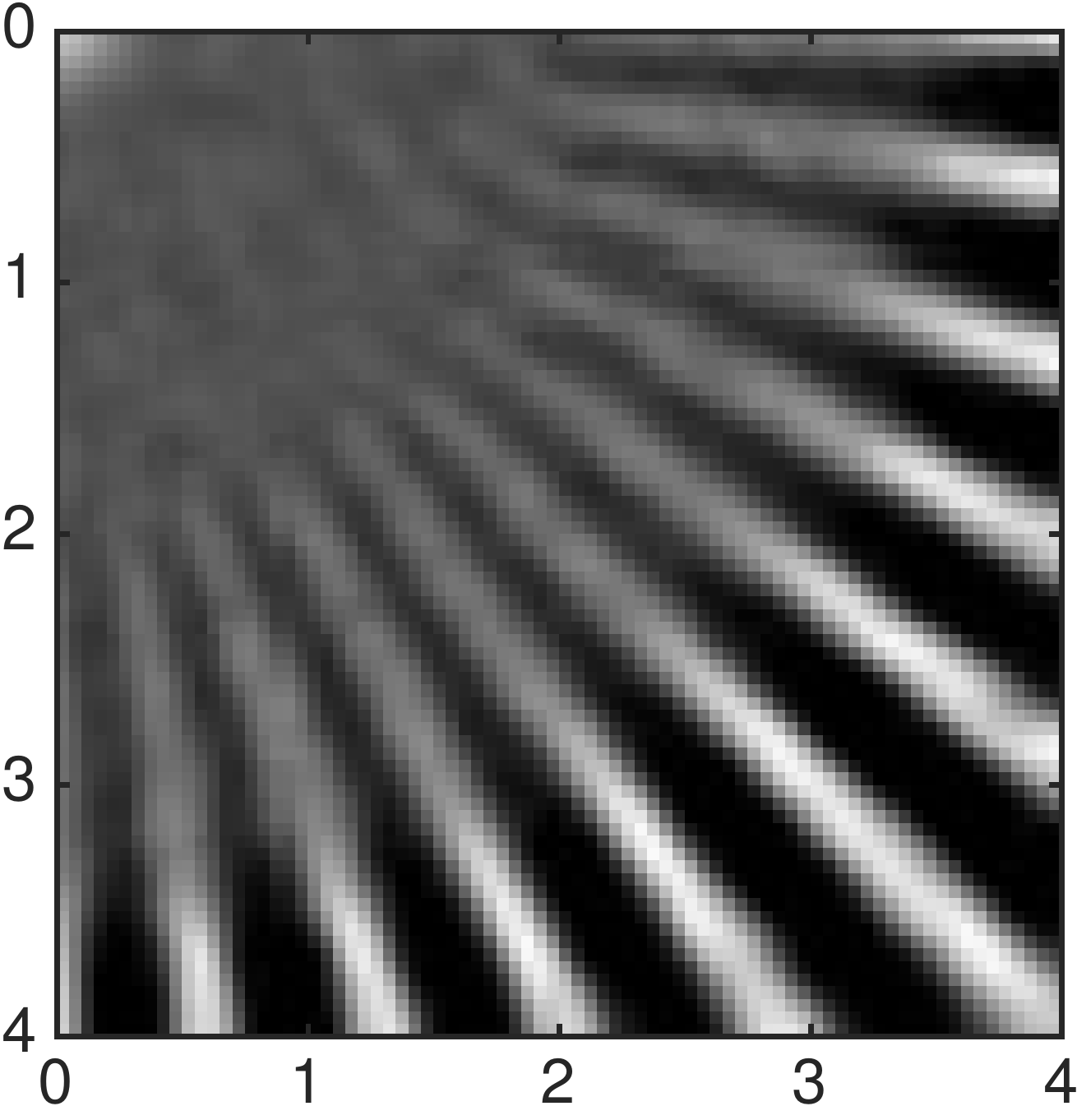}
    &
      \ylabel{\footnotesize   (f) PPDS 1000      it.}&\figc[height=0.19\textwidth]{
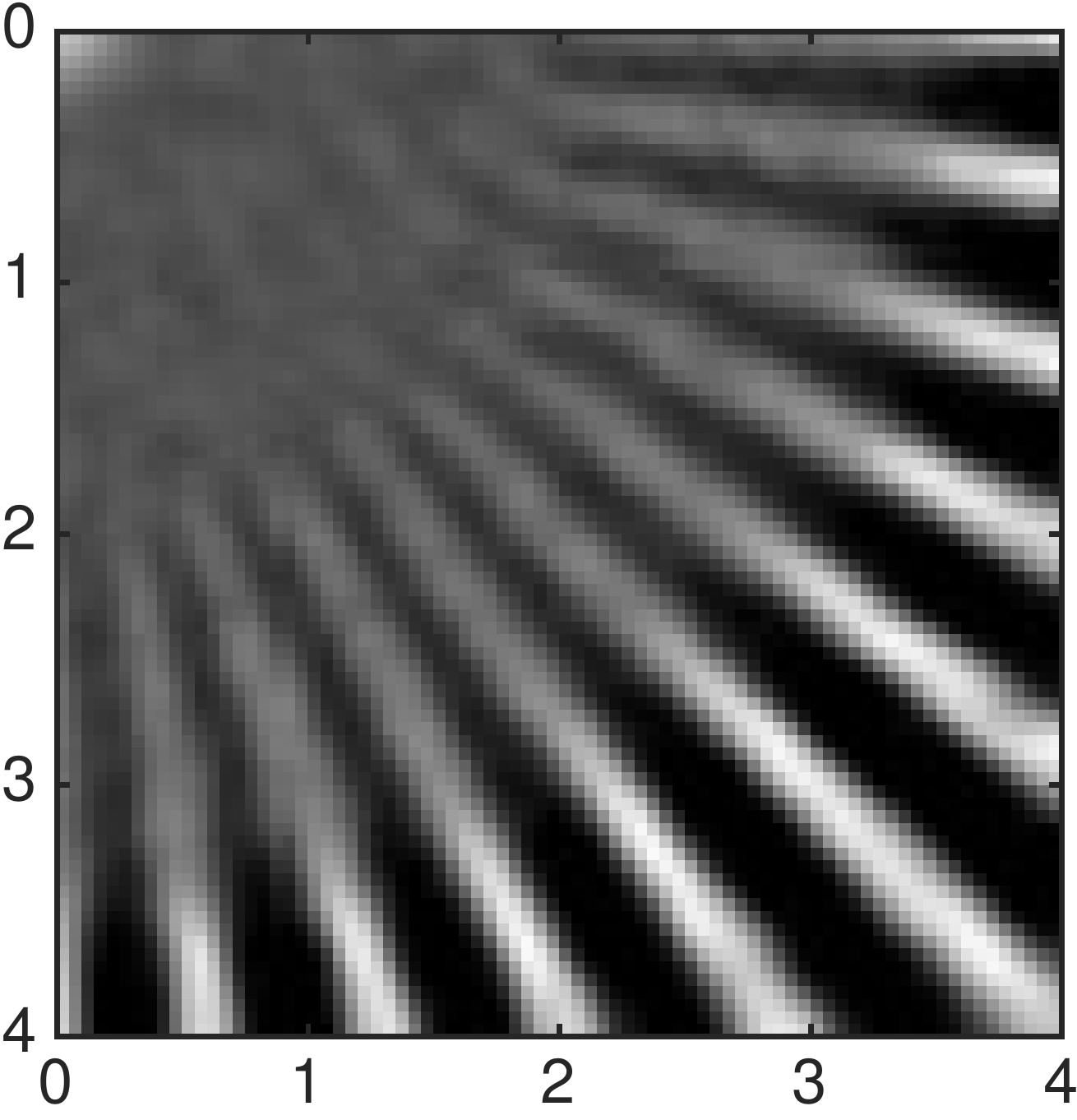}
  \end{tabular}
  \caption{Harmonic joint Blind-SIM reconstruction of the fluorescence
    pattern achieved by the minimization of the criterion 
    \eqref{critere3} with 10, 50 or 1000 FISTA (abc)   or PPDS (def) 
    iterations. 
    For all these simulations, the
    initial-guess is $\qb^{(0)}=\textbf{0}$ and the
    regularization parameters is set to    $(\alpha=0.3,\beta=10^{-6})$.
    The PPDS iteration implements the preconditioner given in \eqref{Precondbis}
    with $\Cb=\Hb^t \Hb$ and $a=1$, see Sec~\ref{blind-SIM_PPDS} for details.
 }
  \label{fig:Algo1}
\end{figure}
\medskip

\subsubsection{Solution of the $m$-th joint Blind-SIM sub-problem}
For the penalized joint Blind-SIM 
problem considered in this paper, the minimization problem 
\eqref{critere3} [equipped with the penalty \eqref{hyperbolic}]  takes 
the form \eqref{critereProx} with  
\begin{subequations}
  \label{critereGen}
  \begin{align}
    \label{critereGen1}
    g(\qb) & \,=\,  || \yb - \Hb \qb||^2 + \beta ||\qb||^2 \\
    \label{critereGen2}  
    h(\qb) & \, = \, \alpha \sum_n  \phi(q_n)
    \intertext{where $\phi:\eR \rightarrow \eR\cup \{+\infty\}$ is such that}
   \label{critereGen3}  
    \phi (u) & \pardef \left\{
    \begin{tabular}{@{\kern1pt}ll}
      $u$ & \quad if $u\geq 0$. \\[-.2em]
      $+\infty$ & \quad otherwise.
    \end{tabular}
  \right.
  \end{align}
\end{subequations}
The gradient of the regular part in the splitting 
\begin{equation}  
  \label{eq:grad}
  \boldsymbol{\nabla} g(\qb) = 2 \left[ \Hb^t(\Hb \qb - \yb)  + \beta \qb\right]
\end{equation}
is $L$-Lipschitz-continuous with 
%
$L = 2 \left(\lambda_{\text{max}}(\Hb^t \Hb) + \beta\right)$
%
where $\lambda_{\text{max}}(\Ab)$ denotes the highest eigenvalue of
the matrix $\Ab$.
Furthermore, the proximity operator \eqref{eq:prox1} with $h$ defined by
\eqref{critereGen2} leads to the well-known \textit{soft-thresholding}
rule  \cite{Moulin99,Figueiredo03}
\begin{equation}  
  \label{eq:prox2}
  \prox{\gamma h}  (\qb) =  \textbf{vect}\left(\max\{q_n - \gamma \alpha,0\}\right). 
\end{equation}
From a practical perspective, both the basic iteration \eqref{algoGP} and its accelerated 
counterpart \eqref{algoFISTA} are easily implemented at a very low computational 
cost\footnote{%
Since $\Hb$ is a convolution matrix, 
the computation of the gradient \eqref{eq:grad} can
be performed by fast Fourier transform and vector dot-products, see for instance
\cite[Sec. 5.2.3]{Vogel02}.
} 
from equations \eqref{eq:grad} and \eqref{eq:prox2}. For our
penalized joint Blind-SIM approach, however, we observed that both 
algorithms exhibit similar convergence behavior in terms of visual
aspect of the current estimate. The convergence speed  is also significantly slow:
several hundreds of iterations are usually required for  solving the $M=200$ sub-problems involved 
in the joint Blind-SIM reconstruction shown in
Fig.~\ref{fig:fig4bis}(B). In addition, Fig.~\ref{fig:Algo1}(a-c)
shows the reconstruction built with ten, fifty and  one thousand FISTA
iterations.  Clearly, we would like that this latter
quality of reconstruction is reached in a reasonable amount of 
time. The next subsection introduces a  \textit{preconditioned 
primal-dual splitting} strategy that achieves a much higher 
convergence speed, as illustrated by Fig.~\ref{fig:Algo1}(right).

\subsection{Preconditioned primal-dual splitting}
\label{PPDS_sec}

{The preconditioning technique \cite[p. 69]{Bertsekas99} is formally equivalent to addressing the initial minimization 
problem \eqref{critereProx} \textit{via} a linear transformation  
}
%
  $\qb : = \Pb \vb$,
%
where  $\Pb \in \eR^{N \times N}$ is a symmetric positive-definite 
matrix. 
There is no formal difficulty in defining a preconditioned
version of the  proximal iteration \eqref{algoGP}. However, if  one
excepts the special case of  diagonal matrices
$\Pb$ \cite{Bonettini09,Pustelnik12,Becker12,RaguetL15},  the
proximity operator of $\Hc(\vb) :=  h(\Pb\vb)$ cannot be obtained
explicitly  and needs to be computed approximately. 
 As a result,
 solving  a nested optimization problem is required \textit{at 
each  iteration}, hence increasing the overall computational cost of 
the algorithm and raising a convergence issue since the sub-iterations 
must be truncated in practice \cite{Becker12,Chouzenoux13}. Despite 
this difficulty, the preconditioning is widely accepted as a very effective 
way for accelerating proximal iterations. In the sequel, the 
versatile primal-dual splitting technique introduced in 
\cite {Condat13,Vu2013,Combettes2012} is used 
to propose a new preconditioned proximal iteration, without any nested optimization problem. 

\begin{figure}[t]
  \centering
    \includegraphics[width=7cm]{
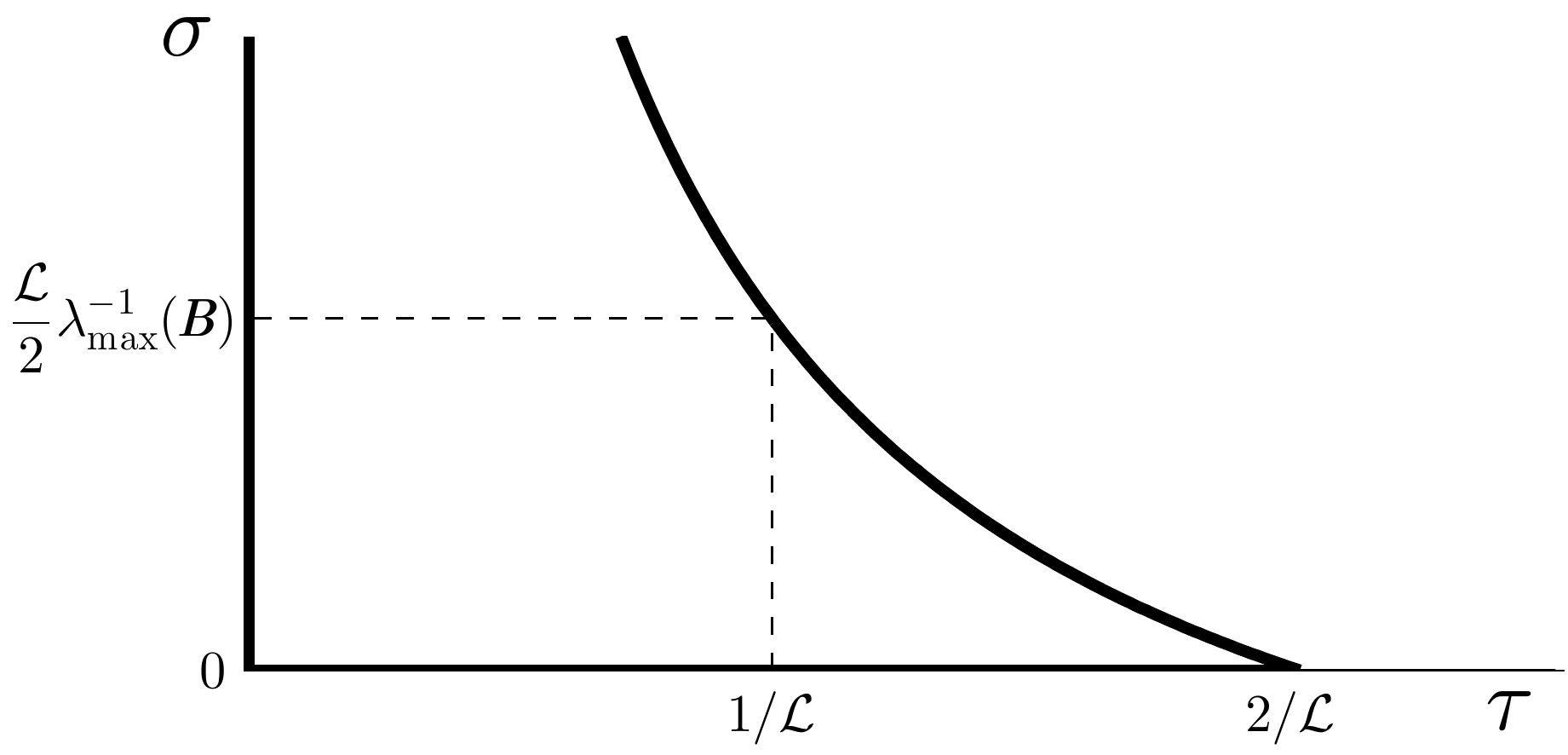}
  \caption{Admissible domain for $(\tau, \sigma)$ ensuring 
    the global  convergence of the PPDS iteration with $\theta
    \in (0;1)$, see Equation \eqref{borne_tau}.}
  \label{domain}
\end{figure}

This new preconditioning technique is now presented for the generic 
 problem \eqref{critereProx}. At first, we express the criterion $f$ with 
respect to the transformed variables
\begin{align}
  \label{criterion_newbasis}
  f(\Pb \vb)  & =  \Gc(\vb) + h(\Pb\vb) 
\end{align}
with $\Gc(\vb) := g(\Pb\vb)$. 
Since the criterion above is a particular case of the form considered 
in \cite[Eq. (45)]{Condat13}, it can be optimized by a primal-dual iteration 
\cite[Eq. (55)]{Condat13} that reads
\begin{subequations}
  \label{iteration1Gen}
    \begin{align}
    \label{iteration1Gena}
      \vb^{(k+1)} & \longleftarrow \vb^{(k)} - \theta \tau \db^{(k)} \\
      \label{iteration1Genb}
      \omegab^{(k+1)} & \longleftarrow  \omegab^{(k)} + \theta \Deltab^{(k)}
    \end{align}
\end{subequations}
with
\begin{subequations}
  \label{iteration2Gen}
    \begin{align}
    \label{iteration2Gena}
      \db^{(k)}  &:= \boldsymbol{\nabla} \Gc(\vb^{(k)}) +  \Pb \omegab^{(k)}\\
      \label{iteration2Genb}
      \Deltab^{(k)}
                 &:= \prox{\sigma h^\star} \big(
    \omegab^{(k)}  + \sigma \Pb (\vb^{(k)} - 2 \tau \db^{(k)})      \big)        - \omegab^{(k)}                
    \end{align}
\end{subequations}
where the proximal mapping applied to $h^\star$, the Fenchel conjugate function for  $h$, 
is easily obtained from
\begin{equation}
\label{iteration4Gen}
    \prox{\sigma h^\star} (\omegab) = \omegab - \sigma \prox{h/\sigma} (\omegab/\sigma).
\end{equation}
The primal update \eqref{iteration1Gena} can also be expressed with respect to 
the \textit{untransformed} variables $\qb$:
\begin{align}
  \label{iterationfinala_oldbasis}
\qb^{(k+1)} & \longleftarrow  \qb^{(k)} - \theta \tau \Bb \zetab^{(k)} 
\end{align}
with $\zetab^{(k)} := \boldsymbol{\nabla} g(\qb^{(k)}) +
\omegab^{(k)}$ and $\Bb := \Pb\Pb$.
Since the update \eqref{iterationfinala_oldbasis} is a \textit{preconditioned} 
primal step, we expect that a clever choice of the preconditioning matrix $\Bb$ will 
provide  a significant acceleration of the primal-dual  procedure.  In addition,  we 
note that the quantity $\ab^{(k)} := \omegab^{(k)}  + \sigma \Pb (\vb^{(k)} - 2 \tau \db^{(k)} )$
involved in the dual step \textit{via} \eqref{iteration2Genb} also  reads 
\begin{align}
\label{les_a}
    \ab^{(k)} &:= \omegab^{(k)} + \sigma ( \qb^{(k)} - 2 \tau    \Bb\zetab^{(k)}).
\end{align}
Hereafter, the primal-dual updating pair   \eqref{iteration1Genb} and \eqref{iterationfinala_oldbasis}  is called a \textit{preconditioned 
primal-dual splitting} (PPDS) iteration. 
Following  \cite[Theorem 5.1]{Condat13}, the convergence of these PPDS
iterations is granted if the following 
conditions are met for the parameters
$(\theta,\tau,\sigma)$:
\begin{subequations}
  \label{conditions}
    \begin{align}
    \label{conditionsa}
    \sigma>0,~\tau>0, ~\theta & >0\\
    \label{conditionsd}  
     \gamma_{\tau,\sigma}       & \in [1;2) \\ 
    \label{conditionsc}  
      \gamma_{\tau,\sigma}  & >\theta 
  \end{align}
\end{subequations}
with
%
   $\gamma_{\tau,\sigma} := 2-\tau \left[ 1 - \tau \sigma \lambda_{\text{max}}(\Bb)\right]^{-1} \Lc/2$, 
%
where $\Lc$ is the Lipschitz-continuity constant of $\boldsymbol{\nabla}
\Gc$, see Eq. \eqref{criterion_newbasis}. Within the convergence
domain ensured by \eqref{conditions}, the practical tuning  of the
parameter set  $(\theta,\tau,\sigma)$ is tedious as it may
impair the convergence speed. 
We propose the following tuning strategy, which appeared to be very
efficient. At first, we note that the step length $\tau$ relates only to the primal 
update \eqref{iteration1Gena} whereas $\sigma$ relates only to the
dual update \eqref{iteration1Genb} \textit{via} $\Deltab^{(k)}$. In addition,  
the relaxation parameter $\theta$ scales both the primal and the dual
steps \eqref{iteration1Gen}.  Considering only under-relaxation  
(\textit{i.e.,} $\theta <1$), \eqref{conditionsc} is
unnecessary and \eqref{conditionsd} is equivalent to 
the following bound 
\begin{equation}
  \label{borne_tau}
  \sigma \leq \overline{\sigma}\quad \text{with} \quad \overline{\sigma} := \big(1/\tau - \Lc/2\big) \, \lambda^{-1}_{\text{max}}(\Bb).
\end{equation}
This relation  defines an admissible domain for $(\tau, \sigma)$
under the condition $\theta < 1$, see Fig.~\ref{domain}. 
Our strategy defines $\tau$ as the \textit{single tuning parameter} of our PPDS
iteration, the parameter $\sigma$ being adjusted so that the dual step
is maximized:
\begin{equation}
\label{borne_rho}
0 <\tau < \overline{\tau}, \quad \sigma = \overline{\sigma} \quad \text{and}
\quad \theta = 0.99,
\end{equation}
with $\overline{\tau}:=2/\Lc$. We set $\theta$ arbitrary close to 1
since practical evidence indicates that under-relaxing $\theta$ slows 
down the convergence rate. The numerical evaluation of the bounds
$\overline{\tau}$ and $\overline{\sigma}$ is application-dependent
since they depend on $\Lc$ and $\lambda_{\text{max}}(\Bb)$. 
%

\subsection{Resolution of the joint Blind-SIM sub-problem}
\label{blind-SIM_PPDS}
For our specific problem, the implementation of the PPDS iteration 
requires first the conjugate function \eqref{iteration4Gen}: with $h$ defined by \eqref{critereGen2}, the Fenchel conjugate 
 is easily found and reads 
\begin{equation}  
  \label{eq:prox3}
  \prox{\sigma h^\star}  (\omegab) =  \textbf{vect} \left(\text{min}\left\{\omega_n ,  \alpha\right\}\right).
\end{equation}
The updating rule for the PPDS iteration then reads
\begin{subequations}
  \label{iterationfinal}
    \begin{align}
    \label{iterationfinala}
    &   \qb^{(k+1)} \longleftarrow \qb^{(k)} - \theta \tau  \Bb \zetab^{(k)}\\ 
    \label{iterationfinalb}  
    &  \omegab^{(k+1)} \longleftarrow \omegab^{(k)} +\theta \Deltab^{(k)}
  \end{align}
\end{subequations}
with $\Deltab^{(k)} =  \textbf{vect}\big(\text{min} \{a_n^{(k)},  \alpha\}\big) - \omegab^{(k)} $
and $a_n^{(k)}$ the $n$th component of the vector $\ab^{(k)}$ defined in \eqref{les_a}.
We note that the positivity constraint is not enforced in the primal update
\eqref{iterationfinala}. Primal feasibility (\textit{i.e.} positivity) therefore
occurs \textit{only} asymptotically thanks to the global 
convergence of the sequence \eqref{iterationfinal} toward 
the minimizer of the functional 
\eqref{critereGen}. Compared to FISTA, this behavior may be 
considered as a drawback of the PPDS iteration. However, we do believe that the ability of the 
PPDS iteration to ``transfer'' the hard constraint from the primal 
to the dual step is precisely the cornerstone of the acceleration
provided by preconditioning. Obviously, such an acceleration 
requires that the preconditioner $\Bb$ is wisely chosen. For our 
joint Blind-SIM problem, the preconditioning matrix  is derived 
from Geman and Yang \textit{semi-quadratic} construction
\cite{Geman95}, \cite[Eq. (6)]{Allain06b} 
\begin{equation}
  \label{Precondbis}
  \Bb = \frac{1}{2}\left( \Cb + \beta\,\Ib_d/a\right)^{-1} 
\end{equation}
where $\Ib_d$ is the identity matrix and $a>0$ is a free parameter 
of the preconditioner. We choose $\Cb$ in the class of positive
semi-definite matrix with a BCCB structure \cite[Sec. 5.2.5]{Vogel02}. This choice enforces that
$\Bb$ is also  BCCB, which considerably helps in reducing the 
computational burden: ($i$) $\Bb$ can be stored efficiently\footnote{%
Any BCCB matrix $\Bb$ reads $\Hb=\Fb^\dagger 
\Lambdab \Fb$ with $\Fb$ the unitary \textit{discrete Fourier
  transform} matrix, '$\dagger$' the transpose-conjugate operator,
and $\Lambdab:=\textbf{Diag}(\widetilde{\bb})$
where  $\widetilde{\bb}:=\textbf{vect}(\widetilde{b}_n)$ are the
eigenvalues of $\Bb$,  see for instance  \cite[Sec. 5.2.5]{Vogel02}. 
As a result, the storage requirement reduces to the storage of $\widetilde{\bb}$.    
} and ($ii$) the matrix-vector product $\Bb\zetab^{(k)}$ in
\eqref{iterationfinala} can be computed with $\Oc(N \log N)$
complexity by  the bidimensional fast Fourier transform (\FFT) algorithm.  
Obviously, if the observation model $\Hb$ is  also a BCCB matrix  
built from the discretized OTF, the choice $\Cb = \Hb^t\Hb$ in \eqref{Precondbis}
leads to $\Bb=(\boldsymbol{\nabla}^2 g)^{-1}$ for  $a=1$. 
Such a preconditioner is expected to bring the fastest asymptotic convergence since it
corrects the curvature anisotropies induced by the regular part $g$ in 
the criterion \eqref{critereProx}. 

The PPDS pseudo-code for solving the
joint Blind-SIM problem is  given in Algorithm~\ref{PseudoCode}. 
%
%
%
\begin{algorithm}[t]
  \hrule\vspace{.5em}
  \textit{Given quantities:} \\
   PSF $\hb$, Dataset $\{\yb_m\}_{m=1}^M$,  Average intensity $\Ib_0\in   \eR^N_+$;\\
  Regularization parameters: $\beta,\alpha \in \eR_+$;\\
  PPDS parameters: $a\in \eR_+$; $\theta \in (0,1)$; $\tau  \in
  (0, 2\Lc)$; $k_{\text{max}} \in \eN$;\\
   Initial guesses: $\{\qb_m^{(0)},\omegab_m^{(0)}\}_{m=1}^M$;\\[.5em]
  \hrule
  \vspace{.5em}
  %
  $\widehat{\rhob} \longleftarrow \textbf{0}$; $\sigma \longleftarrow\overline{\sigma}$ [see \eqref{borne_tau}];\\
  $\widetilde{\hb} \longleftarrow \text{\FFT}(\hb)$;
  $\widetilde{\gammab} \longleftarrow \widetilde{\hb}^* \odot \widetilde{\hb}$;
  $\widetilde{\bb} \longleftarrow (2\widetilde{\gammab} + 2\beta/a)$;\\
  \vspace{.5em}
  // \text{The outer loop: processing each view $\yb_m$...}\\
  \For{$m=1 \cdots M$}{
    \vspace{.5em}
    $\widetilde{\yb} \longleftarrow \text{\FFT}(\yb_m)$;
    $\widetilde{\qb}^{(0)} \longleftarrow \text{\FFT}(\qb_m^{(0)})$;  
    $\widetilde{\omegab}^{(0)}\longleftarrow \text{\FFT}(\omegab_m^{(0)})$\;
    \vspace{.5em}
    // \text{The inner loop: PPDS minimization...}\\
    \nllabel{innerloop}
    \For{$k=0 \cdots k_{\text{max}}$ }{
     
     \vspace{.5em}
      // \text{The primal step (Fourier domain)...}\\
      $\widetilde{\db}^{(k)} \longleftarrow 
      \left( \widetilde{\omegab}^{(k)}  - 2( \widetilde{\hb} \odot\widetilde{\yb} 
        - (\widetilde{\gammab} + \beta)\odot \widetilde{\qb}^{(k)}
     )\right) \oslash \widetilde{\bb}$\;\nllabel{primalstep}
      $\widetilde{\qb}^{(k+1)} \longleftarrow \widetilde{\qb}^{(k)} -      \theta \tau  \widetilde{\db}^{(k)} $\;
      \vspace{.5em}
      // \text{The dual step (direct domain)...}\\
      $\ab^{(k)} \longleftarrow \text{\FFT}^{-1}\left(\widetilde{\omegab}^{(k)} + \sigma ( \widetilde{\qb}^{(k)} - 2 \tau \widetilde{\db}^{(k)})\right)$\;
      $\omegab^{(k+1)} \longleftarrow (1-\theta) \,\omegab^{(k)}      +\theta \,\textbf{vect}(\text{min} \{a_n^{(k)},  \alpha\})$\;
      \vspace{.5em}
     // \text{Prepare next PPDS iteration...}\\
     $\widetilde{\qb}^{(k)} \longleftarrow \widetilde{\qb}^{(k+1)}$; 
      $\widetilde{\omegab}^{(k)}\longleftarrow \text{\FFT}(\omegab^{(k+1)})$\;
    }
    \vspace{.5em}
    // \text{Building-up the joint Blind-SIM estimate...}\\
    $\widehat{\rhob} \longleftarrow \widehat{\rhob} + \frac{1}{M}\text{\FFT}^{-1}  ( \widetilde{\qb}^{(k)} ) \oslash \Ib_0$\;
  }\vspace{.5em}
  \hrule\vspace{.5em}
  \textit{Final result:} 
  The joint Blind-SIM estimate is stored in $\widehat{\rhob}$\\[.5em]
  \hrule
\vspace{1em}
 \caption{Pseudo-Code of the joint Blind-SIM PPDS algorithm, assuming
   that $\Hb$ is a BCCB matrix and   $\Cb=\Hb^t\Hb$. The symbols $\odot$ and $\oslash$   
   are the component-wise product and division, respectively. For the sake of simplicity, this pseudo-code implements 
a very simple stopping rule based on a maximum number of minimizing 
steps, see line~\ref{innerloop}. In practice, a more 
elaborated stopping rule could be used by monitoring the 
norm $||\zetab^{(k)}||$ defined by \eqref{iterationfinala_oldbasis} since it 
tends towards 0 as $\qb^{(k)}$ asymptotically reaches the constrained 
minimizer of the $m$-th nested problem.}
\label{PseudoCode}
\end{algorithm}
This pseudo-code 
requires that $\Lc$ and $\lambda_{\text{max}}(\Bb)$  are given for 
the tuning \eqref{borne_rho}: we get
\begin{subequations}
  \label{tuningBSJ}
    \begin{align}
      \label{specRadius}
   \lambda_{\text{max}}(\Bb) \,= \, 1/\lambda_{\text{min}}(\Bb^{-1}) \, =  \, a\,(2\beta)^{-1} 
    \intertext{since $\Hb$ is \textit{rank deficient} in our context,
      and the Lipschitz constant that reads $\Lc =
      \lambda_{\text{max}}(\Bb  \, \boldsymbol{\nabla}^2 g)$ can be further simplified as}
  \label{LipschitzPPDS}
    \Lc = \left\{
  \begin{array}{ll}
     a & \text{if $a\geq 1$}\\
     {\displaystyle 
     (\widetilde{\gamma}_{\text{max}} +    \beta)(\widetilde{\gamma}_{\text{max}} + \beta/a)^{-1}
    } & \text{otherwise},
  \end{array}
  \right.
  \end{align}
\end{subequations}
with $\widetilde{\gamma}_{\text{max}}$ the maximum of the square magnitude 
of the OTF components. 
From the pseudo-code, we also note that the computation of the primal update \eqref{iterationfinala} 
remains in the Fourier domain during the PPDS iteration, see line~\ref{primalstep}. 
With this strategy (possible because $\boldsymbol{\nabla} g$ is a
linear function), the computational  burden per PPDS 
iteration\footnote{The MATLAB  implementation of the PPDS pseudo-code 
 Algorithm~\ref{PseudoCode} requires less than 6 ms per iteration on a
 standard laptop (Intel Core M 1.3~GHz). For the sake of comparison,
 one FISTA iteration takes almost 5 ms on the same laptop.} is
dominated by one single forward/inverse \FFT pair, \textit{i.e.,} 
PPDS and FISTA have equivalent computational burden per 
iteration. 
\begin{figure}[t]
  \centering
  \begin{tabular}{c@{ }c}
    \ylabel{\small  $f\left(\qb^{(k)}\right) -f\left(\widehat{\qb}\right)$}&
                                                                             \figc[width=7.5cm]{
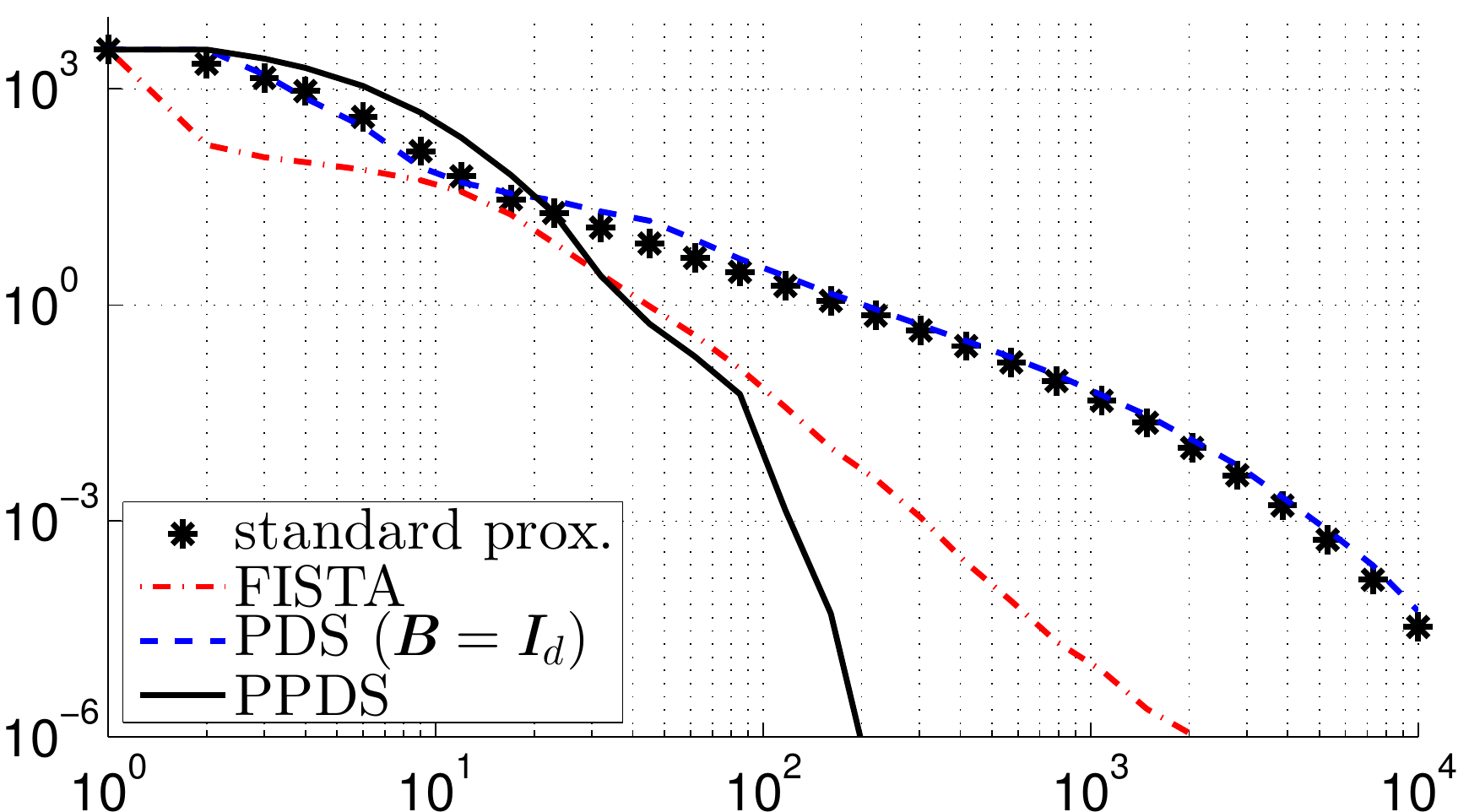}
    \\
    \ylabel{\small  $||\qb^{(k)} -\widehat{\qb}||$}&
                                                     \figc[width=7.5cm]{
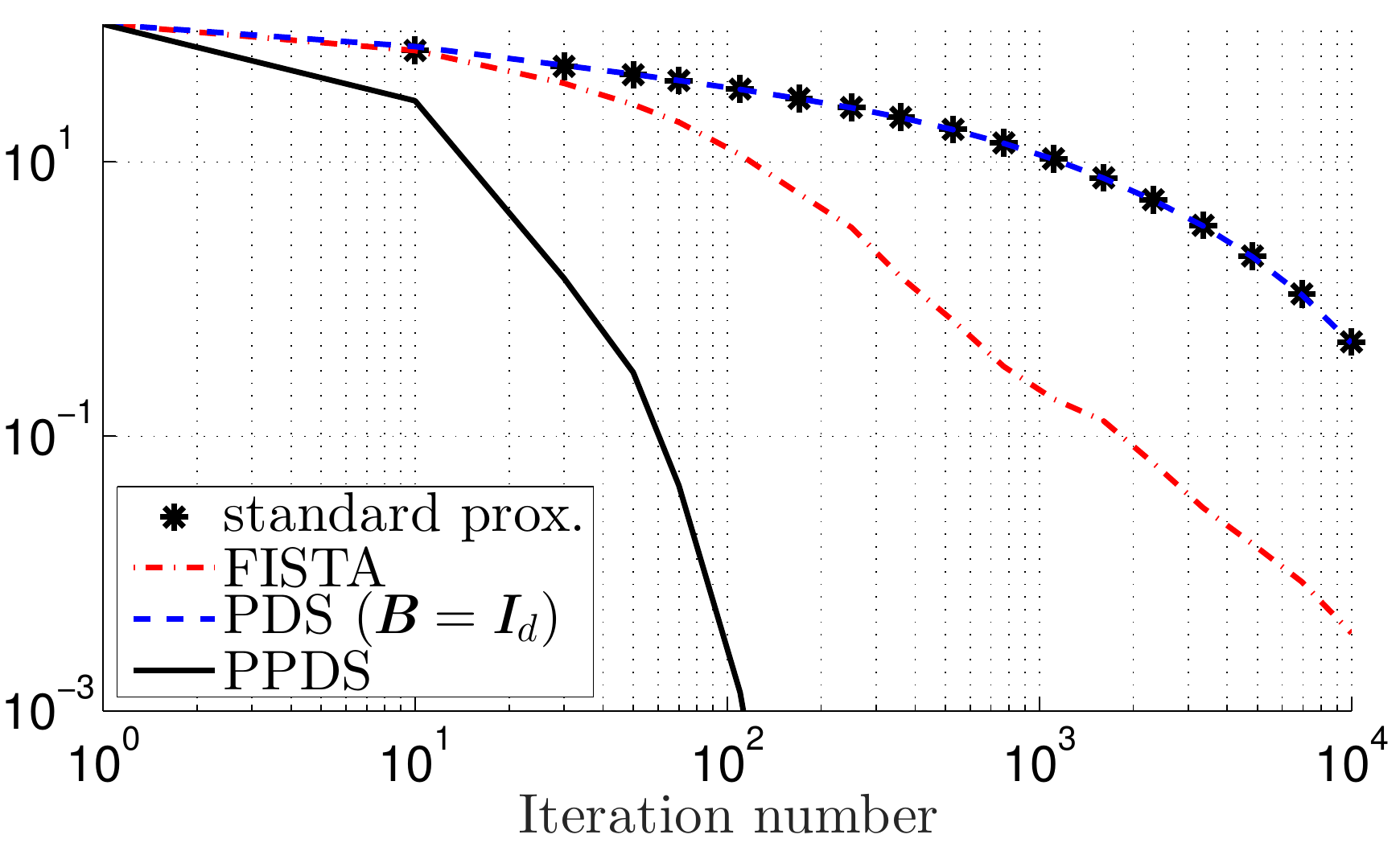}
  \end{tabular}
  \caption{%
  Criterion value (upper plots) and distance to the minimizer (lower
  plots) as a function of the PPDS iterations for the reconstruction 
  problem considered in Fig.~\ref{fig:Algo1}.
   The   chosen initial-guess is $\qb^{(0)}=\textbf{0}$ for the primal 
   variables and $\omegab^{(0)}=-\boldsymbol{\nabla} g(\qb^{(0)})$ for
   the dual variables. The preconditioning parameter is set to  $a=1$  
   and $(\theta,\tau,\sigma)$ were set according to the tuning rule 
   \eqref{borne_rho}.
   For the sake of completeness, the curve of the FISTA iterations and 
   the PDS iterations (\textit{i.e.,} the PPDS equipped  with the
   identity preconditioning matrix $\Bb=\Ib_d$) 
   are also reported. 
}
  \label{fig:precondEffect}
\end{figure}

We now illustrate the performance of the PPDS iterations for minimizing the penalized criterion involved in the joint Blind-SIM reconstruction problem shown
in Fig.~\ref{fig:Algo1}-(right). These simulations were performed with
a standard MATLAB implementation of the pseudo-code shown in
Algorithm~\ref{PseudoCode}. We set $a=1$ so that the preconditioner 
$\Bb$ is the inverse of the Hessian of $g$ in \eqref{critereGen}. With this 
tuning, we expect that the PPDS iterations exhibit a very
favorable convergence rate as long as the set of active 
constraints is correctly identified.  Starting from initial guess $\qb^{(0)}=\textbf{0}$ (the dual variables being set 
accordingly to $\omegab^{(0)}=-\boldsymbol{\nabla} g(\qb^{(0)})$,
see for instance \cite[Sec. 3.3]{Bertsekas99}), the criterion value of the PPDS
iteration depicted in Fig.~\ref{fig:precondEffect} exhibits an asymptotic convergence rate that can be
considered as super-linear. Other tunings for $a$ (not shown here) were 
   tested and found to slow down the convergence speed.  
The pivotal role of the preconditioning in the convergence speed is
also underlined since the PPDS algorithms becomes as slow as the 
standard proximal iteration when we set $\Bb=\Ib_d$, see the ``PDS'' curve in Fig.~\ref{fig:precondEffect}.
In addition, one can  note from the reconstructions shown in Fig.~\ref{fig:Algo1} 
that the high-frequency components (\textit{i.e.,} the SR effect)
are brought in the very early iterations. Actually, once PPDS is properly 
tuned, we always found that it offers a substantial acceleration with 
respect to the FISTA (or the standard proximal) iterates. 

Finally, let us remind that the numerical simulations were performed with a BCCB convolution matrix $\Hb$. 
In some cases, the implicit periodic boundary assumption\footnote{%
Let us recall that the matrix-vector multiplication $\Hb\qb$ with 
$\Hb$ a BCCB matrix corresponds to the \textit{circular} convolution 
of $\qb$ with the convolution kernel that defines $\Hb$.} enforced 
by such matrices is not appropriate and a convolution model 
with a zero boundary assumption is preferable, 
which results in a matrix $\Hb$ with a BTTB structure. 
In such a case, the product of any vector by $\Hb^t\Hb$ can still be performed efficiently in $O(N\log N)$ \textit{via} the
FFT algorithm, see for instance \cite[Sec. 5.2.3]{Vogel02}. This
applies to the computation of $\boldsymbol{\nabla} g(\qb^{(k)})$ in
the primal step \eqref{iterationfinala_oldbasis}, according to
\eqref{eq:grad}. In contrast, exact system solving as required by
\eqref{iterationfinala_oldbasis} cannot be implemented in $O(N\log N)$
anymore if matrix \Hb is only BTTB (and not BCCB). In such a
situation, one can define $\Cb$ as a BCCB approximation of 
$\Hb^t\Hb$, so that the preconditioning matrix $\Bb=(\Cb + \beta
\Ib_d)^{-1}$ remains BCCB, while ensuring that $\Bb (\Hb^t\Hb + \beta
\Ib_d)$ has a clustered spectrum around 1 as the size $N$ increases 
\cite[Th. 4.6]{Chan96}.

Finally, another practical issue arises from the numerical evaluation of $\Lc$. No direct extension of \eqref{LipschitzPPDS} is available when \Hb is BTTB but not BCCB. However, according to \eqref{borne_rho}, global convergence of the PPDS iterations is still granted if $\tau<2/\widehat{\Lc}$ with $\Lc\leq \widehat{\Lc}$. For instance, $\widehat{\Lc}\pardef\lambda_{\text{max}}(\Bb) \, (||\Hb||_\infty ||\Hb||_1  +  \beta)$ is an easy-to-compute upper bound of \Lc.
%
%


\section{Conclusion}
\label{conclusion}

The speckle-based fluorescence microscope proposed in
\cite{Mudry12} holds the promise of a super-resolved  optical
imager that is cheap and easy to use.  The SR mechanism 
behind this strategy, that was not explained, is now properly 
linked with the sparsity of the illumination
patterns. This readily relates joint Blind-SIM to localization
microscopy techniques such as PALM\cite{betzig06} where 
the image sparsity is indeed brought by the sample itself. 
This finding also suggests that ``optimized'' random patterns 
can be used to enhance SR, one example  
being the two-photon excitations proposed in this paper.
{Obviously, even with such excitations, the massively 
sparse activation process at work with  PALM/STORM remains 
unparalleled and one may not expect a resolution 
with joint Blind-SIM greater than twice or three times the resolution 
of a wide-field microscope.  
}
We note, however, that this analysis of the SR 
mechanism is only valid when the sample and the illumination 
patterns are \textit{jointly} retrieved. In other words, this article 
does not tell anything about the SR obtained 
from \textit{marginal} estimation techniques that estimates 
the sample only, see for instance 
\cite{Min13,Oh13,Chaigne16}. Indeed, the SR properties of 
such ``marginal'' techniques are rather distinct  
\cite{Idier15a}.

From a practical perspective, the joint Blind-SIM strategy should be 
tested shortly with experimental datasets.  One expected difficulty arising in the processing of real data 
is the strong background level induced in the focal plane by the 
out-of-focus light. This phenomenon prevents the local extinction of the excitation
intensity, hence destroying the expected SR in joint Blind-SIM. 
A natural approach would be to solve the reconstruction problem 
in its 3D structure, which is numerically challenging, but remains a 
mandatory step to  achieve 3D speckle SIM reconstructions \cite{Negash16}.     
The modeling of the out-of-focus background with a very smooth
function is possible \cite{Orieux12} and will be considered for a fast 
2D reconstruction of the sample in the focal plane. 

Another important motivation of this work is the reduction of the
computational time in joint Blind-SIM reconstructions. The
reformulation of the original (large-scale) minimization 
problem is a first pivotal step as it leads to $M$ sub-problems, all 
sharing the same structure, see Sec.~\ref{Reformulationbis}. The 
new preconditioned proximal iteration proposed in Sec.~\ref{PPDS_sec} 
is also decisive as it efficiently tackles each sub-problem. In our
opinion, this ``preconditioned primal-dual splitting'' (PPDS) technique is
of general interest as it yields preconditioned proximal iterations
that are easy to implement and provably convergent.  For our specific problem, the criterion 
values are found to converge much faster 
with the PPDS iteration than with the standard proximal iterations 
(\textit{e.g.}, FISTA). We do believe, however, that PPDS deserves 
further investigations, both from the theoretical and the experimental 
viewpoints. This minimization strategy should be tested with other 
observation models and prior models. For example, as a natural
extension of this work,  we will consider shortly the Poisson
distribution in the case of image  acquisitions with low 
photon counting rates. The global and local convergence properties 
of PPDS should be explored extensively, in particular when the preconditioning
matrix varies over the iterations. This issue is of importance if one 
aims at defining quasi-Newton proximal iterations with PPDS in 
a general context.

\section*{acknowledgments}
{The authors are grateful to the anonymous reviewers for their 
      valuable comments, and to Christophe Leterrier for the STORM image used in
      Section~\ref{penalize}.}

\bibliographystyle{ieeeji}
\bibliography{fiche}

\end{document}